\documentclass[reqno,11pt]{amsart}
\usepackage[utf8]{inputenc}
\usepackage{enumitem}
\usepackage{graphicx}
\usepackage{amscd}
\usepackage{slashed}
\usepackage{amssymb}
\usepackage{mathtools} 
\usepackage{pstricks}
\usepackage[mathscr]{eucal}
\numberwithin{equation}{section}
\allowdisplaybreaks[1]

\title[Linearized Fields for Causal Variational Principles]{Linearized Fields for Causal Variational Principles: Existence Theory and Causal Structure}

\author[C.\ Dappiaggi]{Claudio Dappiaggi}
\address{Dipartimento di Fisica \\ Universit{\`a} degli Studi di Pavia\\ and INFN, Sezione di Pavia \\ Via Bassi, 6 --  I-27100 Pavia \\ Italy}
\email{claudio.dappiaggi@unipv.it}

\author[F.\ Finster]{Felix Finster \\ \\ November 2018 / November 2019}
\address{Fakult\"at f\"ur Mathematik \\ Universit\"at Regensburg \\ D-93040 Regensburg \\ Germany}
\email{finster@ur.de}

\newtheorem{Def}{Definition}[section]
\newtheorem{Thm}[Def]{Theorem}
\newtheorem{Prp}[Def]{Proposition}
\newtheorem{Lemma}[Def]{Lemma}
\newtheorem{Remark}[Def]{Remark}
\newtheorem{Corollary}[Def]{Corollary}
\newtheorem{Example}[Def]{Example}

\newcommand{\Thanks}{\vspace*{.5em} \noindent \thanks}
\newcommand{\beq}{\begin{equation}}
\newcommand{\eeq}{\end{equation}}
\newcommand{\Proof}{\begin{proof}}
\newcommand{\QED}{\end{proof} \noindent}
\newcommand{\QEDrem}{\ \hfill $\Diamond$}

\newcommand{\la}{\langle}
\newcommand{\ra}{\rangle}
\newcommand{\bra}{\mathopen{<}}
\newcommand{\ket}{\mathclose{>}}

\newcommand{\C}{\mathbb{C}}
\newcommand{\R}{\mathbb{R}}
\newcommand{\1}{\mbox{\rm 1 \hspace{-1.05 em} 1}}

\newcommand{\N}{\mathbb{N}}
\renewcommand{\H}{\mathscr{H}}

\newcommand{\F}{{\mathscr{F}}}

\newcommand{\D}{{\mathscr{D}}}

\renewcommand{\L}{{\mathcal{L}}}
\newcommand{\Sact}{{\mathcal{S}}}
\newcommand{\shield}{s}
\newcommand{\s}{{\mathfrak{s}}}

\newcommand{\loc}{\text{\rm{loc}}}
\renewcommand{\sc}{\text{\rm{sc}}}
\newcommand{\fsc}{\text{\rm{fsc}}}
\newcommand{\psc}{\text{\rm{psc}}}
\newcommand{\scrM}{\myscr M}
\newcommand{\scrN}{\myscr N}
\newcommand{\J}{\mathfrak{J}}
\newcommand{\K}{{\mathfrak{K}}}
\newcommand{\Jtest}{\mathfrak{J}^\text{\rm{\tiny{test}}}}
\newcommand{\Jvary}{\mathfrak{J}^\text{\rm{\tiny{vary}}}}
\newcommand{\Jlin}{\mathfrak{J}^\text{\rm{\tiny{lin}}}}
\newcommand{\Ctest}{C^\text{\rm{\tiny{test}}}}
\newcommand{\Gdiff}{\Gamma^\text{\rm{\tiny{diff}}}}
\newcommand{\Gtest}{\Gamma^\text{\rm{\tiny{test}}}}
\newcommand{\Jdiff}{\mathfrak{J}^\text{\rm{\tiny{diff}}}}
\newcommand{\tmax}{{t_{\max}}}
\newcommand{\tmin}{{t_{\min}}}
\newcommand{\bitem}{\begin{itemize}[leftmargin=2.5em]}
\newcommand{\eitem}{\end{itemize}}
\newcommand{\itemD}{\item[{\raisebox{0.125em}{\tiny $\blacktriangleright$}}]}

\DeclareMathOperator{\norm}{| \hspace*{-0.1em}| \hspace*{-0.1em}|}

\DeclareFontFamily{OT1}{rsfso}{}
\DeclareFontShape{OT1}{rsfso}{m}{n}{ <-7> rsfso5 <7-10> rsfso7 <10-> rsfso10}{}
\DeclareMathAlphabet{\myscr}{OT1}{rsfso}{m}{n}

\DeclareMathOperator{\supp}{supp}

\renewcommand{\u}{\mathfrak{u}}
\renewcommand{\v}{\mathfrak{v}}
\newcommand{\w}{\mathfrak{w}}

\begin{document}

\maketitle

\begin{abstract}
The existence theory for solutions of the linearized field equations
for causal variational principles is developed.
We begin by studying the Cauchy problem locally in lens-shaped regions,
defined as subsets of space-time which admit foliations by surface layers
satisfying hyperbolicity conditions.
We prove existence of weak solutions and show uniqueness
up to vectors in the orthogonal complement of the jets used for testing.
The connection between weak and strong solutions is analyzed.
Global solutions are constructed by exhausting space-time by lens-shaped regions.
We construct advanced and retarded Green's operators and study their properties.
\end{abstract} 

\tableofcontents

\section{Introduction}
This paper is concerned with the initial value problem in the theory of causal fermion systems
(for a general introduction to causal fermion systems and the physical context see the textbook~\cite{cfs}
or the survey article~\cite{dice2014}).
The basic object in this theory is the {\em{universal measure}},
being a measure on a set of linear operators on a Hilbert space.
The physical equations are formulated via a variational principle
for this measure, the {\em{causal action principle}}.
Accordingly, the {\em{initial value problem}} consists in finding
a minimizing measure subject to the constraints imposed by the initial data.
Since the universal measure describes space-time as well as all structures therein,
this initial value problem can be understood in analogy to general relativity:
it involves constructing the space-time geometry and the matter fields in one step.
Due to the nonlinearity of the interaction as described by the causal action
principle, this problem is very difficult. The only results in this direction
are the existence and uniqueness theorems in~\cite{cauchy}
which, however, seem too abstract
for getting a direct connection or seeing the analogy to the initial value problem for hyperbolic 
partial differential equations (PDEs).

Here we are more modest and merely consider the initial value problem for the
{\em{linearized}} field equations of a causal fermion system.
In the analogous setting of general relativity, this linearization corresponds
to studying the initial value problem for linearized fields 
(like for example the Maxwell field, linearized gravity or the Dirac field) in a given space-time geometry.
In the setting of causal fermion systems, such a linearized field is described by a so-called
{\em{jet}}~$\v=(b,v)$,
which consist of a scalar function~$b$ and a vector field~$v$
(for details see~\eqref{Jdef} in Section~\ref{secEL} below). The jet formalism was introduced in~\cite{jet}
for {\em{causal variational principles}}, which are a mathematical generalization of
the causal action principle. For convenience and for the sake of larger generality,
here we also work in the setting of causal variational principles
(the necessary preliminaries will be given in Section~\ref{secprelim}).
The main objective of the present paper is to show that energy methods for hyperbolic
PDEs can be adapted to the setting of causal variational principles such as to obtain
existence and uniqueness results for solutions to the initial value problem
for linearized fields. Moreover, we prove that the linearized fields propagate with finite
speed. We also analyze the resulting causal structure.

We note that our methods and results for linearized fields are also a good starting point for
tackling the nonlinear problem. Namely, the existence theory for the linearized field equations
opens the door for also adapting nonlinear methods from PDEs
(like fixed-point methods for nonlinear symmetric hyperbolic systems~\cite{taylor3}
or methods developed for the Einstein equations~\cite{choquet, rendall, ringstroem}).
Moreover, our results for linearized fields can be applied directly to the general perturbation expansion
for the universal measure as developed in~\cite{perturb}.
Indeed, this perturbative description, which resembles the Feynman diagram expansion of quantum field theory,
relies heavily on Green's operators for the linearized fields.
In the present paper, we shall prove under general assumptions that these Green's operators exist
and have all the properties needed for the perturbative treatment.

The analogy between linear hyperbolic PDEs and the linearized field equations
for causal variational principles deserves a few general words.
The linearized field equations take the form
\beq \label{Delvw}
\Delta \v = \w \:,
\eeq
where~$\w$ is a given inhomogeneity, and the operator~$\Delta$ is defined by
\[ \Delta \v(x) = \nabla \bigg( \int_M \big( \nabla_{1, \v} + \nabla_{2, \v} \big) \L(x,y)\: d\rho(y) - \nabla_\v \,\s \bigg) \:, \]
where~$\L$ is the Lagrangian of the causal variational principle,
space-time~$M$ is defined as the support of the universal measure~$\rho$
(for details see Section~\ref{seccvp}), $\s$ is a positive parameter,
and the jet derivative~$\nabla$ is a combination of
multiplication and directional derivative (for details see Sections~\ref{secEL} and~\ref{seclinfield}).
One should keep in mind that these equations are not differential equations,
but instead they are nonlocal equations involving space-time integrals of specific integral kernels.
The reason why, despite these major structural differences, methods of hyperbolic PDEs
are applicable is that there are positive energies which can be controlled in time by suitable
energy estimates. Once these positive energies have been identified, we can closely follow
the procedure for linear symmetric hyperbolic systems as introduced in~\cite{friedrichs} 
(see also the textbooks~\cite{courant+hilbert2, john, taylor3, intro}
or similarly in globally hyperbolic space-times~\cite{rendall, ringstroem, baergreen}).

We now explain our constructions and results more concretely.
Recall that, in order to set up the initial value problem for a linear hyperbolic PDE in a Lorentzian
space-time~$(\scrM, g)$, one chooses a smooth family~$(\scrN_t)_{t \in [t_0, \tmax]}$
of space-like hypersurfaces, which can be thought of as the surfaces of constant time~$t$
of an observer. Given initial data on~$\scrN_{t_0}$, one seeks for
a solution of the linear hyperbolic equation in the space-time
region~$L := \cup_{t \in [t_0, \tmax]} \scrN_t \subset \scrM$
(see for example the textbooks~\cite[Section~5.3]{john}, \cite[Section~16]{taylor3},
\cite[Section~8.3]{rendall} or~\cite[Chapter~11]{intro}).
The family~$(\scrN_t)_{t \in [t_0, \tmax]}$ is sometimes referred to as a foliation
of the lens-shaped region~$L$.
In the setting of causal variational principles, the situation is more intricate for two reasons:
First, space-time could be discrete, in which case an above foliation does not exist.
Second, it is not clear what an integral over a hypersurface should be, making it impossible
to work with function spaces at fixed times.
The method to overcome these difficulties is to replace hypersurfaces by
so-called surface layers, as we now explain. In the above example of a Lorentzian space-time,
we can introduce functions~$\eta_t$ as the characteristic functions of the past of~$\scrN_t$.
Then their derivative~$\partial_t \eta_t$ is a $\delta$-distribution supported on
the surface~$\scrN_t$. Likewise, integrals over~$\scrN_t$ can be written as
space-time integrals involving the distribution~$\partial_t \eta_t$.
In the setting of causal variational principles, on the other hand,
space-time~$M$ is by definition the support of the
universal measure~$\rho$. We choose 
a family of non-negative functions~$(\eta_t)_{t \in [t_0, \tmax]}$
defined in a space-time region~$U \subset M$. These functions should be
equal to one in the past and equal to zero in the future, interpolating smoothly
between these two values in a neighborhood of a hypersurface.
Moreover, we assume that the ``time'' derivative~$\theta_t := \partial_t \eta_t$ exists
and is non-negative. Then the function~$\theta_t$ is supported in a neighborhood of the hypersurface.
Using a notion first introduced in~\cite{noether}, we refer to the support of~$\theta_t$
as a {\em{surface layer}}. The integral
\[ \int_U \theta_t(x)\: \cdots\: d\rho(x) \]
should be thought of as the generalization of an integral over a hypersurface
to the setting of causal variational principles. The integral is not localized on a hypersurface,
but instead it is ``smeared out'' in a small ``time strip'' around the hypersurface.
This picture is made precise by the notion of a {\em{local foliation}}
by surface layers (see Definition~\ref{deflocfoliate}).

Working with surface layers is well-suited to our problem also because the above ``time strips''
reflect the nonlocality of the linearized field equations. Moreover, integral estimates in
``time strips'' harmonize with the conservation laws for {\em{surface layer integrals}}
as found in~\cite{jet, osi} (for basics see Section~\ref{secosi}). In order for these structures
to fit together even better, we here write the surface layer integrals as
\[ \int_U \eta_t(x)\: d\rho(x) \int_U \big(1-\eta_t(y) \big)\: d\rho(y) \; \big( \cdots \big) \L(x,y)\:, \]
where~$(\cdots)$ is a differential operator involving the jets.

Working in the above setup with suitable energies and imposing corresponding {\em{hyperbolicity conditions}},
we obtain energy estimates which in turn give rise to the desired existence and uniqueness results.
We consider two alternative energies. The first energy is the surface layer inner product~$(.,.)^t$
introduced in~\cite{osi} (see Section~\ref{secenes1}). This choice is motivated from the physical
applications in which~$(.,.)^t$ gives the scalar product of quantum theory (see~\cite{action, fockbosonic}).
The second energy, denoted by~$\la .,. \ra_{[t_0, t]}$,
arises in the study of second variations~\cite{positive} (see Section~\ref{secenes2}).
While it does not have an immediate physical interpretation, it has the advantage that it is positive
as a consequence of the mathematical structure of causal variational principles.
In both cases, the hyperbolicity condition is stated as a positivity property of the
respective energy (see Definitions~\ref{defhypcond} and~\ref{defhypcond2}).
A {\em{lens-shaped region}}~$L$ is defined as a subset of space-time which
admits a local foliation by surface layers which satisfies one of the alternative hyperbolicity conditions
(see Definition~\ref{deflens}).
In a lens-shaped region, we set up the Cauchy problem and prove uniqueness
(see Proposition~\ref{prpunique}).
Moreover, we introduce the notion of a {\em{weak solution}}, defined by the equation
\[ \la \Delta \u, \v \ra_{L^2(L)} = \la \u, \w \ra_{L^2(L)} \]
which must hold for all test jets~$\u$ in a suitable jet space (for details see Section~\ref{secweak}).
We prove existence of weak solutions (Theorems~\ref{thmexist} and~\ref{thmexistpm}).

Our uniqueness statement for weak solutions requires an explanation. As mentioned above,
we want to allow for the possibility that space-time is discrete or has some other, yet unknown
microstructure. In such situations, the hyperbolicity conditions mentioned above are typically known to be
satisfied only on the macroscopic scale, i.e.\ for jets which are almost constant on the microscopic scale
and thus do not ``see'' the unknown microstructure.
This concept can be made precise in the weak formulation by testing only with jets which are almost
constant on the microscopic scale. In order to allow for such situations, we do not specify the
jet space used for testing. In particular, we do not assume that the test jets are dense in~$L^2(L)$.
As a consequence, the weak equation determines the solution only up to vectors in the orthogonal
complement of the test jets. Except for this obvious freedom, weak solutions are unique (see
Proposition~\ref{prpnonunique}).

With the methods and results explained so far, one can solve the Cauchy problem ``locally'' in a lens-shaped region.
In order to construct global solutions, one must extend local
solutions and prove that the resulting globally defined jets satisfy the linearized field equations.
Our method for extending a solution~$\v$ from~$L$ to~$\tilde{L} \supset L$ is to enlarge the test space
to jets supported in the bigger space-time region~$\tilde{L}$.
In view of the above-mentioned freedom in modifying weak solutions, the extension will coincide
with~$\v$ in~$L$ only up to a jet in the orthogonal complement of the test jets in~$L$.
This is a delicate point which we handle using the concept of {\em{shielding}}
(see Definitions~\ref{defshield} and~\ref{defshieldc} as well as the shielding condition~\eqref{Gs}).
We thus succeed in proving existence of global weak solutions under general assumptions
(Theorem~\ref{thmshield} and Corollary~\ref{corshield}).

In view of the fact that the solution of the Cauchy problem for zero initial data vanishes identically
in the whole lens-shaped region, lens-shaped regions tell us about the speed of propagation of
linearized solutions. Using this information systematically, we construct future cones
(see Definition~\ref{defcausal}). The relation ``lies in the future of'' induced by the open future cones
is transitive (Theorem~\ref{thmtransitive}). Moreover, the cone structure is compatible with
the causal propagation speed (as is made precise in Theorem~\ref{thmspeed}).
Combining all the assumptions needed for our constructions leads to the notion
of globally hyperbolic space-times (see Definition~\ref{defglobhyp}).

We finally construct advanced and retarded Green's operators~$S^\wedge$ and~$S^\vee$
(see~\eqref{Swedge} and Corollary~\ref{corshield}).
The difference of these Green's operators~$G$ maps to the homogeneous linearized solutions
(see~\eqref{Kdef}).
We show that the operators~$\Delta$ and~$G$ have useful properties which are summarized in the
exact sequence
\[ 0 \rightarrow \Jtest_0 \overset{\Delta}{\longrightarrow} \J^*_0
\overset{G}{\longrightarrow} \J_\sc
\overset{\Delta}{\longrightarrow} \J^*_\sc \rightarrow 0 \:, \]
where~$\Jtest_0$ and~$\J_0^*$ are spaces of compactly supported jets, whereas~$\J_\sc$
and~$\J^*_\sc$ have spatially compact support (see Theorem~\ref{thmexact}).

The paper is organized as follows. In Section~\ref{secprelim}
we provide the necessary preliminaries.
Section~\ref{sechypsubset} is devoted to the Cauchy problem in a lens-shaped region.
In Section~\ref{secglobhypfol} the causal structure of linearized fields is worked out, and
it is analyzed how and under which assumptions one can construct global solutions.
In Section~\ref{secgreen} causal Green's operators are introduced, and their properties are analyzed.
in Section~\ref{secoutlook} we conclude with a discussion and an outlook on open problems.

\section{Preliminaries} \label{secprelim}
\subsection{Causal Variational Principles in the Non-Compact Setting} \label{seccvp}
We consider causal variational principles in the non-compact setting as
introduced in~\cite[Section~2]{jet}. Thus we let~$\F$ be a (possibly non-compact)
smooth manifold of dimension~$m \geq 1$
and~$\rho$ a (positive) Borel measure on~$\F$ (the {\em{universal measure}}).
Moreover, we are given a non-negative function~$\L : \F \times \F \rightarrow \R^+_0$
(the {\em{Lagrangian}}) with the following properties:
\bitem
\item[(i)] $\L$ is symmetric: $\L(x,y) = \L(y,x)$ for all~$x,y \in \F$.\label{Cond1}
\item[(ii)] $\L$ is lower semi-continuous, i.e.\ for all sequences~$x_n \rightarrow x$ and~$y_{n'} \rightarrow y$,
\[ \L(x,y) \leq \liminf_{n,n' \rightarrow \infty} \L(x_n, y_{n'})\:. \]\label{Cond2}
\eitem
The {\em{causal variational principle}} is to minimize the action
\beq \label{Sact} 
\Sact (\rho) = \int_\F d\rho(x) \int_\F d\rho(y)\: \L(x,y) 
\eeq
under variations of the measure~$\rho$, keeping the total volume~$\rho(\F)$ fixed
({\em{volume constraint}}).
The notion {\em{causal}} in ``causal variational principles'' refers to the fact that
the Lagrangian induces on~$M$ a causal structure given by
\beq \label{basiccausal}
\text{$x,y \in M$ are } \left\{ \begin{array}{c} \text{timelike} \\ \text{spacelike} \end{array} \right\}
\text{ separated if } \left\{ \begin{array}{c} \L(x,y)>0 \\ \L(x,y)=0 \end{array} \right\} \:.
\eeq
The connection between this notion of causality and the causal structure of linearized fields
will be discussed in Section~\ref{secoutlook}.
An important example of causal variational principles is the {\em{causal action principle}} for
{\em{causal fermion systems}} (for the connection see~\cite[Section~2]{jet}). In this case, the
Lagrangian is even continuous. The more general lower semi-continuous setting
arises when optimizing the hypothesis needed for getting a mathematically
well-posed problem. Moreover, lower semicontinuity arises in the context of causal fermion systems
when integrating out degrees of freedom and applying Fatou's lemma (as explained for
static causal fermion systems in~\cite[Section~3.3]{pmt}).

If the total volume~$\rho(\F)$ is finite, one minimizes~\eqref{Sact}
over all regular Borel measures with the same total volume.
If the total volume~$\rho(\F)$ is infinite, however, it is not obvious how to implement the volume constraint,
making it necessary to proceed as follows.
We need the following additional assumptions:
\bitem
\item[(iii)] The measure~$\rho$ is {\em{locally finite}}
(meaning that any~$x \in \F$ has an open neighborhood~$U$ with~$\rho(U)< \infty$).\label{Cond3}
\item[(iv)] The function~$\L(x,.)$ is $\rho$-integrable for all~$x \in \F$, giving
a lower semi-continuous and bounded function on~$\F$. \label{Cond4}
\eitem
Given a regular Borel measure~$\rho$ on~$\F$, we then vary over all
regular Borel measures~$\tilde{\rho}$ with
\[ %\label{totvol}
\big| \tilde{\rho} - \rho \big|(\F) < \infty \qquad \text{and} \qquad
\big( \tilde{\rho} - \rho \big) (\F) = 0 \]
(where~$|.|$ denotes the total variation of a measure).
These variations of the causal action are well-defined.
The existence theory for minimizers is developed in~\cite{noncompact}.

We point out that, since a manifold is by definition locally compact and separable,
$\F$ is a {\em{$\sigma$-compact}} topological space.
\label{sigmacompact}
As a consequence, every
closed subset of~$\F$ is also $\sigma$-compact; this fact will be used
later on.

\subsection{The Euler-Lagrange Equations and Jet Spaces} \label{secEL}
A minimizer of the causal variational principle
satisfies the following {\em{Euler-Lagrange (EL) equations}}:
For a suitable value of the parameter~$\s>0$,
the lower semi-continuous function~$\ell : \F \rightarrow \R_0^+$ defined by
\beq \label{elldef}
\ell(x) := \int_\F \L(x,y)\: d\rho(y) - \s
\eeq
is minimal and vanishes on space-time~$M:= \supp \rho$,
\beq \label{EL}
\ell|_M \equiv \inf_\F \ell = 0 \:.
\eeq
The parameter~$\s$ can be understood as the Lagrange parameter
corresponding to the volume constraint. For the derivation and further details we refer to~\cite[Section~2]{jet}.

The EL equations~\eqref{EL} are nonlocal in the sense that
they make a statement on the function~$\ell$ even for points~$x \in \F$ which
are far away from space-time~$M$.
It turns out that for the applications we have in mind, it is preferable to
evaluate the EL equations only locally in a neighborhood of~$M$.
This leads to the {\em{weak EL equations}} introduced in~\cite[Section~4]{jet}.
Here we give a slightly less general version of these equations which
is sufficient for our purposes. In order to explain how the weak EL equations come about,
we begin with the simplified situation that the function~$\ell$ is smooth.
In this case, the minimality of~$\ell$ implies that the derivative of~$\ell$
vanishes on~$M$, i.e.\
\beq \label{ELweak}
\ell|_M \equiv 0 \qquad \text{and} \qquad D \ell|_M \equiv 0
\eeq
(where~$D \ell(p) : T_p \F \rightarrow \R$ is the derivative).
In order to combine these two equations in a compact form,
it is convenient to consider a pair~$\u := (a, u)$
consisting of a real-valued function~$a$ on~$M$ and a vector field~$u$
on~$T\F$ along~$M$, and to denote the combination of 
multiplication of directional derivative by
\beq \label{Djet}
\nabla_{\u} \ell(x) := a(x)\, \ell(x) + \big(D_u \ell \big)(x) \:.
\eeq
Then the equations~\eqref{ELweak} imply that~$\nabla_{\u} \ell(x)$
vanishes for all~$x \in M$.
The pair~$\u=(a,u)$ is referred to as a {\em{jet}}.

In the general lower-continuous setting, one must be careful because
the directional derivative~$D_u \ell$ in~\eqref{Djet} need not exist.
Our method for dealing with this problem is to restrict attention to vector fields
for which the directional derivative is well-defined.
Moreover, we must specify the regularity assumptions on~$a$ and~$u$.
To begin with, we always assume that~$a$ and~$u$ are {\em{smooth}} in the sense that they
have a smooth extension to the manifold~$\F$. Thus the jet~$\u$ should be
an element of the jet space
\beq \label{Jdef}
\J := \big\{ \u = (a,u) \text{ with } a \in C^\infty(M, \R) \text{ and } u \in \Gamma(M, T\F) \big\} \:,
\eeq
where~$C^\infty(M, \R)$ and~$\Gamma(M,T\F)$ denote the space of real-valued functions and vector fields
on~$M$, respectively, which admit a smooth extension to~$\F$.

Clearly, the fact that a jet~$\u$ is smooth does not imply that the functions~$\ell$
or~$\L$ are differentiable in the direction of~$\u$. This must be ensured by additional
conditions which are satisfied by suitable subspaces of~$\J$
which we now introduce.
First, we let~$\Gdiff$ be those vector fields for which the
directional derivative of the function~$\ell$ exists,
\[ \Gdiff = \big\{ u \in C^\infty(M, T\F) \;\big|\; \text{$D_{u} \ell(x)$ exists for all~$x \in M$} \big\} \:. \]
This gives rise to the jet space
\[ \Jdiff := C^\infty(M, \R) \oplus \Gdiff \;\subset\; \J \:. \]
For the jets in~$\Jdiff$, the combination of multiplication and directional derivative
in~\eqref{Djet} is well-defined. 
We choose a linear subspace~$\Jtest \subset \Jdiff$ with the property
that its scalar and vector components are both vector spaces,
%\beq\label{Gammatest}
\[ \Jtest = \Ctest(M, \R) \oplus \Gtest \;\subseteq\; \Jdiff \:, \]
and the scalar component is nowhere trivial in the sense that
\beq \label{Cnontriv}
\text{for all~$x \in M$ there is~$a \in \Ctest(M, \R)$ with~$a(x) \neq 0$}\:.
\eeq
Then the {\em{weak EL equations}} read (for details cf.~\cite[(eq.~(4.10)]{jet})
\beq \label{ELtest}
\nabla_{\u} \ell|_M = 0 \qquad \text{for all~$\u \in \Jtest$}\:.
\eeq
The purpose of introducing~$\Jtest$ is that it gives the freedom to restrict attention to the portion of
information in the EL equations which is relevant for the application in mind.
For example, if one is interested only in the macroscopic dynamics, one can choose~$\Jtest$
to be composed of jets pointing in directions where the 
microscopic fluctuations of~$\ell$ are disregarded.

Before going on, we point out that the weak EL equations~\eqref{ELtest}
do not hold only for minimizers, but also for critical points of
the causal action. With this in mind, all methods and results of this paper 
(except for the constructions using second variations in Sections~\ref{secsecond} and~\ref{secenes2})
do not apply only to
minimizers, but more generally to critical points of the causal variational principle.
For brevity, we also refer to a measure with satisfies the weak EL equations~\eqref{ELtest}
as a {\em{critical measure}}.

We conclude this section by introducing a few other jet spaces which will be needed
later on. It is useful to define the differentiability properties of the jets by corresponding
differentiability properties of the Lagrangian.
When considering higher derivatives, we always choose 
charts and work in components.
For ease in notation, we usually omit all vector and tensor indices.
But one should keep in mind that, from now on, we always work in suitably chosen charts.
We first introduce the jet spaces~$\J^\ell$, where~$\ell \in \N \cup \{\infty\}$ can be
thought of as the order of differentiability if the derivatives act simultaneously on
both arguments of the Lagrangian:
\begin{Def} \label{defJvary}
For any~$\ell \in \N_0  \cup \{\infty\}$, the jet space~$\J^\ell \subset \J$
is defined as the vector space of jets with the following properties:
\begin{itemize}[leftmargin=2.5em]
\item[\rm{(i)}] For all~$y \in M$ and all~$x$ in an open neighborhood of~$M$,
the directional derivatives
\beq \label{derex}
\big( \nabla_{1, \v_1} + \nabla_{2, \v_1} \big) \cdots \big( \nabla_{1, \v_p} + \nabla_{2, \v_p} \big) \L(x,y)
\eeq
(computed componentwise in charts around~$x$ and~$y$)
exist for all~$p \in \{1, \ldots, \ell\}$ and all~$\v_1, \ldots, \v_p \in \J^\ell$. Here the subscripts $1,2$ refer to the derivatives acting on the first and on the second argument of $\L(x,y)$ respectively.
\item[\rm{(ii)}] The functions in~\eqref{derex} are $\rho$-integrable
in the variable~$y$, giving rise to locally bounded functions in~$x$. More precisely,
these functions are in the space
\[ L^\infty_\text{\rm{loc}}\Big( M, L^1\big(M, d\rho(y) \big); d\rho(x) \Big) \:. \]
\item[\rm{(iii)}] Integrating the expression~\eqref{derex} in~$y$ over~$M$
with respect to the measure~$\rho$,
the resulting function (defined for all~$x$ in an open neighborhood of~$M$)
is continuously differentiable in the direction of every jet~$\u \in \Jtest$.
\eitem
\end{Def}
Here and throughout this paper, we use the following conventions for partial derivatives and jet derivatives:
\bitem
\itemD Partial and jet derivatives with an index $i \in \{ 1,2 \}$, as for example in~\eqref{derex}, only act on the respective variable of the function $\L$.
This implies, for example, that the derivatives commute,
\beq \label{ConventionPartial}
\nabla_{1,\v} \nabla_{1,\u} \L(x,y) = \nabla_{1,\u} \nabla_{1,\v} \L(x,y) \:.
\eeq
\itemD The partial or jet derivatives which do not carry an index act as partial derivatives
on the corresponding argument of the Lagrangian. This implies, for example, that
\[ \nabla_\u \int_\F \nabla_{1,\v} \, \L(x,y) \: d\rho(y) =  \int_\F \nabla_{1,\u} \nabla_{1,\v}\, \L(x,y) \: d\rho(y) \:. \]
\eitem
We point out that, in contrast to the method and conventions used in~\cite{jet},
{\em{jets are never differentiated}}.

We denote the $\ell$-times continuously differentiable test jets by~$\Jtest \cap \J^\ell$.
Moreover, compactly supported jets are denoted by a subscript zero, like for example
\beq \label{J0def}
\Jtest_0 := \{ \u \in \Jtest \:|\: \text{$\u$ has compact support} \} \:.
\eeq
In order to make sure that surface layer integrals exist (see Section~\ref{secosi} below), one needs
differentiability conditions of a somewhat different type (for details see~\cite[Section~3.5]{osi}):
\begin{Def} \label{defslr}
The jet space~$\Jtest$ is {\bf{surface layer regular}}
if~$\Jtest \subset \J^2$ and
if for all~$\u, \v \in \Jtest$ and all~$p \in \{1, 2\}$ the following conditions hold:
\begin{itemize}[leftmargin=2.5em]
\item[\rm{(i)}] The directional derivatives
\beq \nabla_{1,\u} \,\big( \nabla_{1,\v} + \nabla_{2,\v} \big)^{p-1} \L(x,y) \label{Lderiv1}
\eeq
exist.
\item[\rm{(ii)}] The functions in~\eqref{Lderiv1} are $\rho$-integrable
in the variable~$y$, giving rise to locally bounded functions in~$x$. More precisely,
these functions are in the space
\[ L^\infty_\text{\rm{loc}}\Big( L^1\big(M, d\rho(y) \big), d\rho(x) \Big) \:. \]
\item[\rm{(iii)}] The $\u$-derivative in~\eqref{Lderiv1} may be interchanged with the $y$-integration, i.e.
\[ \int_M \nabla_{1,\u} \,\big( \nabla_{1,\v} + \nabla_{2,\v} \big)^{p-1} \L(x,y)\: d\rho(y)
= \nabla_\u \int_M \big( \nabla_{1,\v} + \nabla_{2,\v} \big)^{p-1} \L(x,y)\: d\rho(y) \:. \]
\eitem
\end{Def} \noindent
The precise regularity assumptions needed for our applications will be specified below
whenever we need them.

We finally introduce the space of {\em{dual jets}}~$(\Jtest)^*$.
To this end, we denote the
continuous global one-jets taking values in the cotangent bundle restricted to~$M$ by
\[ \J^* := C^0(M, \R) \oplus C^0(M, T^*\F) \:. \]
We let~$(\Jtest)^*$ be the quotient space
\begin{align*}
(\Jtest)^* &:= \J^* \Big/  \big\{ (g,\varphi) \in \J^* \:\big|\: g(x) \,a(x) + \la \varphi(x), u(x) \ra = 0 \\
&\qquad\qquad\qquad\qquad\quad  \text{ for all~$\u=(a,u) \in \Jtest$ and all~$x \in M$} \big\} \:,
\end{align*}
where~$\la .,. \ra$ denotes the dual pairing of~$T^*_x\F$ and~$T_x\F$.
Here we take equivalence classes simply because it is
convenient to disregard dual jets which are trivial on~$\Jtest$.

\subsection{The Linearized Field Equations} \label{seclinfield}
In simple terms, the {\em{homogeneous linearized field equations}}
describe variations of the universal measure which preserve the EL equations.
More precisely, we consider variations where we multiply~$\rho$ by a non-negative
function and take the push-forward with respect to a mapping from~$M$ to~$\F$.
Thus we consider families of measures~$(\tilde{\rho}_\tau)_{\tau \in (-\delta, \delta)}$ 
of the form
\beq \label{rhotau}
\tilde{\rho}_\tau = (F_\tau)_* \big( f_\tau \, \rho \big) \:,
\eeq
where~$f$ and~$F$ are smooth,
\[ f \in C^\infty\big((-\delta, \delta) \times M \rightarrow \R^+ \big) \qquad \text{and} \qquad
F \in C^\infty\big((-\delta, \delta) \times M \rightarrow \F \big) \:, \]
and have the properties~$f_0(x)=1$ and~$F_0(x) = x$ for all~$x \in M$
(here the push-forward measure is defined
for a subset~$\Omega \subset \F$ by~$((F_\tau)_*\mu)(\Omega)
= \mu ( F_\tau^{-1} (\Omega))$; see for example~\cite[Section~3.6]{bogachev}).
If we demand that~$(\tilde{\rho}_\tau)_{\tau \in (-\delta, \delta)}$ is a family of minimizers,
the EL equations~\eqref{EL} hold for all~$\tau$, i.e.
\beq \label{elltau}
\tilde{\ell}_\tau|_{M_\tau} \equiv \inf_\F \ell_\tau = 0 
\qquad \text{with} \qquad \tilde{\ell}_\tau(x) := \int_\F \L(x,y)\: d\tilde{\rho}_\tau(y) - \s \:,
\eeq
where~$M_\tau$ is the support of the varied measure,
\[ M_\tau := \supp \tilde{\rho}_\tau = \overline{F_\tau(M)} \:. \]
In~\eqref{elltau} we can express~$\tilde{\rho}$ in terms of~$\rho$. Moreover,
it is convenient to rewrite this equation as an equation on~$M$ and to multiply
by~$f_\tau(x)$. We thus obtain the equivalent equation
\[ \ell_\tau|_M \equiv \inf_\F \ell_\tau = 0 \]
with
\[ \ell_\tau(x) :=
\int_\F f_\tau(x) \,\L\big(F_\tau(x),F_\tau(y) \big)\: f_\tau(y)\: d\tilde{\rho}_\tau(y) - f_\tau(x)\: \s \]
In analogy to~\eqref{ELtest} we write the corresponding weak EL equations as
\beq \label{weaktau}
\nabla_{\u} \ell_\tau|_M = 0 \qquad \text{for all~$\u \in \Jtest$}
\eeq
(for details on why the jet space does not depend on~$\tau$ we refer to~\cite[Section~4.1]{perturb}).
Since this equation holds by assumption for all~$\tau$, we can differentiate it with respect to~$\tau$.
Denoting the infinitesimal generator of the variation by~$\v$, i.e.
\beq \label{vvary}
\v(x) := \frac{d}{d\tau} \big( f_\tau(x), F_\tau(x) \big) \Big|_{\tau=0} \:,
\eeq
we obtain the linearized field equations
\beq \label{Lapdef}
0 = \la \u, \Delta \v \ra(x) := 
\nabla_\u \bigg( \int_M \big( \nabla_{1, \v} + \nabla_{2, \v} \big) \L(x,y)\: d\rho(y) - \nabla_\v \,\s \bigg) \:,
\eeq
which are to be satisfied for all~$\u \in \Jtest$ and all~$x \in M$
(for details see~\cite[Section~3.3]{perturb}).
Since these equations hold pointwise in~$x$, we
here refer to these equations as the {\em{strong}} equations
(in distinction of the weak equations obtained by testing and integrating; see
Section~\ref{secweak}). Regarding the brackets~$\la.,.\ra(x)$ in~\eqref{Lapdef} as a dual pairing,
the operator~$\Delta$ is a mapping to the dual jets,
\[ \Delta \::\: \Jtest \rightarrow (\Jtest)^* \:. \]

The corresponding inhomogeneous equation arises for example in the 
perturbation expansion~\cite{perturb}. It reads
\[ \la \u, \Delta \v \ra = \la \u, \w \ra \qquad \text{for all~$\u \in \Jtest$}\:, \]
where~$\w \in (\Jtest)^*$ is a given inhomogeneity.
In order to avoid confusion, we point out that this equation
is again evaluated pointwise for~$x \in M$,
and therefore we refer to it as the {\em{strong}} linearized field equations.
For brevity, sometimes we leave out the pointwise testing and write 
this equation in the shorter form~\eqref{Delvw}.

In~\cite{osi} higher $\tau$-derivatives of~\eqref{weaktau} are computed.
Here we only need the operator~$\Delta_2 \::\: \Jtest \times \Jtest \rightarrow \J^*$ defined by
\begin{align}
\big\la &\u, \Delta_2[\v_1, \v_2] \big\ra(x) \notag \\
&= \frac{1}{2} \:\nabla_{\u} \bigg( \int_{M} \big( \nabla_{1, \v_1} + \nabla_{2, \v_1} \big) \big( \nabla_{1, \v_2} + \nabla_{2, \v_2} \big) \L(x,y)\: d\rho(y)\
-\nabla_{\v_1} \nabla_{\v_2} \s\bigg) \:. \label{Lap2def}
\end{align}
Here we always use the convention that the ``partial jet derivatives'' do not act on
jets contained in other derivatives, so that for example
\[ \big( \nabla_{\v_1} \nabla_{\v_2} \, \s\big)(x) = b_1(x) \,b_2(x) \, \s \:, \]
where~$b_1$ and~$b_2$ denote the scalar components of~$\v_1$ and~$\v_2$, respectively.

\subsection{Surface Layer Integrals} \label{secosi}
{\em{Surface layer integrals}} were first introduced in~\cite{noether}
as double integrals of the general form
\beq \label{osi}
\int_\Omega \bigg( \int_{M \setminus \Omega} (\cdots)\: \L(x,y)\: d\rho(y) \bigg)\, d\rho(x) \:,
\eeq
where $(\cdots)$ stands for a suitable differential operator formed of jets.
A surface layer integral generalizes the concept of a surface integral over~$\partial \Omega$
to the setting of causal fermion systems.
The connection can be understood most easily in the case when~$\L(x,y)$ vanishes
unless~$x$ and~$y$ are close together. In this case, we only get a contribution to~\eqref{osi}
if both~$x$ and~$y$ are close to the boundary of~$\Omega$.
A more detailed explanation of the idea of a surface layer integrals is given in~\cite[Section~2.3]{noether}.

In~\cite{noether, jet, osi}, {\em{conservation laws}} for surface layer integrals were derived.
The statement is that if~$\v$ describes a symmetry of the system or if~$\v$ satisfies the
linearized field equations, then suitable surface layer integrals~\eqref{osi} vanish for every
compact~$\Omega \subset M$.
The significance of these conservation laws for our problem lies in the fact that
it~$\v$ is {\em{not}} a solution of the linearized field equations, then the surface layer integral
still is conserved approximately in the sense that its change in time can be controlled
by~$\Delta \v$. For this reason, these surface layer integrals are very useful for getting
estimates, which we refer to as {\em{energy estimates}}.
More specifically, the following surface layer integrals are important for
developing energy estimates and will (in a slightly modified form) play a crucial
role in our analysis: The {\em{symplectic form}} $\sigma^\Omega$ defined by
(for details see~\cite[Section~4.3]{jet})
\beq
\sigma^\Omega(\u,\v) := \int_\Omega d\rho(x) \int_{M \setminus \Omega} d\rho(y)\;
\big( \nabla_{1, \u} \nabla_{2, \v} - \nabla_{1, \v} \nabla_{2, \u} \big) \, \L(x,y) \label{I2asymm}
\eeq
and the {\em{surface layer inner product}}~$(\u, \v)^\Omega$
(for details see~\cite[Theorem~1.1 and Corollary~3.11]{osi})
\beq
(\u, \v)^\Omega := \int_\Omega d\rho(x) \int_{M \setminus \Omega} d\rho(y)\;
\big( \nabla_{1, \u} \nabla_{1, \v}  - \nabla_{2, \u} \nabla_{2, \v} \big) \L(x,y) \:. \label{I2symm}
\eeq
In~\cite{action}, these surface layer integrals were computed for Dirac systems
in the presence of an electromagnetic potential in Minkowski space.

\subsection{Positive Functionals Arising from Second Variations} \label{secsecond}
For the alternative energy estimates of Section~\ref{secenes2} we will work with
positive functionals which arise in the analysis of second variations~\cite{positive}.
We now recall a few concepts and results (the reader who prefers to work with the
energy estimates of Section~\ref{secenes1} may skip this section).

Clearly, if~$\rho$ is a {\em{minimizing}} measure, then second variations are non-negative.
For our purposes, it again suffices to consider variations of the form~\eqref{rhotau},
where for simplicity we assume that~$f_\tau$ and~$F_\tau$ are trivial outside a compact set.
Under these assumptions, it is proven in~\cite[Theorem~1.1]{positive} that
\[ \int_M d\rho(x) \int_M d\rho(y) \:\nabla_{1,\v} \nabla_{2,\v} \L(x,y) 
+ \int_M \nabla^2 \ell|_x(\v,\v)\: d\rho(x) \geq 0 \:, \]
where jet~$\v$ is again the infinitesimal generator of the variation~\eqref{vvary}.
For our purposes, it is preferable to write this inequality as
\beq \label{posint}
\frac{1}{2} \int_M d\rho(x) \int_M d\rho(y) \:
\big( \nabla_{1,\v} + \nabla_{2,\v} \big)^2 \L(x,y) - (\nabla^2 \s)(\v,\v)\geq 0 \:.
\eeq
Then it is obvious that the integrals are well-defined if we assume that~$\u,\v \in \J^2$
(see Definition~\ref{defJvary}).
Moreover, using~\eqref{elldef} and~\eqref{Lapdef}, the inequality can be written in the compact form
\beq \label{vvpositive}
\la \v, \Delta \v \ra_M \geq 0 \:,
\eeq
where we used the notation
\beq \label{uLapv}
\la \u, \Delta \v \ra_M := \int_M \la \u, \Delta \v \ra(x)\: d\rho(x) \:.
\eeq
In other words, the operator~$\Delta$ is {\em{positive semi-definite}}.
This might come as a surprise, because the analogous inequality for the wave operator
in Minkowski space is violated. Instead, this inequality holds (up to an irrelevant sign)
for the Laplacian in the {\em{Riemannian}} setting. These facts are not a contradiction if
one keeps in mind that the operator~$\Delta$ has a structure which is
very different from a PDE. The basic reason why~\eqref{vvpositive} holds is that, in the
setting of causal variational principles, we consider minimizers. In contrast, the Dirichlet
energy in the hyperbolic setting is unbounded from below, making it necessary to work merely with
critical points.

\section{Hyperbolic Subsets of Space-Time} \label{sechypsubset}
\subsection{Local Foliations by Surface Layers} \label{seclocfol}
\begin{figure}
% \usepackage[usenames,dvipsnames]{pstricks}
% \usepackage{epsfig}
% \usepackage{pst-grad} % For gradients
% \usepackage{pst-plot} % For axes
% \usepackage[space]{grffile} % For spaces in paths
% \usepackage{etoolbox} % For spaces in paths
% \makeatletter % For spaces in paths
% \patchcmd\Gread@eps{\@inputcheck#1 }{\@inputcheck"#1"\relax}{}{}
% \makeatother
% 
\psscalebox{1.0 1.0} % Change this value to rescale the drawing.
{
\begin{pspicture}(-2.5,-1.9044089)(14.139044,1.9044089)
\definecolor{colour1}{rgb}{0.8,0.8,0.8}
\definecolor{colour0}{rgb}{0.6,0.6,0.6}
\definecolor{colour2}{rgb}{0.4,0.4,0.4}
\pspolygon[linecolor=colour1, linewidth=0.02, fillstyle=solid,fillcolor=colour1](1.1190436,-0.09187103)(1.7090436,-0.101871036)(2.3290436,-0.09187103)(3.1790435,-0.13187103)(3.8890436,-0.15187103)(4.4990435,-0.23187104)(5.1490436,-0.23187104)(5.7290435,-0.28187102)(5.3790436,-0.09187103)(4.8790436,0.23812896)(4.3590436,0.61812896)(3.7590437,0.92812896)(3.2690437,0.97812897)(2.7290435,0.81812894)(2.1090436,0.48812896)
\rput[bl](6.7190437,1.358129){\normalsize{$U \subset M := \supp \rho$}}
\psbezier[linecolor=black, linewidth=0.04](6.389044,1.298129)(5.585711,1.8936595)(2.0168202,2.0884824)(1.0690436,1.5881289672851562)(0.12126687,1.0877756)(-0.35428908,-0.8663405)(0.44904357,-1.461871)(1.2523762,-2.0574017)(5.331267,-1.9822243)(6.179044,-1.451871)(7.02682,-0.92151767)(7.192376,0.70259845)(6.389044,1.298129)
\psbezier[linecolor=colour0, linewidth=0.08](0.07904358,-0.15187103)(1.6590469,-0.024428569)(2.4390447,-0.15040691)(3.4390435,-0.15187103271484376)(4.4390426,-0.15333515)(5.51902,-0.4087202)(6.909044,-0.16187103)
\rput[bl](0.7290436,0.79812896){\normalsize{$\eta_t \equiv 0$}}
\rput[bl](0.8090436,-1.101871){\normalsize{$\eta_t \equiv 1$}}
\psbezier[linecolor=colour0, linewidth=0.08](0.7390436,-0.111871034)(1.4801531,-0.16686673)(2.449648,1.0228906)(3.4490435,0.9881289672851562)(4.448439,0.95336735)(5.2490635,-0.3373064)(5.9590435,-0.25187102)
\psbezier[linecolor=colour2, linewidth=0.08](1.0590435,-0.071871035)(1.870874,-0.09410611)(2.4199824,0.2114487)(3.4190435,0.16812896728515625)(4.4181046,0.124809235)(4.629186,-0.1793354)(5.7390437,-0.25187102)
\rput[bl](3.7290435,-0.93187106){\normalsize{$\supp \theta_t$}}
\psbezier[linecolor=black, linewidth=0.02, arrowsize=0.05291667cm 2.0,arrowlength=1.4,arrowinset=0.0]{->}(3.6090436,-0.791871)(3.0911489,-0.8228388)(3.0490437,-0.41154847)(2.8890436,0.16812896728515625)
\rput[bl](3.2790437,0.528129){\normalsize{$L$}}
\end{pspicture}
}
\caption{A local foliation.}
\label{figlocfol}
\end{figure}
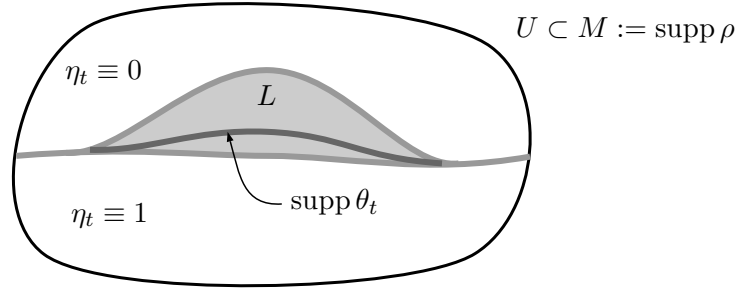

Following the procedure for hyperbolic partial differential equations, our first
goal is to analyze the initial value problem ``locally'' in an open subset~$U$ of space-time~$M$.
In analogy to the usual procedure of choosing a local time function~$t$ (like for example the
time coordinate of a local observer) and considering the foliation by the hypersurfaces~$t=\text{const}$,
we here want to choose a foliation of a compact subset~$L \subset U$ by surface layers.
This motivates the following definition.
\begin{Def} \label{deflocfoliate}
Let~$U \subset M$ be an open subset of space-time and~$I \subset \R$ a compact interval.
Moreover, we let~$\eta \in C^\infty(I \times U, \R)$ be a function with~$0 \leq \eta \leq 1$ which for
all~$t \in I$ has the following properties:
\bitem
\item[{\rm{(i)}}] The function~$\theta(t,.) := \partial_t \eta(t,.)$ is non-negative and compactly supported in~$U$.
\item[{\rm{(ii)}}] For all~$x \in \supp \theta(t,.)$ and all~$y \in M \setminus U$,
the function~$\L(x,y)$ as well as its first and second derivatives in the direction of~$\Jtest_0$ vanish.
\eitem
We also write~$\eta(t,x)$ as~$\eta_t(x)$ and~$\theta(t,x)$ as~$\theta_t(x)$.
We refer to~$(\eta_t)_{t \in I}$ as a {\bf{local foliation}} inside~$U$.
\end{Def} \noindent
The situation in mind is shown in Figure~\ref{figlocfol}.
The parameter~$t$ can be thought of as the time of a local observer
and will often simply be referred to as {\em{time}}.
The support of the function~$\theta_t$ is a {\em{surface layer}}.
The function~$\eta_t$ should be thought of as being equal to one in the past
and equal to zero in the future of this surface layer
(where the distinction between future and past will become clear later; see
the last paragraph of Section~\ref{seclens} below).
The condition~(i) implies that the set~$L$ defined by
\beq \label{Ldef}
L := \bigcup_{t \in I} \supp \theta_t
\eeq
is compact. It is the region of space-time described by the local foliation.
The condition~(ii) has the purpose to ensure that the dynamics in the region~$L$
does not depend on the jets outside~$U$, making it possible to restrict
attention to the space-time region~$U$.
Sometimes, we refer to this property that~$L$ is {\em{$\L$-localized}} in~$U$.
\label{cldef}%
One way of satisfying~(ii) is to simply choose~$U=M$.
However, in the applications it may be desirable to ``localize'' the problem
for example by choosing~$U$ as the domain of a coordinate chart.
In applications when~$\L(x,y)$ is of short range (as introduced in~\cite[Section~2.3]{jet}),
the condition~(ii) can be arranged easily by choosing~$U$ to be
relatively compact and sufficiently large.
When constructing global solutions, it will be useful to assume that~$U$ is relatively compact
(see Definition~\ref{deffiniterange} in Section~\ref{secglobal}).

For the following constructions, it will be useful to combine the
functions~$\eta_t$ and~$\theta_t$ with the measure~$\rho$ such as to form new measures:
The measure
\beq
d\rho_t(x) := \theta_t(x)\: d\rho(x) \label{rhot}
\eeq
with~$t \in I$ is supported in the surface layer at time~$t$. Likewise, the measures
\[ \eta_t\: d\rho \qquad \text{and} \qquad 
\big(1-\eta_t)\: d\rho \]
are supported in the past respectively future of the surface layer at time~$t$.
For the measures supported in
a space-time strip, we use the notation
\beq \label{etabracket}
\eta_{[t_0, t_1]}\: d\rho \qquad \text{with} \qquad 
\eta_{[t_0,t_1]} := \eta_{t_1} - \eta_{t_0} \in C^\infty_0(U) \:,
\eeq
where we always choose~$t_0,t_1 \in I$ with~$t_0 \leq t_1$.
Note that the function~$\eta_{[t_0, t_1]}$ is supported in~$L$.

\subsection{Energy Estimates Using the Surface Layer Inner Product} \label{secenes1}
For the analysis of local foliations we shall make use of 
class of surface layer integral, which we now introduce.
In preparation, we need to specify the class of jets to work with.
In order to have differentiability and regularity properties, it is a good idea
to restrict attention to test jets. But, depending on the application, it might be
necessary to restrict the jet space even further
(the crucial point is that one must satisfy the hyperbolicity conditions in Definition~\ref{defhypcond}
below). In order to have the largest possible flexibility, we shall work with a subspace
\beq \label{Jvarydef}
\Jvary \subset \Jtest \:,
\eeq
which we can choose arbitrarily (in particular, the scalar component of~$\Jvary$
does not need to be nontrivial in the sense~\eqref{Cnontriv}; this will be
discussed in Example~\ref{excalzero} at the end of this paper).
Similar to~\eqref{J0def}, $\Jvary_0$ denotes the compactly supported jets in~$\Jvary$.
We let~$\J_U$ be the restriction of these jets to~$U$,
\[ %\beq \label{JUdef}
\J_U := \big\{\u|_U \:\big|\:  \u \in \Jvary_0 \big\} \:.
\] %\eeq
For any~$t \in I$ we introduce the bilinear form
\begin{align}
I_2^t(.,.) \:&:\: \J_U \times \J_U \rightarrow \R \:, \notag \\
I_2^t(\u,\v) &= \int_U d\rho(x)\: \eta_t(x) \int_U d\rho(y)\: \big(1-\eta_t(y)\big) \notag \\
&\qquad\qquad \times
\: \big( \nabla_{1,\u} - \nabla_{2,\u} \big) \big( \nabla_{1,\v} + \nabla_{2,\v} \big)\L(x,y) \:.
\label{I2def}
\end{align}
In order to ensure that the integrals are well-defined, we assume throughout
this section that~$\Jtest$ is {\em{surface layer regular}} (see Definition~\ref{defslr}).
Symmetrizing and anti-symmetrizing gives the bilinear forms
\begin{align}
(\u, \v)^t &= \frac{1}{2}\: \big( I_2^t(\u,\v) + I_2^t(\v, \u) \big) \notag \\
&= \int_U d\rho(x)\: \eta_t(x) \int_U d\rho(y)\: \big(1-\eta_t(y)\big)
\: \Big( \nabla_{1,\u} \nabla_{1,\v} - \nabla_{2,\u} \nabla_{2,\v} \Big) \L(x,y) \label{jipdef} \\
\sigma^t(\u,\v) &= \frac{1}{2}\: \big( I_2^t(\u,\v) - I_2^t(\v, \u) \big) \notag \\
&= \int_U d\rho(x)\: \eta_t(x) \int_U d\rho(y)\: \big(1-\eta_t(y)\big)
\: \Big( \nabla_{1,\u} \nabla_{2,\v} - \nabla_{1,\v} \nabla_{2,\u} \Big) \L(x,y) \label{sympdef} \:,
\end{align}
referred to as the {\em{surface layer inner product}} and the {\em{symplectic form}}, respectively.
These surface layer integrals are ``softened versions'' of the surface layer integrals~\eqref{I2symm}
and~\eqref{I2asymm} mentioned in Section~\ref{secosi}, where
the characteristic functions~$\chi_\Omega$ and~$\chi_{M \setminus \Omega}$ are
replaced by the smooth cutoff functions~$\eta_t$ and~$1-\eta_t$, respectively.

The quantity~$(\u, \u)^t$ will be of central importance in the following constructions.
It plays the role of the energy used in our energy estimates
(for its physical interpretation see the paragraph after Definition~\ref{defhypcond}).
In preparation of these estimates, we derive an energy identity:
\begin{Lemma} {\bf{(energy identity)}} \label{lemmaenid}
For any jet~$\v =(b,v) \in \J_U$,
\beq \begin{split}
\frac{d}{dt}\: (\v, \v)^t 
&= 2 \int_U \la \v, \Delta \v \ra(x) \: d\rho_t(x) \\
&\quad\: -2 \int_U \Delta_2[\v, \v] \: d\rho_t(x)
+ \s \int_U b(x)^2\:d\rho_t(x) \:.
\end{split} \label{enid}
\eeq
\end{Lemma}
\Proof We first derive the identity by a formal computation and
give the analytic justification afterward. Differentiating~\eqref{jipdef} 
with respect to~$t$ gives
\begin{align}
\frac{d}{dt}\: (\v, \v)^t
&= \int_U d\rho(x)\: \theta_t(x) \int_U d\rho(y)\: \big(1-\eta_t(y)\big)
\: \big( \nabla^2_{1,\v} - \nabla^2_{2,\v} \big) \L(x,y) \notag \\
&\qquad -\int_U d\rho(x)\: \eta_t(x) \int_U d\rho(y)\: \theta_t(y) 
\: \big( \nabla^2_{1,\v} - \nabla^2_{2,\v} \big) \L(x,y) \notag \\
&= \int_U d\rho(x)\: \theta_t(x) \int_U d\rho(y)\: \big( \nabla^2_{1,\v} - \nabla^2_{2,\v} \big) \L(x,y) \:. \label{vvt}
\end{align}
Next, for all~$x \in L$ we may use Definition~\ref{deflocfoliate}~(ii) to change the integration range
in~\eqref{Lapdef} from~$M$ to~$U$,
\[ \la \v, \Delta \v \ra(x) = \int_U \nabla_{1,\v} \big( \nabla_{1,\v} + \nabla_{2,\v} \big)\: \L(x,y)\: d\rho(y)
- \s\:b(x)^2 \:. \]
Multiplying by~$\theta_t$ and integrating, we obtain
\begin{align*}
0&= \int_U \theta_t(x)\:\la \v, \Delta \v \ra(x) \: d\rho(x) + \s \int_U \theta_t(x) \: b(x)^2\:d\rho(x) \\
&\quad\:
-\int_U d\rho(x)\: \theta_t(x)\: \int_U d\rho(y)\: \big( \nabla_{1,\v}^2 + \nabla_{1,\v} \nabla_{2,\v} \big)\: \L(x,y) \:.
\end{align*}
We multiply this equation by two and add~\eqref{vvt}. This gives
\begin{align*}
\frac{d}{dt}\: (\v, \v)^t 
&= -\int_U d\rho(x)\: \theta_t(x) \int_U d\rho(y)\: \big( \nabla_{1,\v} + \nabla_{2,\v} \big)^2 \L(x,y) \\
&\quad\: +2 \int_U \theta_t(x)\:\la \v, \Delta \v \ra(x) \: d\rho(x) + 2 \s \int_U \theta_t(x) \: b(x)^2\:d\rho(x) \:.
\end{align*}
Using the property in Definition~\ref{deflocfoliate}~(ii), in the $y$-integral we may
replace the integration range~$U$ by~$M$, making it possible to apply~\eqref{Lap2def}.
Rewriting the obtained integrals using the notation~\eqref{rhot} gives~\eqref{enid}.

It remains to give a rigorous justification of taking the time derivative of~\eqref{jipdef}.
To this end, we first take the difference quotient and rewrite it as
\begin{align*}
&\frac{1}{\Delta t} \Big( (\v, \v)^{t+\Delta t} - (\v, \v)^t \Big) \\
&= \int_U d\rho(x)\: \frac{\eta_{t+\Delta t}(x) - \eta_t(x)}{\Delta t} \int_U d\rho(y)\:
\Big( \nabla_{1,\v} \nabla_{1,\v} - \nabla_{2,\v} \nabla_{2,\v} \Big) \L(x,y) \\
&\quad -\int_U d\rho(x)\: \frac{\big(\eta_{t+\Delta t} - \eta_t \big)(x)}{\Delta t} \int_U d\rho(y)\: \big(\eta_{t+\Delta t}
+ \eta_t \big)(y)\: \Big( \nabla_{1,\v} \nabla_{1,\v} - \nabla_{2,\v} \nabla_{2,\v} \Big) \L(x,y)
\end{align*}
Since~$\Jtest$ is assumed to be surface layer regular, we know from Definition~\ref{defslr}~(ii)
and Definition~\ref{defJvary}~(ii) (both evaluated for~$p=2$) that the
above jet derivatives exist and are in~$L^\infty_\text{\rm{loc}}( L^1M, d\rho(y)), d\rho(x))$.
Therefore, the above~$y$-integrals can be all be bounded uniformly in~$\Delta t$ by the function
\[ 2 \int_M \Big| \big( \nabla_{1,\v} \nabla_{1,\v} - \nabla_{2,\v} \nabla_{2,\v} \big) \L(x,y) \Big| \: d\rho(y)
\in L^1_\loc(M, d\rho)\:. \]
Clearly, the factor~$\eta_{t+\Delta t}+ \eta_t$ converges pointwise to~$2 \eta_t$.
Moreover, the difference quotient~$(\eta_{t+\Delta t} - \eta_t)/\Delta t$ has uniformly compact
support and converges pointwise to~$\theta_t$. Therefore, we can take the limit~$\Delta t \rightarrow 0$
with the help of Lebesgue's dominated convergence theorem.
\QED

In order to make use of this energy identity, we need to impose a condition
which we call hyperbolicity condition. This notion can be understood as follows.
In the theory of hyperbolic partial differential equations, the hyperbolicity of the
equations (as expressed for example by the notions of
normally hyperbolic operators or symmetric hyperbolic systems) gives rise to a positive
energy. In our setting, we clearly have no partial differential equation.
Instead, we take a positivity condition for the energy to {\em{define}} hyperbolicity.
As we shall see, this condition is precisely what is needed in order to
obtain existence and uniqueness of solutions.
We first define the hyperbolicity condition and explain it afterward.

For all~$x \in M$ we choose the subspace of the tangent space spanned by the test jets,
\[ \Gamma_x := \big\{ u(x) \:|\: u \in \Gtest \big\} \;\subset\; T_x\F\:. \]
We introduce a Riemannian metric~$g_x$ on~$\Gamma_x$. This Riemannian metric also induces a pointwise scalar product on the jets. Namely, setting
\[ \J_x := \R \oplus \Gamma_x \:, \]
we obtain the scalar product on~$\J_x$
\beq
\la .,. \ra_x \,:\, \J_x \times \J_x \rightarrow \R \:,\qquad
\la \v, \tilde{\v} \ra_x := b(x)\, \tilde{b}(x) + g_x \big(v(x),\tilde{v}(x) \big) \:. \label{vsprod}
\eeq
We denote the corresponding norm by~$\|.\|_x$.

\begin{Def} \label{defhypcond}
The local foliation~$(\eta_t)_{t \in I}$ inside~$U$
satisfies the {\bf{hyperbolicity condition}}
if there is a constant~$C>0$ such that for all~$t \in I$,
\beq
(\v, \v)^t \geq \frac{1}{C^2} \int_U \Big( \|\v(x)\|_x^2\: + \big|\Delta_2[\v, \v]\big| \Big) \: d\rho_t(x) 
\qquad \text{for all~$\v \in \J_U$} \:. \label{hypcond}
\eeq
\end{Def}
Let us explain the hyperbolicity condition.
The inner product~$(.,.)^t$ was first introduced in~\cite{osi} in a slightly different form where the
smooth cutoff function~$\eta_t$ is replaced by the characteristic function of a set~$\Omega$.
In~\cite{action} it was shown by longer explicit computations that for Dirac sea configurations in Minkowski
space and choosing~$\eta_t$ as a characteristic function being identically equal to one in the
past of the hypersurface~$t=\text{const}$,
the inner product~$(.,.)^t$ reduces to a (positive definite) scalar product on Dirac wave functions and on the
Maxwell field tensor. With this in mind, it is physically sensible to assume that~$(\v,\v)^t$ is positive.

The lower bound in~\eqref{hypcond} is a stronger and more quantitative version of positivity.
Again for Dirac sea configurations in Minkowski space and for~$\theta_t$ replaced by a characteristic function
of the past of the surface~$t=\text{const}$, this inequality is satisfied
in view of the explicit formulas in~\cite{action}. In more general situations, the inequality~\eqref{hypcond}
is not obvious and must be verified in all applications.
More specifically, in the applications one can use the freedom in choosing the jet spaces~$\Jtest$ and~$\Jvary$,
the Riemannian metric in the scalar product~\eqref{vsprod} and the functions~$\eta_t$
in order to arrange that~\eqref{hypcond} holds.
Clearly, the smaller the jet space~$\Jvary$ is chosen, the easier it is to satisfy~\eqref{hypcond}.
The drawback is that the Cauchy problem will be solvable for more restrictive initial data
(as will be made precise in Section~\ref{secstrong}).

We now explain how the above hyperbolicity condition can be used to
derive energy estimates. We let~$L$ be a lens-shaped region inside~$U$
with the local foliation~$(\eta_t)_{t \in I})$.
We denote the norm corresponding to the jet scalar product by~$\|\v\|^t := \sqrt{(\v,\v)^t}$.
We begin with a simple estimate of the energy identity in Lemma~\ref{lemmaenid}.

\begin{Lemma} \label{lemmaenes}
Assume that the hyperbolicity condition of Definition~\ref{defhypcond} holds.
Then for every~$t \in I$ and all~$\v \in \J_U$,
\beq \frac{d}{dt}\: \|\v\|^t 
\leq C\:\|\Delta \v\|_{L^2(U, d\rho_t)}  + c\: \|\v\|^t \label{enes0}
\eeq
with
\[ c := C^2 + \frac{C^2 \,\s}{2} \:. \]
\end{Lemma}
\Proof Applying~\eqref{hypcond} in~\eqref{enid}, we obtain
\begin{align*}
\frac{d}{dt}\: (\v, \v)^t 
&\leq 2 \int_U \la \v, \Delta \v \ra_x \: d\rho_t(x)
-2 \int_U \Delta_2[\v, \v] \: d\rho_t(x) + \s \int_U b(x)^2\: d\rho_t(x) \\
&\leq 2 \int_U \la \v, \Delta \v \ra_x \: d\rho_t(x)
+\Big( 2 C^2 + C^2 \,\s \Big)\: (\v,\v)^t \\
&\leq 2 \,\|\v\|_{L^2(U, d\rho_t)} \: \|\Delta \v\|_{L^2(U, d\rho_t)}
+2c\: (\v,\v)^t \\
&\leq 2 C\,\|\v\|^t\: \|\Delta \v\|_{L^2(U, d\rho_t)}
+2c\: (\v,\v)^t \:,
\end{align*}
where in the last line we applied~\eqref{hypcond}.
Using the relation~$\partial_t \|\v\|^t = \partial_t (\v, \v)^t / (2 \|\v\|^t)$ gives the result.
\QED

Applying Gr\"onwall-type estimates, the inequality~\eqref{enes0} shows that~$\|\v\|^t$ grows at most
exponentially in time, provided that~$\Delta \v$ decays in time sufficiently fast.
We here make this statement precise by estimates in Hilbert spaces
of jets with zero initial values.
In the lens-shaped region~$L$ we work with the $L^2$-scalar product
\beq \label{L2prod}
\la \u, \v \ra_{L^2(L)} := \int_L \la \u(x),\v(x) \ra_x\: \eta_I(x) \, d\rho(x) \:,
\eeq
which, according to~\eqref{rhot} and~\eqref{etabracket}, can also be written in terms of a time integral,
\beq \label{L2prodtime}
\la \u, \v \ra_{L^2(L)} = \int_{t_0}^{\tmax} \la\u,\v\ra_{L^2(U, d\rho_t)}\: dt \:.
\eeq
The corresponding norm is denoted by~$\| . \|_{L^2(L)}$.
\begin{Prp} {\bf{(energy estimate)}} \label{prpenes}
Assume that the hyperbolicity condition of Definition~\ref{defhypcond} holds.
Then, choosing
\beq \label{Gammachoose}
\Gamma = 2\,C\, e^{2c\,(\tmax-t_0)}\: (\tmax-t_0)\:,
\eeq
the following estimate holds,
\[ %\beq \label{enes1}
\|\v\|_{L^2(L)} \leq \Gamma\: \|\Delta \v\|_{L^2(L)} \qquad \text{for all~$\v \in \J_U$ with~$\|\v\|^{t_0}=0$}\:.
\] %\eeq
\end{Prp}
\Proof  We write the energy estimate of Lemma~\ref{lemmaenes} as
\[ \frac{d}{dt}\: \big( e^{-2c t}\, (\v, \v)^t \big) \leq 2 \: e^{-2c t} \:C\: \|\v\|^t\: \|\Delta \v\|_{L^2(U, d\rho_t)} \:. \]
Integrating over~$t$ from~$t_0$ to some~$t \in I$ and using the hyperbolicity condition~\eqref{hypcond}, we obtain
\begin{align*}
e^{-2c t}\, (\v, \v)^t &=  \int_{t_0}^t \frac{d}{dt'} \: \big( e^{-2c t'} (\v, \v)^{t'} \big)\: dt' \\
&\leq 2\,C\, \int_{t_0}^{t} e^{-2c t'} \: \|\v\|^{t'}\: \|\Delta \v\|_{L^2(U, d\rho_{t'})} \: dt' \:.
\end{align*}
Multiplying by~$e^{2c t}$ gives the inequality
\begin{align*}
(\v, \v)^t &\leq 2\,C\, \int_{t_0}^{t} e^{2c \,(t-t')} \: \|\v\|^{t'}\: \|\Delta \v\|_{L^2(U, d\rho_{t'})} \: dt' \\
&\leq 2\,C\, e^{2c \,(\tmax-t_0)} \: \int_{t_0}^{\tmax} \|\v\|^{t'}\: \|\Delta \v\|_{L^2(U, d\rho_{t'})} \: dt' \\
&\leq 2\,C\, e^{2c \,(\tmax-t_0)} \: \|\Delta \v\|_{L^2(L)}\:
\bigg( \int_{t_0}^{\tmax} (\v, \v)^{t'} \:dt' \bigg)^\frac{1}{2} \:,
\end{align*}
where in the last step we used the Schwarz inequality and~\eqref{L2prodtime}.
Integrating once again over~$t$ from~$t_0$ to~$\tmax$ gives
\beq \label{intfinal}
\bigg( \int_{t_0}^{\tmax}(\v, \v)^{t} \:dt \bigg)^\frac{1}{2} \leq 2\,C\, e^{2c \,(\tmax-t_0)}\:(\tmax-t_0) \: \|\Delta \v\|_{L^2(L)} \:.
\eeq

Finally, we apply the hyperbolicity condition~\eqref{hypcond} in~\eqref{L2prodtime},
\[ \|v\|_{L^2(L)} = \bigg( \int_{t_0}^{\tmax}  \|\v \|_{L^2(U, d\rho_{t})}^2 \:dt \bigg)^\frac{1}{2}
\leq C \:\bigg( \int_{t_0}^{\tmax}  (\v,\v)^t \:dt \bigg)^\frac{1}{2} \:. \]
Combining this inequality with~\eqref{intfinal} gives the result.
\QED

\subsection{Alternative Energy Estimates Using Second Variations} \label{secenes2}
The energy estimates of the previous section were based on the hyperbolicity condition
of Definition~\ref{defhypcond}. Working with the surface layer inner product~$(.,.)^t$
has the advantage that it has a clear physical interpretation and significance
(in particular, it gives rise to the scalar product of quantum theory~\cite{fockbosonic}).
Also, it can be verified in important examples that the hyperbolicity condition~\eqref{hypcond}
is indeed satisfied. But one should keep in mind that the positivity of the surface
layer inner product is a physical assumption which needs to be verified in all applications.
From the mathematical point of view, it would be more convincing to work with
quantities which are positive as a consequence of the mathematical
structure of the causal variational principle.
Such positive quantities were obtained in~\cite{positive} by considering second variations
(for basics see Section~\ref{secsecond}).
We now show that positive quantities obtained from second variations
can indeed be used for energy estimates, giving an alternative to the energy estimates
in the previous section. The corresponding hyperbolicity condition
(see Definition~\ref{defhypcond2} below) is more natural from the mathematical point of view.
The energy estimates in this section shed new light on the mathematical structure of causal variational principles.
The reader who prefers to work with the surface layer inner product and the
hyperbolicity condition of Definition~\ref{defhypcond} may skip this section.

Throughout this section, we assume that~$\rho$ is a {\em{minimizing}} measure
and that~$\Jtest$ is {\em{surface layer regular}}
(see Definition~\ref{defslr}).
Then, according to Definition~\ref{defJvary}~(ii) for~$p=2$, we know that for all~$\u, \v \in \Jtest$,
\[ \big( \nabla_{1, \u} + \nabla_{2, \u} \big) \big( \nabla_{1, \v} + \nabla_{2, \v} \big) \L(x,y)
\;\in\; L^\infty_\text{\rm{loc}}\Big( L^1\big(M, d\rho(y) \big), d\rho(x) \Big) \:. \]
Using that the function~$\eta_{[t_0,t]}$ has compact support (see~\eqref{etabracket}), 
it follows that the following expression is well-defined,
\beq \label{posuv}
\begin{split}
&\int_U d\rho(x) \int_U d\rho(y) \:
\big( \nabla_{1,\eta_{[t_0,t]}\,\u} + \nabla_{2,\eta_{[t_0,t]} \u} \big)
\big( \nabla_{1,\eta_{[t_0,t]}\,\v} + \nabla_{2,\eta_{[t_0,t]} \v} \big) \L(x,y) \\
&-2 \int_U \eta_{[t_0,t]}(x)^2\: \big(\nabla^2 \s \big)(\v,\v)\big|_x\: d\rho(x) \:.
\end{split}
\eeq
Exactly as explained after~\eqref{posint}, this expression can be written in the more compact
form~$\la \eta_{[t_0,t]} \u, \Delta (\eta_{[t_0,t]} \v ) \ra_M$, giving rise to a bilinear form
\beq
\la \u, \v \ra_{[t_0,t]} \::\: \J_U \times \J_U \rightarrow \R \:, \qquad
\la \u, \v \ra_{[t_0,t]} := \la \eta_{[t_0,t]} \,\u, \Delta (\eta_{[t_0,t]} \,\v ) \ra_M \:. \label{uvsprod}
\eeq
According to~\eqref{vvpositive}, this inner product is positive semi-definite, i.e.
\beq \label{vvplus}
\la \v, \v \ra_{[t_0,t]} \geq 0 \qquad \text{for all~$\v \in \J_U$} \:.
\eeq
We denote the corresponding semi-norm by~$\| . \|_{[t_0,t]}$.
Before going on, we point out that the jet~$\eta_{[t_0,t]} \u$
will in general {\em{not}} lie in~$\J^2$, because the condition~(iii) in Definition~\ref{defJvary}
may be violated. Therefore, one should always keep in mind our jets are in~$\J_U$ only
before multiplying by the cutoff function~$\eta_{[t_0,t]}$.

We again begin with an energy identity.
\begin{Lemma} {\bf{(energy identity)}} \label{lemmaenid2}
For all~$\v \in \J_U$,
\beq \label{enid2}
\frac{d}{dt} \la \v, \v \ra_{[t_0,t]} =
2 \:\big\la \eta_{[t_0,t]} \v, \Delta ( \theta_t\, \v) \big\ra_M
\eeq
(where we again used the notation~\eqref{uLapv}).
\end{Lemma}
\Proof The identity is obtained immediately by formally differentiating~\eqref{uvsprod}
and using the symmetry of the bilinear form~$\la ., \Delta . \ra_M$.
Therefore, the only task is to justify the differentiation and the product rule.
To this end, similar as in the proof of Lemma~\ref{lemmaenid}, we 
analyze the difference quotient in the limit~$\Delta t \rightarrow 0$.
In the last integral in~\eqref{posuv}, this is straightforward because the integrand
converges pointwise and has uniformly compact support.
Therefore, it remains to consider the first line in~\eqref{posuv} for~$\u=\v$. 
Using the symmetry in the arguments~$x$ and~$y$, we can write the
difference quotient as
\begin{align*}
&2 \int_U d\rho(x) \int_U d\rho(y) \:
\frac{1}{\Delta t} \Big( \nabla_{1,\eta_{[t_0,t+\Delta t]}\, \v} - \nabla_{1,\eta_{[t_0,t]} \,\v} \Big) \\
&\qquad \times \Big( \big( \nabla_{1,\eta_{[t_0,t+\Delta t]}\,\v} + \nabla_{2,\eta_{[t_0,t + \Delta t]} \v} \big)
+ \big( \nabla_{1,\eta_{[t_0,t]}\,\v} + \nabla_{2,\eta_{[t_0,t]} \v} \big) \Big) \L(x,y) \:.
\end{align*}
Since~$\Jtest$ is surface layer regular, we know from Definition~\ref{defslr}~(i)
that the derivatives exist. Moreover, the $y$-integral can be estimated
uniformly in~$\Delta t$ by
\[ 2 \int_M \Big| \nabla_{1,\v} \big( \nabla_{1,\v} + \nabla_{2,\v}\big) \L(x,y) \Big| \: d\rho(y)
\in L^1_\loc(M, d\rho)\:. \]
Now we can take the limit~$\Delta t \rightarrow 0$ exactly as in the proof of
Lemma~\ref{lemmaenid}.
\QED

\begin{Def} \label{defhypcond2}
The local foliation~$(\eta_t)_{t \in I=[t_0, \tmax]}$ inside~$U$
satisfies the {\bf{alternative hyperbolicity condition}}
if there exists a constant~$C>0$ and~$\underline{t} \in I$ such that, for all~$t \in [\underline{t}, \tmax]$
and all~$\v \in \J_U$,
\begin{align}
\la \v, \v \ra_{[t_0,t]} &\geq \frac{1}{C^2}\: \int_U \eta_{[t_0, t]}(x)\: \la \v(x), \v(x) \ra_x\: d\rho(x) \label{hypalt} \\
\Big| \big\la \eta_{[t_0,t]}\, \v, \Delta ( \theta_t\, \v) \big\ra_M \Big| &\leq
C\, \Big( \|\v\|_{[t_0,t]}^2 + \big| \big\la \eta_{[t_0,t]} \,\v, \Delta \big( (1- \eta_{[t_0,t]})\,  \v \big) \big\ra_M \big|
\Big)  \:. \label{hypcond2}
\end{align}
\end{Def}
We now explain these conditions and compare them to the previous hyperbolicity condition
of Definition~\ref{defhypcond}. Both inequalities~\eqref{hypalt} and~\eqref{hypcond2}
strengthen and quantify the positivity property~\eqref{vvplus}. The explicit
computations in~\cite[Section~6]{action} show that these inequalities are satisfied
for Dirac sea configurations in Minkowski space in the presence of Dirac currents
and a Maxwell field. Compared to the hyperbolicity condition in Definition~\ref{defhypcond},
there are several major structural differences: First, as already mentioned at the beginning of this
section, the positivity~\eqref{vvplus} is not a physical assumption, but it follows already from the
structure of the causal variational principle. Second, in contrast to~\eqref{hypcond},
the energy identity~\eqref{enid2}
and consequently also the inequalities~\eqref{hypalt} and~\eqref{hypcond2} do not
involve the quadratic correction~$\Delta_2$ to the linearized field equations.
This is remarkable because it means that we do not need to control the nonlinear
corrections in the energy estimates.
A third difference is that, in contrast to the surface layer integral~$(.,.)^t$,
the energy~$\la .,.\ra_{[t_0,t]}$ in~\eqref{vvplus} involves an integral over the
time strip~$[t_0, t]$. As a consequence, this inner product typically tends to zero
in the limit~$t \rightarrow t_0$, making it difficult to satisfy the inequalities~\eqref{hypalt}
and~\eqref{hypcond2}. This is the reason why in Definition~\ref{defhypcond2}
we merely assume that that these inequalities hold
for all~$t \in [\underline{t}, \tmax]$.
We finally remark that, in contrast to~$(.,.)^t$, the energy~$\la ., .\ra_{[t_0,t]}$
does not distinguish a direction of time; this will be discussed further in Section~\ref{seclens} below.

By combining the above energy identity with the hyperbolicity condition, we
now derive energy estimates.
\begin{Lemma} \label{lemmaenes2}
Assume that the alternative hyperbolicity condition of Definition~\ref{defhypcond2} holds.
Then for every~$t \in  [\underline{t}, \tmax]$ and all~$\v \in \J_U$,
\[ \frac{d}{dt} \, \|\v\|_{[t_0,t]} \leq C^2\, \|\Delta \v\|_{L^2(L)}
+ c \: \|\v\|_{[t_0,t]} \:, \]
where~$c=2C$.
\end{Lemma}
\Proof We estimate~\eqref{enid2} using~\eqref{hypcond2} to obtain
\begin{align*}
\Big| \frac{d}{dt} \la \v, \v \ra_{[t_0,t]} \Big|
&\leq 2C \:\Big( \|\v\|_{[t_0,t]}^2 +  \big| \big\la \eta_{[t_0,t]}\, \v, \Delta \big( (1- \eta_{[t_0,t]})\, \v \big) \big\ra_M 
\Big) \\
&\leq 4C \: \|\v\|_{[t_0,t]}^2 +  2C\, \big| \big\la \eta_{[t_0,t]}\, \v, \Delta \v \big\ra_M  \big| \\
&= 4C \: \|\v\|_{[t_0,t]}^2
+ 2C\, \Big| \big\la \v, \Delta \v \big\ra_{L^2 \big(L, d\overline{\underline{\rho}}_\tmin^t \big)}  \Big| \:.
\end{align*}
Applying the Schwarz inequality as well as~\eqref{hypalt} gives
\begin{align*}
\Big| \frac{d}{dt} \la \v, \v \ra_{[t_0,t]} \Big| &\leq 4C \: \|\v\|_{[t_0,t]}^2
+ 2C\, \| \v \|_{L^2 \big(L, d\overline{\underline{\rho}}_\tmin^t \big)} \: \|\Delta \v\|_{L^2(L)} \\
&\leq 4C \: \|\v\|_{[t_0,t]}^2 + 2C^2\, \| \v \|_{[t_0, t]}  \: \|\Delta \v\|_{L^2(L)} \:.
\end{align*}
Using the relation~$\partial_t \|\v\|_{[t_0,t]} = \partial_t \la \v, \v \ra_{[t_0,t]} / (2 \|\v\|_{[t_0,t]})$ gives the result.
\QED

\begin{Prp} \label{prpenes2} {\bf{(energy estimate)}}
Assume that the alternative hyperbolicity condition of Definition~\ref{defhypcond2} holds.
Then, choosing
\[ %\beq \label{Gammachoose2}
\Gamma = \frac{C^2}{c}\:\big(e^{c\, (\tmax - \underline{t})} - 1 \big) \:,
\] %\eeq
the following estimate holds,
\beq \label{enes2}
\|\v\|_{L^2(L)} \leq \Gamma\: \|\Delta \v\|_{L^2(L)} \qquad \text{for all~$\v \in \J_U$ with~$\|\v\|_{[t_0,
\underline{t}]}=0$}\:.
\eeq
\end{Prp}
\Proof We write the energy estimate of Lemma~\ref{lemmaenes2} as
\[ \frac{d}{dt} \, \big( e^{-ct} \, \|\v\|_{[t_0,t]} \big)
\leq C^2\, e^{-ct}\: \|\Delta \v\|_{L^2(L)} \:. \]
Integrating from~$\underline{t}$ to~$\tmax$ and using that the initial data vanishes gives
\[ e^{-c \tmax} \, \|\v\|_I \leq \frac{C^2}{c}\:\big(e^{-c \underline{t}} - e^{-c\tmax} \big)\: \|\Delta \v\|_{L^2(L)} \:. \]
Multiplying by~$e^{c \tmax}$ gives the result.
\QED

We finally motivate the hyperbolicity conditions of Definition~\ref{defhypcond2}
and clarify the connection between the norms~$\|\v\|_I$ and~$\|\v\|_{L^2(L)}$.
We first show that, under general assumptions on the Lagrangian,
the norm~$\|\v\|_I$ can be estimated from above by the $L^2$-norm.
\begin{Lemma} Assume that the Lagrangian satisfies the condition
\[ C_\L^2 := \|\eta_I\, \nabla^2 \ell \|_{L^\infty(L)} +
\sup_{x \in M} \big\| \nabla_1 \nabla_2 \L(x,y) \big\|_{L^1(L)} < \infty \]
(where, similar to the notation in~\eqref{L2prod}, $L^2(L)$ refers to the
measure~$\eta_I\, d\rho$). Then for any~$\v \in \J_U$,
\beq \label{vhin}
\|\v\|_I \leq C_\L\, \|\v\|_{L^2(L)} \:.
\eeq
\end{Lemma}
\Proof We first note that, according to~\eqref{uLapv},
\begin{align*}
\la \v, \v \ra_I &= \la \eta_I \,\v, \Delta \big( \eta_I \,\v \big) \ra_M \\
&= \int_M d\rho(x) \:\nabla_{\eta_I \v} \bigg( \int_M \big( \nabla_{1, \eta_I \v} + \nabla_{2, \eta_I \v} \big) \L(x,y)\: 
d\rho(y) - \nabla_{\eta_I \v} \,\s \bigg) \\
&= \int_M \eta^2_I(x)\:\nabla^2_{\v} \ell(x)\:\: d\rho(x) \\
&\qquad + \int_M \eta_I(x)\: d\rho(x) \int_M \eta_I(y) \: d\rho(y)\;
\nabla_{1, \v} \nabla_{2, \v} \L(x,y) \:.
\end{align*}
We estimate the first integral by
\[ \bigg| \int_M \eta^2_I(x)\:\nabla^2_{\v} \ell(x)\:\: d\rho(x) \bigg|
\leq \|\v\|_{L^2(L)}^2\: \|\eta_I\, \nabla^2 \ell\|_{L^\infty(M)}\:. \]
The second integral, on the other hand, can be estimated by
\begin{align*}
&\bigg| \int_M \eta_I(x)\: d\rho(x) \int_M \eta_I(y) \: d\rho(y)\;
\nabla_{1, \v} \nabla_{2, \v} \L(x,y) \bigg| \\
&\leq \int_M d\rho(x) \int_M d\rho(y)\: f(x,y)\, f(y,x)\: K(x,y) \:,
\end{align*}
where we introduced the abbreviations
\[ f(x,y) = \sqrt{\eta_I(x)}\: \|\v(x)\|_x\: \sqrt{\eta_I(y)} \qquad \text{and} \qquad
K(x,y) = \big\| \nabla_1 \nabla_2 \L(x,y) \big\| \:. \]

The last integral can be estimated as follows,
\begin{align*}
&\int_M d\rho(x) \int_M d\rho(y)\: f(x,y)\, f(y,x)\: K(x,y) \\
&\leq \frac{1}{2} \int_M d\rho(x) \int_M d\rho(y)\: \Big( f(x,y)^2 + f(y,x)^2 \Big) \: K(x,y) \\
&= \int_M d\rho(x) \int_M d\rho(y)\: f(x,y)^2 \: K(x,y) \\
&\leq \bigg( \int_M \|\v(x)\|_x^2\: \eta_I(x)\: d\rho(x) \bigg) \; \sup_{z \in M} \int_M |K(z,y)| \:
\eta_I(y)\: d\rho(y) \\
&= \|\v\|_{L^2(L)}^2 
\:\sup_x \big\| K(x,.)\|_{L^1(L)} \:.
\end{align*}
Combining the terms gives the result.
\QED
We finally note that~\eqref{hypalt} is the converse inequality to~\eqref{vhin}.
The inequality~\eqref{hypalt} in general does not hold for~$L^2$-jets.
The same is true for the inequality~\eqref{hypcond2}.
This is why in Definition~\ref{defhypcond2} we restrict attention to jets in~$\J_U$.

\subsection{Lens-Shaped Regions and Time Orientation} \label{seclens}
We combine the previous concepts in the following useful notion:
\begin{Def} \label{deflens}
A compact set~$L \subset M$ is a {\bf{lens-shaped region}} inside~$U$
if there is a local foliation~$(\eta_t)_{t \in I}$ inside~$U$
satisfying~\eqref{Ldef} which satisfies the {\bf{hyperbolicity conditions}}
of either Definition~\ref{defhypcond} or Definition~\ref{defhypcond2}.
\end{Def}

We now discuss the question of {\em{time orientability}}.
A local foliation~$(\eta_t)_{t \in I}$ distinguishes the future (the region where~$\eta_t \equiv 0$)
from the past (where~$\eta_t \equiv 1$). 
But the time orientation was arbitrary; we could just as well have chosen
a local foliation with the opposite time orientation.
Indeed, the hyperbolicity condition of Definition~\ref{defhypcond} removes this arbitrariness, because
it {\em{distinguishes a direction of time}}. In order to explain how this comes about, we note that
changing the time direction corresponds to the replacement~$\eta_t \rightarrow (1-\eta_t)$.
In the above surface layer integrals, this corresponds to interchanging~$x$ and~$y$,
which in~\eqref{jipdef} gives rise to a minus sign. Consequently, if we changed the time orientation,
the inner product~$(\v,\v)^t$ in~\eqref{hypcond} would become negative.
Therefore, a lens-shaped region which satisfies the hyperbolicity condition of Definition~\ref{defhypcond}
always comes with a distinguished time orientation.
If we assume that~$M$ can be covered by lens-shaped regions (as is made precise by
the notion of local hyperbolicity in Definition~\ref{defcompacthyp} below),
we automatically obtain a global time orientation.

The alternative hyperbolicity condition of Definition~\ref{defhypcond2}, however,
does {\em{not}} distinguish a time direction.
Therefore, when working with this hyperbolicity condition, we must always assume that
space-time can be oriented in the sense that we can distinguish lens-shaped regions
with mutually compatible time directions. For brevity, we do not formalize this assumption.

\subsection{The Cauchy Problem and Uniqueness of Strong Solutions}
We want to study the Cauchy problem to the future (the solution to the future and past will be studied
in Section~\ref{secfuturepast} below).
Therefore, we assume that we are given a local foliation with~$I=[t_0, \tmax]$
of a lens-shaped region~$L$ inside~$U$,
where~$t_0$ and~$\tmax$ are the initial and final times, respectively.

In preparation of setting up the initial value problem,
we need to specify what we mean by ``$\v$ vanishes in the past of~$t_0$.''
The obvious notion is to demand that~$\v$ vanishes identically in the region where~$\eta_{t_0}$
is strictly positive, i.e.\ that~$\eta_{t_0}\, \v \equiv 0$.
This condition is quite strong, because it also implies that~$\v$ vanishes inside the
surface layer at time~$t_0$. Nevertheless, this  condition is not strong enough for 
two reasons. First, if working with the alternative hyperbolicity condition of Definition~\ref{defhypcond2},
the jet~$\v$ should vanish even in the past of~$\underline{t}$. For this reason, we always demand
that~$\eta_{\underline{t}} \,\v= 0$, and in case we do {\em{not}} work with the alternative hyperbolicity condition,
we simply choose~$\underline{t}=\tmin$.
Second, if working with the hyperbolicity condition of Definition~\ref{defhypcond}, 
we need in addition that the norm~$\|\v\|^{t_0}$ vanishes. Moreover, it will
be useful to also impose that the symplectic form vanishes in the sense
that~$\sigma^{t_0}(\u,\v)=0$ for all~$\v \in \J_U$.
For convenience, we combine the last two conditions for the surface layer inner product and the
symplectic form by expressing them in terms of the surface layer integral~$I_2^{t_0}$ in~\eqref{I2def}.
This motivates the definition of the jet space
\beq \label{uJtestdef}
\underline{\J_U}_{t_0} := \big\{ \u \in \J_U \:\big|\: \eta_{\underline{t}}\, \u \equiv 0 \quad \text{and} \quad
I_2^{t_0}(\u,\v)=0 \text{ for all~$\v \in \J_U$} \big\} \:.
\eeq
Similarly, we define the space of jets which vanish at time~$\tmax$ by
\beq \label{oJtestdef}
\overline{\J_U}^\tmax := \big\{ \u \in \J_U \:\big|\: \big(1-\eta_{\overline{t}}\big)\, \u \equiv 0 \quad \text{and} \quad
I_2^{\tmax}(\u,\v)=0 \text{ for all~$\v \in \J_U$} \big\} \:,
\eeq
where~$\overline{t} \in [t_0, \tmax]$ is chosen equal to~$\tmax$
in case we do not work with the alternative hyperbolicity condition.

A {\em{strong solution}} of the Cauchy problem is a jet~$\v \in \J_U$ which satisfies the equations
\beq \label{cauchystrong}
\Delta \v = \w \quad \text{in~$L$} \qquad \text{and} \qquad \v-\v_0 \in \underline{\J_U}_{t_0} \:,
\eeq
where~$\v_0 \in \J_U$ is the initial data and~$\w$ is the inhomogeneity.
According to~\eqref{Lapdef}, the inhomogeneity~$\w \in (\Jtest)^*$ is a dual jet.
Having the scalar product~\eqref{vsprod} at our disposal, we can identify
jets with dual jets. For technical simplicity, we here choose~$\w \in \J_U$.

\begin{Prp} \label{prpunique} {\bf{(uniqueness of strong solutions)}}
If~$L$ is a lens-shaped region inside~$U$ with foliation~$(\eta_t)_{t \in I}$, then the Cauchy problem~\eqref{cauchystrong} with~$\v_0, \w \in \J_U$ has at most one solution~$\v$ in~$L$.
\end{Prp}
\Proof Let~$\v$ be the difference of two solutions. Then~$\v$ is a solution of the homogeneous equation
with zero initial data. Applying Lemma~\ref{lemmaenes} or Lemma~\ref{lemmaenes2}, we obtain
\[ \Big| \frac{d}{dt}\: \|\v\|^t \Big| \leq c\: \|\v\|^t \qquad \text{and thus} \qquad
\frac{d}{dt}\: \big( e^{-ct} \,\|\v\|^t \big) \leq 0 \:. \]
It follows that~$\|\v\|^t$ vanishes for all~$t$ in the respective interval. 
Using~\eqref{hypcond}, we conclude that~$\v$ vanishes identically in~$L$.
This gives the result.
\QED

\subsection{Weak Solutions of the Cauchy Problem} \label{secweak}
Our goal is to construct solutions of the Cauchy problem~\eqref{cauchystrong}.
As usual, replacing~$\v$ by~$\v-\v_0$ and~$\w$ by~$\w -\Delta \v_0 \in \J_U$,
it suffices to consider the Cauchy problem for zero initial data, i.e.
\beq \label{cauchyzeroinit}
\Delta \v = \w \quad \text{in~$U$} \qquad \text{and} \qquad \v \in \underline{\J_U}_{t_0}\:.
\eeq
In order to derive the notion of a weak solution, we take the inner product with
a test jet~$\u \in \J_U$ and integrate over space-time.
In order to integrate only over~$L$,
we again work with the scalar product~$\la .,.\ra_{L^2(L)}$ introduced in~\eqref{L2prod}.
We thus obtain the equation
\beq \label{strongtest0}
\big\la \u, (\Delta \v - \w) \big\ra_{L^2(L)} = 0 \qquad \text{for all~$\u \in \J_U$} \:.
\eeq
Before going on, we compare this equation with~\eqref{cauchyzeroinit}.
If the space~$\J_U$ is dense in~$L^2(L)$, then these equations are equivalent.
However, as explained after~\eqref{ELtest}, in most situations the space of jets
will not be dense. In this case, equation~\eqref{strongtest0} contains less
information than~\eqref{cauchyzeroinit}. This information loss can be understood
similarly as explained after~\eqref{ELtest} by our wish for restricting attention to
part of the information contained in the linearized field equations.
With this in mind, in what follows we are content with constructing solutions of~\eqref{strongtest0}.

The following Lemma makes it possible to ``integrate by parts.''
\begin{Lemma} {\bf{(Green's formula)}} \label{lemmagreen}
For all~$\u, \v \in \J_U$,
\beq \label{green}
\sigma^{\tmax}(\u, \v) - \sigma^{t_0}(\u, \v) = \la \u, \Delta \v \ra_{L^2(L)} - \la \Delta \u, \v \ra_{L^2(L)} \:.
\eeq
\end{Lemma}
\Proof Using the definitions~\eqref{L2prod} and~\eqref{Lapdef},
\begin{align*}
\la &\u, \Delta \v \ra_{L^2(L)} - \la \Delta \u, \v \ra_{L^2(L)} =
\int_U  \Big( \la \u, \Delta \v \ra - \la \Delta \u, \v \ra \Big)\: \eta_I \,d\rho \\
&= \int_U d\rho(x)\: \eta_I(x)\;
\nabla_{\u} \bigg( \int_M \big( \nabla_{1, \v} + \nabla_{2, \v} \big) \L(x,y)\: d\rho(y) - \nabla_\v \,\s \bigg) \\
&\quad-\int_U d\rho(x)\: \eta_I(x)\;
\nabla_{\v} \bigg( \int_M \big( \nabla_{1, \u} + \nabla_{2, \u} \big) \L(x,y)\: d\rho(y) - \nabla_\u \,\s \bigg) \:.
\end{align*}
Here the space-time point~$x$ is in~$L$. Using Definition~\ref{deflocfoliate}~(ii), we get a contribution to the
integrals only if~$y \in U$. Therefore, we may replace the integration range~$M$ by~$U$.
We thus obtain
\begin{align}
\la &\u, \Delta \v \ra_{L^2(L)} - \la \Delta \u, \v \ra_{L^2(L)} \notag \\
&= \int_U d\rho(x)\: \eta_I(x) \int_U d\rho(y) \big( \nabla_{1,\u} \nabla_{2, \v} 
- \nabla_{2,\u} \nabla_{1, \v} \big) \L(x,y)\: d\rho(y) \:, \label{eq1}
\end{align}
where we used that, following our convention~\eqref{ConventionPartial}, the
second derivatives of the Lagrangian are symmetric.
Using the definition~\eqref{etabracket} as well as
 the anti-symmetry of the integrand, the term~\eqref{eq1} can be rewritten as
\begin{align*}
&\int_U d\rho(x)\: \eta_I(x) \int_U d\rho(y) \big( \nabla_{1,\u} \nabla_{2, \v} 
- \nabla_{2,\u} \nabla_{1, \v} \big) \L(x,y)\: d\rho(y) \\
&=\int_U d\rho(x) \int_U d\rho(y) \:\eta_{t}(x)\: \big( \nabla_{1,\u} \nabla_{2, \v}
- \nabla_{2,\u} \nabla_{1, \v} \big) \L(x,y)\: d\rho(y) \Big|_{t_0}^{\tmax} \\
&=\int_U d\rho(x)\ \int_U d\rho(y) \Big( \eta_{t}(x) - \eta_{t}(x)\: \eta_{t}(y) \Big)\,
\big( \nabla_{1,\u} \nabla_{2, \v} - \nabla_{2,\u} \nabla_{1, \v} \big) \L(x,y)\: d\rho(y) \Big|_{t_0}^{\tmax} \\
&=\int_U d\rho(x)\ \int_U d\rho(y) \:\eta_{t}(x)\: \big(1-\eta_{t}(y) \big)\,
\big( \nabla_{1,\u} \nabla_{2, \v} - \nabla_{2,\u} \nabla_{1, \v} \big) \L(x,y)\: d\rho(y) \Big|_{t_0}^{\tmax} \\
&= \sigma^{\tmax}(\u,\v) - \sigma^{t_0}(\u,\v) \:.
\end{align*}
This gives the result.
\QED

Assume that~$\v$ is a strong solution of the Cauchy problem~\eqref{cauchyzeroinit}.
Then, applying the above Green's formula, we obtain for any~$\u \in \J_U$,
\[ \la \u, \w \ra_{L^2(L)} = \la \u, \Delta \v \ra_{L^2(L)} = 
\la \Delta \u, \v \ra_{L^2(L)} - \sigma^{\tmax}(\u, \v) + \sigma^{t_0}(\u, \v) \:. \]
Having implemented the vanishing initial data by the
condition~$\v \in \underline{\J_U}_{t_0}$, the symplectic form vanishes at time~$t_0$ 
(note that the symplectic form is obtained by anti-symmetrizing the functional~$I_2$
in~\eqref{uJtestdef}.
In order to also get rid of the boundary values at time~$\tmax$, we
restrict attention to test jets which vanish at~$\tmax$. This leads us to the following definition:

\begin{Def} \label{defweak}
A jet~$\v \in L^2(L)$ is a {\bf{weak solution}} of the Cauchy problem~\eqref{cauchyzeroinit} if
\beq \label{weak}
\la \Delta \u, \v \ra_{L^2(L)} = \la \u, \w \ra_{L^2(L)} \qquad \text{for all~$\u \in \overline{\J_U}^\tmax$}\:.
\eeq
\end{Def}

\subsection{Existence of Weak Solutions} \label{secexist}
Our existence proof is inspired by the method invented by K.O.\ Friedrichs for
symmetric hyperbolic systems in~\cite{friedrichs}; see also~\cite[Section~5.3]{john}
and~\cite[Chapter~11]{intro}.

We want to construct a weak solution~\eqref{weak}.
Clearly, the energy estimate of Propositions~\ref{prpenes} or~\ref{prpenes2} also holds
if we exchange the roles of~$\tmax$ and~$t_0$, i.e.\
\beq \label{hyprev}
\|\u\|_{L^2(L)} \leq \Gamma\: \|\Delta \u\|_{L^2(L)} \qquad \text{for all~$\u \in \overline{\J_U}^\tmax$}
\eeq
(where the constant~$\Gamma$ is again given by~\eqref{Gammachoose}).

We introduce the positive semi-definite bilinear form
\[ \bra .,. \ket \::\: \overline{\J_U}^\tmax \times \overline{\J_U}^\tmax \rightarrow \R\:,\qquad
\bra \u, \v \ket = \la \Delta \u, \Delta \v \ra_{L^2(L)} \:. \]
Dividing out the null space and forming the completion, we obtain a Hilbert space
$(\mathcal{H}, \bra .,. \ket)$. The corresponding norm is denoted by~$\norm . \norm$.

We now consider the linear functional~$\la \w, . \ra_{L^2(L)}$ on~$\overline{\J_U}^\tmax$.
Applying the Schwarz inequality and~\eqref{hyprev}, we obtain
\[ \big| \la \w, \u \ra_{L^2(L)} \big| \leq \|\w\|_{L^2(L)} \:  \|\u\|_{L^2(L)} 
\leq \Gamma\:\|\w\|_{L^2(L)} \:  \norm \u \norm \:, \]
proving that the linear functional~$\la \w, . \ra_{L^2(L)}$ on~$\overline{\J_U}^\tmax$
is bounded on~$\mathcal{H}$.
Therefore, it can be extended uniquely to a bounded linear functional on all of~$\mathcal{H}$.
Moreover, by the Fr{\'e}chet-Riesz theorem there is a unique vector~$V \in \mathcal{H}$ with
\[ %\beq \label{FR}
\la \w, \u \ra_{L^2(L)} = \bra V, \u \ket = 
\la \Delta V, \Delta \u \ra_{L^2(L)}\qquad \text{for all~$\u \in \overline{\J_U}^\tmax$}\:.
\] %\eeq
Hence~$\v := \Delta V \in L^2(L)$ is the desired weak solution.
We point out that in the above estimates, the inhomogeneity~$\w$ enters
only via its~$L^2$-norm, making it possible to generalize our methods
to~$\w \in L^2(L)$.
We have obtain the following result:

\begin{Thm} \label{thmexist} Assume that~$L$ is a lens-shaped region inside~$U$
with foliation~$(\eta_t)_{t \in I}$ with~$I=[t_0,\tmax]$. 
Then for every~$\w \in L^2(L)$ there is a weak solution~$\v \in L^2(L)$
of the Cauchy problem~\eqref{weak}. This solution is bounded by
\beq \label{vbound}
\|\v\|_{L^2(L)} \leq \Gamma\, \|\w\|_{L^2(L)} \:.
\eeq
\end{Thm}
\Proof It remains to prove the estimate~\eqref{vbound}. To this end,
we use that the Fr{\'e}chet-Riesz theorem also yields that the norm of~$\v$
equals the sup-norm of the linear functional. Hence
\[ \|\v\|_{L^2(L)} = \|\Delta V \|_{L^2(L)} = \norm v \norm = \| \la \w, . \ra_{L^2(L)} \|_{\H^*} \leq \Gamma\, \|\w\|_{L^2(L)}\:, \]
concluding the proof.
\QED

\subsection{Are Weak Solutions Unique?} \label{secunique}
We now analyze the uniqueness problem for weak solutions.
It is obvious from~\eqref{weak} that a weak solution~$\v \in L^2(L)$
is unique up to vectors which are orthogonal to all vectors~$\Delta \u$
with~$\u \in \overline{\J_U}^\tmax$:
\begin{Prp} \label{prpnonunique} Let~$\v, \tilde{\v} \in L^2(L)$ be two solutions of 
the weak Cauchy problem~\eqref{weak}. Then
\beq \label{nonunique}
\v - \tilde{\v} \in \Big( \Delta \big( \overline{\J_U}^\tmax \big) \Big)^\perp \;\subset\; L^2(L) \:.
\eeq
\end{Prp}
As an immediate consequence, we obtain the following result:
\begin{Corollary} If~$\,\Delta(\overline{\J_U}^\tmax)$ is dense in~$L^2(L)$,
then the weak Cauchy problem~\eqref{weak} has a unique solution.
\end{Corollary} \noindent
Under the made denseness assumption, this corollary gives an alternative
proof of uniqueness of strong solutions (Proposition~\ref{prpunique}).
However, this result is only of limited relevance because in most applications,
the space~$\,\Delta(\overline{\J_U}^\tmax)$ will {\em{not}}
be dense in~$L^2(L)$. This corresponds to our general concept explained after~\eqref{ELtest}
that by choosing~$\Jtest$ we want to restrict attention to the portion of
information in the EL equations which is relevant for the application in mind.
Using notions from information theory, one can say equivalently that~$\Jtest$
determines the bandwidth of the information relevant for our application.
With this in mind, the freedom to modify the weak solution according to~\eqref{nonunique}
is irrelevant to us because it only affects the information which we disregard.
Implementing this point of view mathematically, one could regard the freedom in~\eqref{nonunique}
as an equivalence relation and take the uniquely determined equivalence classes as the
physically relevant solutions.
In order to keep the setting as simple as possible, we here prefer not to form equivalence classes,
but to work instead with solutions in~$L^2(L)$, which are determined only up to the freedom in~\eqref{nonunique}.
Using this freedom, one can try to find solutions which are particularly simple.
For example, the construction of the previous section gives us a {\em{canonical}}
solution~$\v = \Delta V$, which is distinguished by the fact that the $L^2$-norm of~$\v$ is minimal.

\subsection{Weak Solutions in the Future and Past} \label{secfuturepast}
In the previous section we solved the weak Cauchy problem to the future from
the initial time~$t_0$ to the final time~$\tmax$. We now analyze how to construct
a solution also to the past. Thus we consider a local foliation
with~$I=[\tmin, \tmax]$ of a lens-shaped region~$L$ inside~$U$.
Our goal is to construct a weak solution in~$L$ for zero initial data at
time~$t_0 \in I$.

In preparation, we reconsider the solution to the future constructed in the previous section.
Thus setting
\[ I^+ = [t_0, \tmax] \qquad \text{and} \qquad L^+ = \bigcup_{t \in I^+} \supp \theta_t
= \supp \eta_{[t_0, \tmax]} \:, \]
in Theorem~\ref{thmexist} we constructed a solution~$\v \in L^2(L^+)$ of the weak equation
\beq \label{weakp}
\la \Delta \u, \v \ra_{L^2(L^+)} = \la \u, \w \ra_{L^2(L^+)} \qquad \text{for all~$\u \in \overline{\J_U}^\tmax$}\:.
\eeq
We now want to transform this equation with the goal of working
instead of the measure~$\overline{\underline{\rho}}^\tmax_{t_0}$ (see~\eqref{L2prod})
with the measure~$\rho$.
We first note that, by definition of~$\overline{\J_U}^\tmax$ (see~\eqref{oJtestdef}), the jet~$\u$ vanishes
identically unless~$\eta_{\tmax}$ is equal to one. Therefore, the right hand side of~\eqref{weakp} can be
rewritten as
\begin{align*}
\la \u, \w \ra_{L^2(L^+)} &= \int_{L^+} \la \u(x), \w(x) \ra_x\: \eta_{[t_0, \tmax]}\: d\rho \\
&= \int_{L^+} \la \u(x), \w(x) \ra_x\: \big(1 - \eta_{t_0} \big)\: d\rho
= \la \u, \w \ra_{L^2\big(L^+, (1-\eta_{t_0}) d\rho \big)} \:.
\end{align*}
In order to also remove the dependence of the integration measure on~$\eta_{t_0}$, we write
\[ \la \u, \w \ra_{L^2 \big(L^+, (1-\eta_{t_0}) d\rho \big)} = \la \u, \w^+ \ra_{L^2(L^+, d\rho)} \]
with
\beq \label{wpdef}
\w^+ := \big( 1 - \eta_{t_0} \big)\, \w \;\in\; L^2(L^+, d\rho)\:.
\eeq
On the left hand side of~\eqref{weakp}, we rewrite the integral as
\begin{align*}
\la \Delta \u, \v \ra_{L^2(L^+)} &= \int_{L^+} \la (\Delta \u)(x), \v(x) \ra_x\: \eta_{[t_0, \tmax]}\: d\rho
= \la \Delta \u, \v^+ \ra_{L^2(L^+, d\rho)} \:,
\end{align*}
where we set
\beq \label{vpdef}
\v^+ := \eta_{[t_0, \tmax]}\, \v \;\in\; L^2(L^+, d\rho) \:.
\eeq
Thus we can rewrite~\eqref{weak} as
\beq \label{weak2p}
\la \Delta \u, \v^+ \ra_{L^2(L^+, d\rho)} = \la \u, \w^+ \ra_{L^2(L^+, d\rho)}
\qquad \text{for all~$\u \in \overline{\J_U}^\tmax$}\:.
\eeq
In this formulation, the existence result of Theorem~\ref{thmexist} can be stated that for every~$\w^+$
of the form~\eqref{wpdef} there is a weak solution~$\v^+ \in L^2(L^+, d\rho)$ of~\eqref{weak2p}.

Changing the time orientation in an obvious way by reparametrizing~$\eta_t$ by
\[ \eta_t \rightarrow \big(1-\eta_{t'} \big) \qquad \text{with} \quad
t' = \tmax + \tmin - t \]
and flipping the sign in the hyperbolicity condition~\ref{hypcond}, we obtain
similarly a solution~$\v^- \in L^2(L^-, d\rho)$ to the past, i.e.
\beq \label{weak2m}
\la \Delta \u, \v^- \ra_{L^2(L^-, d\rho)} = \la \u, \w^- \ra_{L^2(L^-, d\rho)} \qquad
\text{for all~$\u \in \underline{\J_U}_\tmin$}\:,
\eeq
where in analogy to~\eqref{wpdef} and~\eqref{vpdef} we now set
\begin{align}
\w^- &:= \eta_{t_0}\, \w \label{wmdef} \\
\v^- &:= \eta_{[\tmin, t_0]}\, \v \;\in\; L^2(L^-, d\rho) \:. \label{vmdef}
\end{align}

The interesting point is that, according to~\eqref{weak2p}, \eqref{weak2m} and~\eqref{wpdef}, \eqref{wmdef},
by extending the solutions~$\v^+$ and~$\v^-$ by zero to~$L$ and adding them, 
we get a weak solution in~$L$ for the desired inhomogeneity~$\w$.
We thus obtain the following result:
\begin{Thm} \label{thmexistpm} Assume that~$L$ is a lens-shaped region inside~$U$
with foliation~$(\eta_t)_{t \in I}$ with~$I=[\tmin,\tmax]$. 
Then for every~$\w \in L^2(L, d\rho)$ and every~$t_0 \in I$,
there is a solution~$\hat{\v} \in L^2(L, d\rho)$ of the weak equation
\beq \label{weak3}
\la \Delta \u, \hat{\v} \ra_{L^2(L, d\rho)} = \la \u, \w \ra_{L^2(L, d\rho)} \qquad
\text{for all~$\u \in \overline{\underline{\J_U}}^\tmax_\tmin$} \:,
\eeq
where~$\overline{\underline{\J_U}}^\tmax_\tmin:= \overline{\J_U}^\tmax
\cap \underline{\J_U}_\tmin$. Moreover, the solution~$\hat{\v}$ vanishes at time~$t_0$
in the following sense: There is a decomposition
\[ \hat{\v} = \v^+ + \v^- \qquad \text{with} \qquad \supp \v^\pm \subset L^\pm \]
such that~$\v^+$ and~$\v^-$ are weak solutions of~\eqref{weak2p}
and~\eqref{weak2m}, respectively. 

The solution~$\hat{\v}$ satisfies the energy estimate
\beq \label{vboundpm}
\|\hat{\v}\|_{L^2(L, d\rho)} \leq \Gamma\: \|\w\|_{L^2(L, d\rho)} 
\qquad \text{with} \qquad \Gamma = \sqrt{2}\: \max (\Gamma^+, \Gamma^-) \:,
\eeq
where~$\Gamma^+$ and~$\Gamma^+$ are the constants in the
energy estimate~\eqref{vbound} for the lens-shaped regions~$L^+$ and~$L^-$, respectively,
\end{Thm}
\Proof It remains to prove the energy estimate~\eqref{vboundpm}.
We first consider~$\v^+$ as given by~\eqref{vpdef}. Applying~\eqref{vbound} to~$\v$ gives
\begin{align}
\|\v^+\|_{L^2(L^+, d\rho)}^2 &= \int_{L^+}  \eta_{[t_0, \tmax]}(x)^2\: \|\v(x)\|_x^2\: d\rho(x) \notag \\
&\leq \int_{L^+}  \|\v(x)\|_x^2 \:\eta_{[t_0, \tmax]}(x) \: d\rho(x) \notag \\
&= \|\v\|_{L^2(L^+)}^2 \leq (\Gamma^+)^2\: \|\w\|_{L^2(L^+)}^2 \:. \label{wnorm}
\end{align}
Adding the corresponding inequality for~$\v^-$ gives
\begin{align*}
\|&\v^+\|_{L^2(L^+, d\rho)}^2 + \|\v^-\|_{L^2(L^-, d\rho)}^2 \\
&\leq \big( \max (\Gamma^+, \Gamma^-) \big)^2 \int_{L^+}  \eta_I(x)\: \|\w(x)\|_x^2\: d\rho(x) \\
&\leq \big( \max (\Gamma^+, \Gamma^-) \big)^2 \: \|\w(x)\|_{L^2(L, d\rho)}^2 \:.
\end{align*}
We finally combine this estimate with the inequality
\[ \|\v\|_{L^2(L, d\rho)}^2 = \|\v^+ + \v^-\|_{L^2(L, d\rho)}^2 \leq 2\, \big( 
\|\v^+\|_{L^2(L^+, d\rho)}^2 + \|\v^-\|_{L^2(L^-, d\rho)}^2 \big) \]
and take the square root.
\QED
For clarity, we point out that the energy estimate~\eqref{vboundpm} does {\em{not}} hold for~$\v^+$
separately. Indeed, the norm~$\|\w\|_{L^2(L^+)}$ in~\eqref{wnorm} cannot be bounded from above
by~$\|\w^+\|_{L^2(L^+, d\rho)}$, because the inequality
\begin{align*}
\|\w\|_{L^2(L^+)}^2 &= \int_{L^+} \|\w(x)\|_x^2\: \eta_{[t_0, \tmax]}(x)\: d\rho(x) \\
&\geq \int_{L^+} \eta_{[t_0, \tmax]}(x)^2\: \|\w(x)\|_x^2\: d\rho(x)
= \|\w^+\|_{L^2(L^+, d\rho)}^2
\end{align*}
goes in the wrong direction.

\subsection{Restricting and Extending Weak Solutions} \label{secextend}
We now turn attention to the following questions. Suppose that we are given a weak solution~$\v$
in a lens-shaped region~$L$.
If~$\hat{L}$ is another lens-shaped region contained in~$L$, is the restriction of~$\v$ to~$\hat{L}$
again a weak solution?
Conversely, if~$\hat{L}$ is a lens-shaped region containing~$L$, can~$\v$ be extended to a
weak solution in~$\hat{L}$?

In preparation, we specify what we mean by ``a lens-shaped region is contained in
another lens-shaped region.'' In addition to the obvious inclusion of the lens-shaped regions,
we must also impose that the jet spaces and the initial data surface layers fit together.

\begin{Def} \label{defnested}
Let~$L$ be a lens-shaped region inside~$U$ with foliation~$(\eta_t)_{t \in [\tmin, \tmax]}$,
and~$\tilde{L}$ a lens-shaped region inside~$\tilde{U}$ with
foliation~$(\tilde{\eta}_t)_{t \in [\tmin, \tmax]}$. We say that~$L$ is {\bf{nested}} in~$\tilde{L}$, denoted by
\[ L \prec \tilde{L} \:, \]
if the following conditions are satisfied:
\begin{itemize}[leftmargin=2.5em]
\item[\rm{(i)}] $L \subset \tilde{L}$ and~$U \subset \tilde{U}$
\item[\rm{(ii)}] The jet spaces are contained in each other, i.e.\
\[ \overline{\J_U}^\tmax \subset \overline{\J_{\tilde{U}}}^{\tilde{t}_{\max}}
\qquad \text{and} \qquad \underline{\J_U}_\tmax \subset \underline{\J_{\tilde{U}}}_{\tilde{t}_{\min}} \:, \]
where we extended the jets in~$U$ by zero to~$\tilde{U}$.
\item[\rm{(iii)}] The initial data surfaces layers are compatible in the sense that
for suitable~$t_0 \in [\tmin, \tmax]$ and~$\tilde{t}_0 \in [\tilde{t}_{\min}, \tilde{t}_{\max}]$,
\[ \eta_{t_0} = \tilde{\eta}_{\tilde{t}_0} \big|_U \:. \]
\eitem
\end{Def}

We begin with the restriction problem. Based on the weak formulation of Theorem~\ref{thmexistpm},
this problem has a simple answer:
\begin{Prp} \label{prprestrict} {\bf{(restriction property)}} Let~$\hat{L} \prec L$ be two nested
lens-shaped regions. Moreover, let~$\v \in L^2(L, d\rho)$
be the weak solution of the Cauchy problem for the inhomogeneity~$\w \in L^2(L, d\rho)$
with zero initial data at time~$t_0$ as constructed in Theorem~\ref{thmexistpm}.
Then the jet~$\hat{\v} := \v|_{\hat{L}} \in L^2(\hat{L}, d\rho)$
is a weak solution for the inhomogeneity~$\w$ with zero initial data at time~$\hat{t}_0$.
\end{Prp}
\Proof The result follows immediately from the fact
that the weak equations~\eqref{weak2p}, \eqref{weak2m}
and~\eqref{weak3} remain valid if the jet space used for testing is made smaller.
\QED

The extension problem is more subtle. The basic difficulty can be understood as follows.
Suppose that~$\v$ and~$\hat{\v}$ are weak solutions in~$L$ respectively~$\hat{L}$
with~$L \prec \hat{L}$. Due to the nonlocality of the operator~$\Delta$, we cannot
expect that~$\hat{\v}|_L = \v$
(due to the restriction property of Proposition~\ref{prprestrict},
we know that~$\hat{\v}|_L$ is again a weak
solution in~$L$, but in view of the non-uniqueness result of Proposition~\ref{prpnonunique}
this does not imply that~$\hat{\v}|_L=\v$).
For example for constructing global solution (for details see Section~\ref{secglobal} below), 
it is important to quantify~$\hat{\v}^+|_L - \v^+$ depending on the size of the lens-shaped region~$\hat{L}$.
Since the necessary estimates are a bit technical, we begin with the following simpler question:
Is there an open set~$\Omega \subset L$ in which~$\hat{\v}$ coincides with~$\v$?
The next proposition shows that, under certain conditions, this question has an
affirmative answer. We first state and prove our result and explain it afterward.

\begin{Def} \label{defshield} Let~$A, B \subset L^2(M, d\rho)$ be two (not necessarily closed)
subspaces of the Hilbert space of square-integrable jets.
Moreover let~$V \subset M$ be a subset of space-time.
The subspace~$A$ {\bf{shields}}~$V$ from~$B$
if all jets in~$A^\perp \cap \overline{B}$ vanish identically in~$V$.
\end{Def}

We introduce the jet spaces\footnote{At this stage, it would be sufficient to
define the set~$\K(L^+)$ (and similarly~$\K(L^-)$)
by~$\K(L^+) = \eta_{[t_0, \tmax]}\,\Delta ( \overline{\J_U}^\tmax )$.
The more general definition with~$t \in [t_0, \tmax]$ is of advantage in view of the
constructions in Section~\ref{secspeed}.}
\beq \label{Kpmdef}
\begin{split}
\K(L^+) &:= \text{span} \Big\{  \eta_{[t, \tmax]}\,\Delta \big( \overline{\J_U}^\tmax \big) \:\Big|\:
t \in [t_0, \tmax] \Big\} \\
\K(L^-) &:= \text{span} \Big\{ \eta_{[\tmin, t]}\,\Delta \big( \underline{\J_U}_\tmin \big) \:\Big|\:
t \in [\tmin, t_0] \Big\}
\end{split}
\eeq
(and similarly with hats), where the multiplication of a function in space-time with a jet space
means that all jets are multiplied pointwise by this function.
We extend all jets by zero to all of~$M$ and consider them as vectors in~$L^2(M, d\rho)$.

\begin{Prp} \label{prpextend} {\bf{(extension property)}} 
Let~$L \prec \hat{L}$ be two nested lens-shaped regions.
Moreover, let~$\v \in L^2(L, d\rho)$
be the weak solution of the Cauchy problem with inhomogeneity~$\w := \hat{\w}|_U$
with zero initial data at time~$t_0$ as constructed in Theorem~\ref{thmexistpm}.
Finally, assume that~$\Omega \subset L$ is an open set such that
\beq \label{scond}
\begin{split}
\chi_{L^+}\, \Delta\big( \overline{\J_U}^\tmax \big) \qquad &\text{shields $\Omega$ from} \qquad 
\text{\rm{span}} \big( \K(\hat{L}^+), \K(L^+) \big)  \\
\chi_{L^-}\, \Delta\big( \underline{\J_U}_\tmin \big) \qquad &\text{shields $\Omega$ from} \qquad 
\text{\rm{span}} \big( \K(\hat{L}^-), \K(L^-) \big)
\end{split}
\eeq
(where~$\chi_{L^\pm}$ denote the characteristic functions of~$L^\pm$).
Then the solution~$\hat{\v}$ of the Cauchy problem in~$\hat{L}$
with zero initial data at time~$\hat{t}_0$
constructed in Theorem~\ref{thmexistpm} has the property that it extends~$\v$ in~$\Omega$
in the sense that
\[ %\beq \label{concideV}
\hat{\v}|_\Omega = \v|_\Omega \:.
\] %\eeq
\end{Prp}
\Proof We let~$\v$ and~$\hat{\v}$ be the solutions constructed in Theorem~\ref{thmexistpm}.
It suffices to consider the solutions~$\v^+$ and~$\hat{\v}^+$ in the future
(as defined by~\eqref{weak2p}),
because the solutions to the past are treated analogously.

The first step is to show that
\beq \label{vvhspace}
\v^+ \in \overline{\K(L^+)} \qquad \text{and} \qquad
\hat{\v}^+ \in \overline{\K(\hat{L}^+)}
\eeq
(where the overline denotes the closure in~$L^2(M, d\rho)$).
To this end, we note that the solution constructed in Theorem~\ref{thmexist}
lies in the $L^2(L^+)$-completion of~$\Delta \overline{\J_U}^\tmax$.
The solution~$\hat{\v}^+$ is obtained from this solution by multiplication
with cutoff functions~\eqref{vpdef}. We also saw that this solution is in~$L^2(L, d\rho)$.
Hence it lies in the $L^2$-completion of~$\K(L^+)$ as defined in~\eqref{Kpmdef}.
The argument for~$\K(\hat{L}^+)$ is the same.

According to~\eqref{weak2p}, $\v^+$ and~$\hat{\v}^+$ satisfy the weak equations
\begin{align*}
\la \Delta \u, \v^+ \ra_{L^2(L^+, d\rho)}
&= \la \u, \w^+ \ra_{L^2(L^+, d\rho)} &&\hspace*{-1cm} \text{for all~$\u \in \overline{\J_U}^\tmax$} \\
\la \Delta \u, \hat{\v}^+ \ra_{L^2(\hat{L}^+, d\rho)} &= \la \u, \hat{\w}^+ \ra_{L^2(\hat{L}^+, d\rho)} &&\hspace*{-1cm}
\text{for all~$\u \in \overline{\J_{\hat{U}}}^{\hat{t}_{\max}}$} \:.
\end{align*}
According to Definition~\ref{defnested}~(ii), we may restrict the second equation
to~$\u \in \overline{\J_U}^\tmax$ and combine it with the first equation to obtain
\[ \la \Delta \u, \hat{\v}^+ - \v^+ \ra_{L^2(L^+, d\rho)}
= 0 \qquad \text{for all~$\u \in \overline{\J_U}^\tmax$} \:. \]
In other words, extending $\hat{\v}^+ - \v^+$ by zero to all of~$M$,
this function lies in the orthogonal complement of~$\chi_{L^+} \Delta(\overline{\J_U}^\tmax)$.
Moreover, from~\eqref{vvhspace} we know that the extension of~$\hat{\v}^+ - \v^+$ lies in the
completion of the span of~$\K(\hat{L}^+)$ and~$\K(L^+)$.
The shielding property implies that~$\hat{\v}^+ - \v^+$ vanishes identically in~$\Omega$.
This concludes the proof.
\QED

We now explain the concept of shielding and discuss if the conditions~\eqref{scond} are
sensible assumptions for the applications in mind.
Intuitively speaking, the shielding property of Definition~\ref{defshield} means that,
restricting attention to the space-time region~$V$, the jets are described completely
by the jets in~$A$. This intuitive picture is made precise by demanding that all jets
in the orthogonal complement of~$A$ should vanish identically in~$V$.
In order to illustrate the notion of shielding, we now discuss a few examples, for simplicity
for real-valued functions on the real line. In the first example, we choose the Hilbert space~$L^2(\R)$
and the subspaces
\[ A := L^2\big( (0,1) \big) \;\subset\; B := L^2(\R) \:. \]
Moreover, we choose~$V=(0,1)$. In this example, the functions in the space
\beq \label{AperpB}
A^\perp \cap \overline{B} = L^2 \big(\R \setminus (0,1) \big)
\eeq
vanish identically in~$V$. Thus~$A$ shields~$V$ from~$B$.
The situation is similar if we consider smooth functions, like in the example
\[ A := \big\{ \u \in C^\infty_0(\R) \:|\: \supp \u \subset [0,1] \big\} \;\subset\;
B := C^\infty_0(\R)\:. \]
In this example, the set~$A^\perp \cap \overline{B}$ is again given by~\eqref{AperpB},
showing that~$A$ again shields~$V$ in~$B$.

\begin{figure}
% \usepackage[usenames,dvipsnames]{pstricks}
% \usepackage{epsfig}
% \usepackage{pst-grad} % For gradients
% \usepackage{pst-plot} % For axes
% \usepackage[space]{grffile} % For spaces in paths
% \usepackage{etoolbox} % For spaces in paths
% \makeatletter % For spaces in paths
% \patchcmd\Gread@eps{\@inputcheck#1 }{\@inputcheck"#1"\relax}{}{}
% \makeatother
% 
\psscalebox{1.0 1.0} % Change this value to rescale the drawing.
{
\begin{pspicture}(0,-1.31833)(13.220031,1.31833)
\rput[bl](2.4300306,-1.31333){\normalsize{$\delta$}}
\psline[linecolor=black, linewidth=0.04, arrowsize=0.05291667cm 2.0,arrowlength=1.4,arrowinset=0.0]{<-}(1.3050307,1.3266699)(1.3050307,-1.2483301)
\psline[linecolor=black, linewidth=0.04, arrowsize=0.05291667cm 2.0,arrowlength=1.4,arrowinset=0.0]{<-}(6.3300304,-0.8833301)(0.0,-0.87333006)
\psline[linecolor=black, linewidth=0.04, arrowsize=0.05291667cm 2.0,arrowlength=1.4,arrowinset=0.0]{<-}(13.135031,-0.86833006)(6.6850305,-0.86833006)
\psline[linecolor=black, linewidth=0.04, arrowsize=0.05291667cm 2.0,arrowlength=1.4,arrowinset=0.0]{<-}(8.110031,1.3266699)(8.105031,-1.2583301)
\psline[linecolor=black, linewidth=0.04](0.10503067,-0.7633301)(0.10503067,-0.9633301)
\psline[linecolor=black, linewidth=0.04](2.5050306,-0.7783301)(2.5050306,-0.9783301)
\psline[linecolor=black, linewidth=0.04](3.7100306,-0.7783301)(3.7100306,-0.9783301)
\psline[linecolor=black, linewidth=0.04](4.9150305,-0.7783301)(4.9150305,-0.9783301)
\psline[linecolor=black, linewidth=0.04](10.510031,-0.7733301)(10.510031,-0.9733301)
\psline[linecolor=black, linewidth=0.04](9.310031,-0.7933301)(9.310031,-0.99333006)
\psline[linecolor=black, linewidth=0.04](6.9000306,-0.7683301)(6.9000306,-0.9683301)
\psline[linecolor=black, linewidth=0.04](11.70503,-0.7633301)(11.70503,-0.9633301)
\psbezier[linecolor=black, linewidth=0.02](0.24003068,-0.87333006)(0.40091127,-0.8802878)(0.4157056,-1.0459676)(0.55503064,-0.988330078125)(0.6943557,-0.9306926)(0.8564128,1.0842288)(1.3050307,1.0766699)(1.7536485,1.0691111)(2.000564,-0.97913337)(2.1400306,-0.98833007)(2.2794971,-0.99752676)(2.2609444,-0.87863964)(2.4200306,-0.8833301)
\psbezier[linecolor=black, linewidth=0.02](1.4450307,-0.86833006)(1.6059113,-0.87528783)(1.6207056,-1.0409675)(1.7600306,-0.983330078125)(1.8993558,-0.9256926)(2.0614128,1.0892287)(2.5100307,1.0816699)(2.9586484,1.0741111)(3.2055643,-0.9741334)(3.3450308,-0.9833301)(3.484497,-0.99252677)(3.4659445,-0.87363964)(3.6250308,-0.87833005)
\psbezier[linecolor=black, linewidth=0.02](2.6450307,-0.87333006)(2.8059113,-0.8802878)(2.8207057,-1.0459676)(2.9600306,-0.988330078125)(3.0993557,-0.9306926)(3.2614129,1.0842288)(3.7100306,1.0766699)(4.1586485,1.0691111)(4.4055643,-0.97913337)(4.5450306,-0.98833007)(4.6844974,-0.99752676)(4.6659446,-0.87863964)(4.825031,-0.8833301)
\psbezier[linecolor=black, linewidth=0.02](3.8400307,-0.87333006)(4.000911,-0.8802878)(4.0157056,-1.0459676)(4.1550307,-0.988330078125)(4.294356,-0.9306926)(4.456413,1.0842288)(4.9050307,1.0766699)(5.3536487,1.0691111)(5.600564,-0.97913337)(5.740031,-0.98833007)(5.879497,-0.99752676)(5.8609443,-0.87863964)(6.0200305,-0.8833301)
\rput[bl](1.5150306,1.05167){\normalsize{$\u_0$}}
\rput[bl](2.7250307,1.03167){\normalsize{$\u_1$}}
\psbezier[linecolor=black, linewidth=0.02](7.0400305,-0.86333007)(7.245911,-0.87028784)(7.4157057,-0.51596755)(7.5450306,-0.068330078125)(7.6743555,0.37930736)(7.656413,1.0942287)(8.105031,1.0866699)(8.553649,1.0791111)(8.495564,0.41086662)(8.650031,-0.043330077)(8.804497,-0.49752676)(8.925944,-0.87363964)(9.220031,-0.87333006)
\psbezier[linecolor=black, linewidth=0.02](10.640031,-0.87333006)(10.845911,-0.8802878)(11.015706,-0.52596754)(11.145031,-0.078330078125)(11.274356,0.36930737)(11.2564125,1.0842288)(11.70503,1.0766699)(12.153648,1.0691111)(12.095564,0.40086663)(12.2500305,-0.05333008)(12.404497,-0.50752676)(12.525945,-0.88363963)(12.82003,-0.8833301)
\psbezier[linecolor=black, linewidth=0.02](9.440031,-0.87333006)(9.645911,-0.8802878)(9.815705,-0.52596754)(9.94503,-0.078330078125)(10.074356,0.36930737)(10.056413,1.0842288)(10.505031,1.0766699)(10.953649,1.0691111)(10.895564,0.40086663)(11.050031,-0.05333008)(11.204497,-0.50752676)(11.325944,-0.88363963)(11.62003,-0.8833301)
\psbezier[linecolor=black, linewidth=0.02](8.24003,-0.87333006)(8.445911,-0.8802878)(8.6157055,-0.52596754)(8.74503,-0.078330078125)(8.874355,0.36930737)(8.856413,1.0842288)(9.305031,1.0766699)(9.753649,1.0691111)(9.695564,0.40086663)(9.850031,-0.05333008)(10.004498,-0.50752676)(10.125944,-0.88363963)(10.420031,-0.8833301)
\rput[bl](8.430031,1.0066699){\normalsize{$\u_0$}}
\rput[bl](9.66003,0.99666995){\normalsize{$\u_1$}}
\rput[bl](9.24003,-1.31833){\normalsize{$\delta$}}
\rput[bl](6.1700306,-1.19833){\normalsize{$x$}}
\rput[bl](13.025031,-1.1683301){\normalsize{$x$}}
\end{pspicture}
}
\caption{Example with shielding (left) and without shielding (right).}
\label{figshield}
\end{figure}
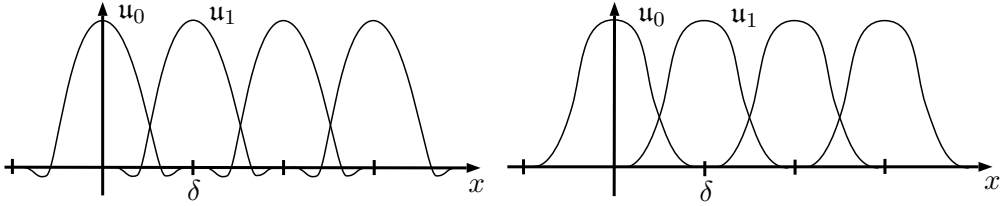

As explained after~\eqref{ELtest}, the purpose of~$\Jtest$ is to restrict attention
to part of the information contained in the EL equations. Of particular interest are situations
when the jets describe only the macroscopic behavior but disregards structures which are
smaller than a microscopic length scale~$\delta$. In order to illustrate this situation in a simple
example, we consider the functions
\[ \u_\ell(x) := \eta(x-\delta \ell) \:,\qquad \ell \in \N_0 \:, \]
where~$\eta \in C^\infty_0((-\delta, \delta))$ (see Figure~\ref{figshield}).
We consider again the Hilbert space~$L^2(\R)$ and choose
\[ A := \text{span} \big( \u_0, \ldots, \u_L \big) \;\subset\; B := \text{span} \big( \u_0, \u_1, \ldots \big) \:, \]
with a parameter~$L>2$. Moreover, we choose~$V=[0,\delta]$. In this example, the shielding property
is more subtle. If the vectors~$\u_\ell$ are orthogonal; i.e.\ by symmetry if
\[ 0 = \la \u_0, \u_1 \ra_{L^2(\R)} = \int_0^\delta \eta(x)\: \eta(\delta-x)\: dx \]
(see the left of Figure~\ref{figshield}), then
\[ A^\perp \cap \overline{B} = \overline{\text{span} \big( \u_{L+1}, \u_{L+2}, \ldots \big)} \:, \]
showing that~$A$ again shields~$V$ from~$B$.
However, if the vectors~$\eta_\ell$ are not orthogonal, then the space~$A^\perp \cap \overline{B}$
contains functions which do {\em{not}} vanish identically in~$V$. Thus the shielding property is violated,
although, intuitively speaking, $A$ does describe the jets in~$V$ completely. The reason for this
seeming inconsistency is that taking the orthogonal complement also involves the behavior of
the functions in~$A$ outside~$V$. More precisely, a short computation shows that
the space~$A^\perp \cap \overline{B}$ is spanned by the vector
\beq \label{overlap}
\u_{L+1} - \kappa\, \u_L + \kappa^2\; \u_{L-1} - \cdots + (-\kappa)^{L+1}\: \u_0
\qquad \text{with} \qquad 
\kappa := \frac{\la \u_0, \u_1 \ra_{L^2(\R)}}{\| u_0\|^2_{L^2(\R)}}
\eeq
as well as the vectors~$\u_{L+2}, \u_{L+3}, \ldots$.
The overlap of the jets~$\u_L \in A$ with~$\u_{L+1} \not\in A$ has the effect
that the vector~\eqref{overlap} does not vanish identically in~$V$.
But at least, equation~\eqref{overlap} shows that the error of shielding decays exponentially if~$L$
is increased, in the sense that for all functions in~$A^\perp \cap \overline{B}$ the inequality
\beq \label{exp}
\| \u \|_{L^2(V)} \leq e^{-\kappa L} \:\|\u\|_{L^2(\R)}
\eeq
holds.

We next consider the typical length scales. As mentioned above, $\delta$ is a microscopic length
scale (which can be thought of as the Planck scale). Therefore, the inequality~\eqref{exp}
shows that shielding takes place on a microscopic length scale. Taking into account that all
other length scales (like for example the constant~$1/c$ in the hyperbolic estimate~\eqref{enes2}
and~\eqref{Gammachoose}) are macroscopic, from the physical point of view the shielding
assumption~\eqref{scond} is an extremely good approximation.

From the mathematical perspective, however, the assumption~\eqref{scond} is too strong
because it is violated in most applications of interest.
Inspecting how~\eqref{scond} enters the proof of Proposition~\ref{prpextend},
one also sees that the extension property does not hold.
Instead, extending a solution necessarily changes the solution slightly in~$V$.
While this effect is not surprising in view of the nonlocality of the causal action principle,
it is a major complication of the mathematical analysis.
For a mathematically convincing treatment, in~\eqref{scond} we must allow for an error
term of a form similar to~\eqref{exp}, and this error term must be controlled in the subsequent
estimates. This method will be introduced when constructing global solutions in Section~\ref{secglobal}.

\subsection{Estimates of the Initial Data} \label{secstrong}
We now analyze if a weak solution is a strong solution. It is most convenient to work with
the weak formulation of Theorem~\ref{thmexistpm}. 
Thus let~$\v \in L^2(L, d\rho)$ be a weak solution of~\eqref{weak3} with zero initial data
at time~$t_0$ (as is made precise in the statement of Theorem~\ref{thmexistpm}).
At this point, it is convenient to work instead of~$\rho$ with the
measure~$\overline{\underline{\rho}}^\tmax_\tmin$. To this end, we divide the
solution in~\eqref{weak3} by the function~$\eta_I$
(which is possible in view of~\eqref{vpdef} and~\eqref{vmdef}).
On the right side of~\eqref{weak3} we may change the integration measure
because the jet~$\u$ vanishes unless~$\eta_I$ is identically equal to one; see~\eqref{uJtestdef}
and~\eqref{oJtestdef}). We thus obtain
a solution of the weak equation
\[ \la \Delta \u, \v \ra_{L^2(L)} = \la \u, \w \ra_{L^2(L)} \qquad
\text{for all~$\u \in \overline{\underline{\J_U}}^\tmax_\tmin$} \:, \]
where we used the abbreviation~$L^2(L) = L^2(L, d\overline{\underline{\rho}}^\tmax_\tmin)$.
After extending~$\v$ by zero to~$U$, we can apply the
transformations of the integrals in the proof of Lemma~\ref{lemmagreen}
and exploit that the symplectic form in~\eqref{green} vanishes in view
of the definition of the jet space~$\overline{\underline{\J_U}}^\tmax_\tmin$ (see~\eqref{uJtestdef} and~\eqref{oJtestdef}). We conclude that~$\la \u, \Delta \v \ra_{L^2(L)}$
is well-defined and satisfies the Green's formula~\eqref{green}
with a vanishing left side. 
As a consequence, the above weak equation can be written as as
\beq \label{strongtest}
\big\la \u, (\Delta \v - \w) \big\ra_{L^2(L)} = 0 \qquad
\text{for all~$\u \in \overline{\underline{\J_U}}^\tmax_\tmin$} \:.
\eeq
This is the strong equation tested with the jet~$\u$. Similar as explained after~\eqref{strongtest0},
this is precisely the equation we are aiming for.

The remaining question is whether and in which sense the weak solution satisfies the
initial conditions. Recall that for the strong solution in~\eqref{cauchyzeroinit},
the trivial initial data was imposed by demanding that~$\v \in \underline{\J_U}_{t_0}$.
In the weak formulation~\eqref{weak}, however, the initial condition is encoded
implicitly by the fact that the test jets~$\u \in \overline{\J_U}^\tmax$ do {\em{not}}
need to vanish at time~$t_0$. But does this equation imply that~$\v$ vanishes at time~$t_0$?
If yes, in which sense?
These questions are rather subtle. In order to understand the basic difficulty, we
``integrate by parts'' in~\eqref{weak} with the help of the Green's formula 
in Lemma~\ref{lemmagreen}. This gives the equation
\beq \label{stronginit}
\la \u, (\Delta \v - \w) \ra_{L^2(L)} = \sigma^{t_0}(\u, \v-\v_0)
\qquad \text{for all~$\u \in \overline{\J_U}^\tmax$}\:.
\eeq
Similar to~\eqref{strongtest}, the left side of this equation is the strong equation tested with~$\u$.
However, this is the formulation where we solve only to the future, making it impossible
to deduce from~\eqref{strongtest} that the left side of~\eqref{stronginit} vanishes.
As a consequence, we cannot conclude that the right side of~\eqref{stronginit} is zero.
In other words, \eqref{stronginit} involves a combination of volume and boundary terms,
making it impossible to read off the boundary data.
The situation does not become easier in Section~\ref{secfuturepast} when constructing solutions to the future
and past, because in the weak formulation of Theorem~\ref{thmexistpm}, the boundary conditions
are encoded only implicitly in the weak equations~\eqref{weak2p} and~\eqref{weak2m}.

In order to clarify the situation, we now give a method for estimating the initial data~$\|\v\|^{t_0}$.
We again use the concept of shielding (see Definition~\ref{defshield}).
The assumption of shielding should be regarded mainly as a technical simplification.
Indeed, in situations where shielding does not hold (as explained at the end of Section~\ref{secextend}),
the following method can still be used if combined in a straightforward way
with quantitative estimates of the error of shielding (see Definition~\ref{defshieldc}
and the proofs of Theorem~\ref{thmshield} in Section~\ref{secglobal} below).

Let~$\v$ be the weak solution constructed in Theorem~\ref{thmexistpm}. Moreover,
choosing a subinterval~$\hat{I} := [\hat{t}_{\min}, \hat{t}_{\max}] \subset I$ and setting
\[ \hat{L} := \bigcup_{t \in \hat{I}} \supp \theta_t \:, \]
the set~$\hat{L}$ is again a lens-shaped region inside~$U$, having the local foliation~$(\eta_t)_{t \in \hat{I}}$.
We again define the sets~$\K(L^\pm)$ by~\eqref{Kpmdef} (and similarly with hats).
\begin{Thm} \label{thmesinit} Assume that
\beq \label{scond2}
\begin{split}
\chi_{\hat{L}^+}\, \Delta\big( \overline{\J_U}^{\hat{t}_{\max}} \big) \qquad &\text{shields $\supp \theta_{t_0}$ from} \qquad 
\text{\rm{span}} \big( \K(\hat{L}^+), \K(L^+) \big)  \\
\chi_{\hat{L}^-}\, \Delta\big( \underline{\J_U}_{\hat{t}_{\min}} \big) \qquad &\text{shields $\supp \theta_{t_0}$ from} \qquad 
\text{\rm{span}} \big( \K(\hat{L}^-), \K(L^-) \big) \:.
\end{split}
\eeq
Then
\[ \|\v\|_{L^2(L, d\rho_{t_0})} \leq \hat{c}\: \|\w\|_{L^2(\hat{L}, d\rho)} \:, \]
where the constant~$\hat{c}$ is given by
\beq \label{hatc}
\hat{c} = 4\,C\, e^{2c\,(\hat{t}_{\max}-\hat{t}_{\min})}\: (\hat{t}_{\max}-\hat{t}_{\min}) \:
\sqrt{ \| \theta_{t_0} \|_{L^\infty} } \:.
\eeq
\end{Thm}
\Proof We first consider the solution~$\v^+$ in~$L^+$ and the corresponding solution~$\hat{\v}^+$
in~$\hat{L}^+$. Applying Proposition~\ref{prpextend}, these solutions coincide in~$\supp \theta_{t_0}$.
Similarly, the solutions~$\v^-$ and~$\hat{\v}^-$ coincide on~$\supp \theta_{t_0}$.
As a consequence, the functions~$\v$ and~$\hat{\v}$ coincide on~$\supp \theta_{t_0}$.
Therefore, it suffices to estimate~$\hat{\v}$. 

The energy estimate~\eqref{vboundpm} gives
\[ \|\hat{\v}\|_{L^2(L, d\rho)} \leq \hat{\Gamma}\: \|\w\|_{L^2(L, d\rho)} \:, \]
where we choose~$\hat{\Gamma} =2 \,(\Gamma^+ + \Gamma^-)$
with~$\Gamma^+$ and~$\Gamma^-$ according to~\eqref{Gammachoose}.
Finally, we estimate the norm on the left by
\[ \|\hat{\v}\|_{L^2(L, d\rho)}^2 = \int_L \|\hat{\v}(x)\|_x^2\: d\rho(x)
\geq \frac{1}{\|\theta_{t_0}\|_{L^\infty}} \int_L \theta_{t_0}(x)\: \|\hat{\v}(x)\|_x^2\: d\rho(x) \:. \]
This gives the result.
\QED

We now explain this result and formulate two corollaries.
The main point of the above estimate is that the constant~$\hat{c}$ in~\eqref{hatc}
becomes small if~$\hat{t}_{\max}-\hat{t}_{\min}$ tend to zero.
This means that the error of the initial values is directly related to the shielding.
In physical applications, shielding occurs on a microscopic scale~$\delta$ (as explained
at the end of Section~\ref{secextend}). Therefore, we can choose~$\hat{t}_{\max}-\hat{t}_{\min} \sim \delta$,
showing that the error in the initial data is extremely small.
Moreover, one sees that an error in the initial data occurs only if~$\w$ does not vanish near
the boundary, as is made precise by the following statement.
\begin{Corollary} Under the assumptions of Theorem~\ref{thmesinit}, the following implication holds:
\[ \w|_{\hat{L}} \equiv 0 \qquad \Longrightarrow \qquad \v|_{\hat{L}} \equiv 0 \:. \]
\end{Corollary}
We finally rewrite the above results for solutions of the Cauchy problem with non-trivial
initial data. To this end, we return to the strong Cauchy problem for non-trivial initial data in~\eqref{cauchystrong}.
The method for solving this equation is to construct a strong solution~$\tilde{\v}$
for the inhomogeneity~$\tilde{\w} = \w - \Delta \v_0$
with trivial initial data~\eqref{cauchystrong} and to set~$\v=\tilde{\v}+\v_0$.
Now suppose that~$\tilde{\v} \in L^2(L)$ is a corresponding weak solution
as constructed in Theorem~\ref{thmexistpm}.
Then the jet~$\v := \tilde{\v}+\v_0$ satisfies in generalization of~\eqref{weak3} the weak equation
\[ \la \Delta \u, (\v-\v_0) \ra_{L^2(L, d\rho)} = \la \u, (\w - \Delta \v_0) \ra_{L^2(L, d\rho)} \qquad
\text{for all~$\u \in \overline{\underline{\J_U}}^\tmax_\tmin$} \:. \]
The equations for~$\v^\pm$ as well as the estimate of Theorem~\ref{thmesinit} are obtained similarly
by the simple replacements
\[ \v \rightarrow \v - \v_0 \qquad \text{and} \qquad \w \rightarrow \w - \Delta \v_0 \:. \]
This gives the following result:
\begin{Corollary} Assume that the shielding property~\eqref{scond2} holds. Then
\[ \|\v - \v_0 \|_{L^2(\hat{L}, d\rho_{t_0})} \leq \hat{c}\: \| \Delta \v_0 - \w \|_{L^2(\hat{L}, d\rho)} \]
with~$\hat{c}$ as in~\eqref{hatc}. Moreover, the following implication holds:
\[ \big( \Delta \v_0 - \w \big) \big|_{\hat{L}} \equiv 0 \qquad \Longrightarrow \qquad
\big(\v - \v_0 \big) \big|_{\hat{L}} \equiv 0 \:. \]
\end{Corollary}
Intuitively speaking, this result can be understood as follows:
If the Cauchy problem for the initial value~$\v_0$ can be solved ``locally'' in a
small lens-shaped region~$\hat{L}$, then it also has a solution in the
larger lens-shaped region~$L$. The size of~$\hat{L}$ is determined
by the shielding of the jets; in physical applications this size will be of the order of the
microscopic length scale~$\delta$. This result fits nicely to our earlier concept
of prescribing the initial data not on a hypersurface, but in a surface layer.
In applications, it seems natural and easiest to choose the width of the surface layers
of the same order as the length scale~$\delta$ of shielding.

\section{Causal Structure and Global Hyperbolicity} \label{secglobhypfol}
\subsection{Causal Cones and Transitive Causal Relations} \label{seccone}
In this section we shall clarify the causal structure of space-time by introducing
causal cones. In particular, we shall get the connection to partially ordered sets.
Our method is to construct cone structures from the lens-shaped regions.
The construction is based on the assumption that there are arbitrarily large
lens-shaped regions, as is made precise by the following notion of compact hyperbolicity:
\begin{Def} \label{defcompacthyp}
Space-time is {\bf{locally hyperbolic}} if every~$x \in M$ has an open
neighborhood~$\Omega$ contained in a lens-shaped region~$L$.
It is {\bf{compactly hyperbolic}} if every compact subset~$K \subset M$ has an open
neighborhood~$\Omega \supset K$ contained in a lens-shaped region~$L$.
\end{Def} \noindent
In what follows, for every lens-shaped region we always choose a corresponding
local foliation~$(\eta_t)_{t \in [0, \tmax]}$ inside a set~$U$ (see Definition~\ref{deflens}).
For ease in notation, the corresponding objects~$(L, U, \eta_t)$ always
carry the same indices, tildes and hats.

Since~$M$ is $\sigma$-compact (see the last paragraph of Section~\ref{seccvp}
on page~\pageref{sigmacompact}), we can choose an
{\em{exhaustion}} of~$M$ by compact sets~$(K_n)_{n \in \N}$, i.e.\
\[ %\beq \label{Kexhaust}
K_1 \subset \overset{\circ}{K}_2 \subset K_2 \subset \overset{\circ}{K}_3 \subset \cdots \qquad \text{and} \qquad
\bigcup_{n \in \N} K_n = M \:.
\] %\eeq
In the following constructions, we will frequently work with such exhaustions.
Clearly, we must always verify that the resulting objects and notions do not depend on the choice of the exhaustion.

\begin{figure}
% \usepackage[usenames,dvipsnames]{pstricks}
% \usepackage{epsfig}
% \usepackage{pst-grad} % For gradients
% \usepackage{pst-plot} % For axes
% \usepackage[space]{grffile} % For spaces in paths
% \usepackage{etoolbox} % For spaces in paths
% \makeatletter % For spaces in paths
% \patchcmd\Gread@eps{\@inputcheck#1 }{\@inputcheck"#1"\relax}{}{}
% \makeatother
% 
\psscalebox{1.0 1.0} % Change this value to rescale the drawing.
{
\begin{pspicture}(0,-1.131183)(7.8363075,1.131183)
\definecolor{colour0}{rgb}{0.8,0.8,0.8}
\definecolor{colour1}{rgb}{0.4,0.4,0.4}
\pspolygon[linecolor=colour0, linewidth=0.02, fillstyle=solid,fillcolor=colour0](0.8664722,-0.3139131)(1.9914721,-0.4739131)(3.3664722,-0.9389131)(3.891472,-1.0339131)(4.756472,-1.0189131)(5.9314723,-0.6739131)(7.041472,-0.48391312)(7.4514723,-0.45391312)(6.711472,-0.2839131)(6.151472,0.14608689)(5.261472,0.77608687)(4.336472,1.0860869)(3.631472,1.0860869)(2.641472,0.6660869)(1.7814722,0.116086885)
\rput[bl](5.3514724,-0.4389131){\normalsize{$\hat{L}$}}
\psbezier[linecolor=colour0, linewidth=0.08](0.001472168,-0.38258958)(1.9131359,-0.45328024)(2.3415318,1.1347313)(4.1507826,1.0710868835449219)(5.9600334,1.0074425)(6.2261634,-0.6253353)(7.8314724,-0.42891312)
\psbezier[linecolor=colour0, linewidth=0.08](0.05147217,-0.38391313)(2.6490345,-0.22647065)(2.894832,-1.127449)(4.2015834,-1.078913116455078)(5.508335,-1.0303773)(5.506106,-0.59076226)(7.791472,-0.4339131)
\psbezier[linecolor=colour1, linewidth=0.08](2.5714722,0.6410869)(2.9968557,0.8328486)(3.367412,1.1010426)(4.1158037,1.0810868835449219)(4.864196,1.0611311)(5.277158,0.77852213)(5.546472,0.61812395)
\psbezier[linecolor=colour1, linewidth=0.08](5.521472,0.6381239)(5.0960884,0.4463622)(4.1955323,-0.3118318)(3.7771404,-0.2618760794180389)(3.3587487,-0.21192035)(2.8607867,0.4806887)(2.5914721,0.6410869)
\pspolygon[linecolor=colour1, linewidth=0.02, fillstyle=solid,fillcolor=colour1](2.6564722,0.6410869)(3.046472,0.27108687)(3.5314722,-0.15391311)(3.8414721,-0.24391311)(4.3514724,-0.058913115)(4.8914723,0.2810869)(5.2064724,0.4660869)(5.501472,0.6460869)(5.3064723,0.7460869)(5.0364723,0.8760869)(4.7464724,0.97608685)(4.341472,1.0410869)(3.9564722,1.0710869)(3.3864722,0.97608685)(2.9664721,0.8260869)
\rput[bl](2.0514722,-0.048913117){\normalsize{$L$}}
\psbezier[linecolor=black, linewidth=0.02, arrowsize=0.05291667cm 2.0,arrowlength=1.4,arrowinset=0.0]{->}(2.3997917,0.044960637)(2.5891407,-0.021133604)(2.7858722,-0.22205666)(3.1081526,0.05721312875879619)
\end{pspicture}
}
\caption{Two lens-shaped regions~$L$ and~$\hat{L}$ with~$L \ll \hat{L}$.}
\label{figpc}
\end{figure}
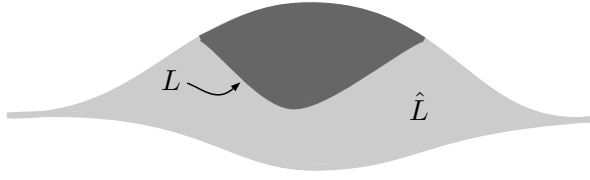

\begin{Def} A lens-shaped region~$L$ is {\bf{past-contained}} in~$\hat{L}$,
denoted by~$L \ll \hat{L}$, if
\beq \label{pc}
U=\hat{U} \qquad \text{and} \qquad \overline{\J_U}^\tmax \supset \overline{\J_{\hat{U}}}^\tmax \:.
\eeq
\end{Def} \noindent
The inclusion in~\eqref{pc} means that~$L$ involves weaker boundary conditions
at~$\tmax$ than~$\hat{L}$, which in turn can be understood intuitively by the condition that
the future boundary of~$L$ must be contained in the future boundary of~$\hat{L}$ (see Figure~\ref{figpc}).

\begin{Def} \label{defcausal}
Let~$(K_n)_{n \in \N}$ be an exhaustion of~$M$ by compact sets.
Given~$x \in M$ and~$N \in \N$, the set~$J_N^\vee(x) \subset M$ is defined as the set of all space-time points~$y$
with the property that for all lens-shaped regions~$\hat{L}$ and~$L$ 
with~$L \ll \hat{L} \supset K_N$ and~$x$ contained in the
interior of~$L$, the point~$y$ lies in~$L$, i.e.\
\beq \label{defJN}
J^\vee_N(x) := \Big\{ y \in M \:\Big|\:
\mathop{\text{\Large $\forall$}}_{\hat{L} \supset K_N} \:\mathop{\text{\Large $\forall$}}_{L \ll \hat{L}} \::\:
x \in \overset{\circ}{L}  \Longrightarrow y \in L \Big\} \:.
\eeq
Its interior is denoted by~$I_N^\vee(x)$,
\beq \label{defIN}
I_N^\vee(x) := \overbrace{J_N^\vee(x)}^{\circ} \:.
\eeq
Moreover, the sets~$J^\vee(x)$ and~$I^\vee(x)$ are defined by taking the union over~$N$,
\beq \label{defIJ}
J^\vee(x) := \bigcup_{N \in \N} J_N^\vee(x) \qquad \text{and} \qquad
I^\vee(x) := \bigcup_{N \in \N} I_N^\vee(x) \:.
\eeq
Finally, for a compact set~$K \subset M$, we set
\beq \label{defIJK}
J^\vee(K) := \bigcup_{x \in K} J^\vee(x) \qquad \text{and} \qquad
I^\vee(K) := \bigcup_{x \in K} I^\vee(x) \:.
\eeq
We refer to~$J^\vee$ as the {\bf{future cone}} and~$I^\vee$ as the {\bf{open future cone}}.
\end{Def}
We note that, being the union of open sets, the open future light cone is indeed an open subset of~$M$.
We next verify that the definitions are independent of the choice of the exhaustion:
\begin{Prp}
The sets~$J^\vee(x)$ and~$I^\vee(x)$
do not depend on the choice of the exhaustion~$(K_n)_{n \in \N}$.
\end{Prp} 
\Proof
Using quantifiers, the sets~$J^\vee(x)$ and~$I^\vee(x)$ can be written as
\begin{align}
J^\vee(x) &= \Big\{ y \in M \:\Big|\: \mathop{\text{\Large $\exists$}}_{N \in \N}\:
\mathop{\text{\Large $\forall$}}_{\hat{L} \supset K_N} \:\mathop{\text{\Large $\forall$}}_{L \ll \hat{L}} \::\:
x \in \overset{\circ}{L}  \Longrightarrow y \in L \Big\} \label{Jquant} \\
I^\vee(x) &= \Big\{ y \in M \:\Big|\: \mathop{\text{\Large $\exists$}}_{N \in \N}\:
\mathop{\text{\Large $\exists$}}_{U_y \ni y}\:
\mathop{\text{\Large $\forall$}}_{\hat{L} \supset K_N} \:\mathop{\text{\Large $\forall$}}_{L \ll \hat{L}} \::\:
x \in \overset{\circ}{L}  \Longrightarrow U_y \subset L \Big\} \:, \label{Iquant}
\end{align}
where~$U_y$ denotes an open neighborhood of~$y$ in~$M$.
Since the sets~$J^\vee_N(x)$ are increasing, i.e.\
\beq \label{increase}
J_1^\vee(x) \subset J_2^\vee(x) \subset \cdots \:,
\eeq
their union~$J^\vee(x)$ is characterized purely by the lens-shaped regions enclosing
large compact sets~$K_N$ for large~$N$. In particular, it is independent of the choice
of the exhaustion. The proof for~$I^\vee(x)$ is similar.
\QED

We also point out that the set~$I^\vee(x)$ in general does not coincide with the interior of~$J^\vee(x)$.
Namely, writing this interior as
\beq \label{Iqbad}
\overbrace{J^\vee(x)}^{\circ} = 
\Big\{ y \in M \:\Big|\: \mathop{\text{\Large $\exists$}}_{U_y \ni y}\;
\mathop{\text{\Large $\forall$}}_{\tilde{y} \in U_y}\:
\mathop{\text{\Large $\exists$}}_{N \in \N}\:
\mathop{\text{\Large $\forall$}}_{\hat{L} \supset K_N} \:\mathop{\text{\Large $\forall$}}_{L \ll \hat{L}} \::\:
x \in \overset{\circ}{L}  \Longrightarrow \tilde{y} \in L \Big\} \:,
\eeq
the parameter~$N$ may depend on~$\tilde{y}$, giving rise to a weaker condition. Therefore,
in general we only have the inclusion
\[ I^\vee(x) \subset \overbrace{J^\vee(x)}^{\circ} \:. \]
This situation resembles similar results in low regularity Lorentzian geometry;
see for example~\cite{chrusciel+grant, grant+kunziger}.
Before coming back to this subtle point (see Section~\ref{secglobhyp}), we prove that the sets~$I^\vee(x)$
induce a transitive causal relation on space-time:
\begin{Thm} \label{thmtransitive}
The partial relation defined by the sets~$I^\vee(x)$ is {\bf{transitive}}, meaning that
\[ y \in I^\vee(x) \quad \text{and} \quad z \in I^\vee(y) \qquad \Longrightarrow \qquad
z \in I^\vee(x) \:. \]
\end{Thm}
\Proof 
Let~$(K_n)_{n \in \N}$ be an exhaustion by compact sets.
Then, by our definition~\eqref{defIJ} and using that the sets~$I_n^\vee$
are increasing in view of~\eqref{increase}, there is~$N$ such that
\[ y \in I_N^\vee(x) \quad \text{and} \quad z \in I_N^\vee(y) \:. \]
Thus, using~\eqref{defIN}, there are open neighborhoods~$U_y$ of~$y$ and~$U_z$ of~$z$ with
\beq \label{inclusions}
U_y \subset J_N^\vee(x) \quad \text{and} \quad U_z \in J_N^\vee(y) \:.
\eeq

We choose any lens-shaped regions~$\hat{L}$ and~$L$ with~$L \ll \hat{L} \supset K_N$ and~$x \in
\overset{\circ}{L}$. Combining the first inclusion in~\eqref{inclusions} with~\eqref{defJN},
it follows that~$U_y \subset L$. Hence~$y \in \overset{\circ}{L}$, and combining
the second inclusion in~\eqref{inclusions} again with~\eqref{defJN}, we conclude that~$U_z \subset L$.
It follows that~$U_z \subset J_N^\vee(x)$ and thus~$z \in I_N^\vee(x)$.
Using again~\eqref{defIJ} implies that~$z \in I^\vee(x)$, concluding the proof.
\QED

The result of this lemma allows us to introduce the relation~$\ll$ on~$M \times M$
by the condition that~$x \ll y$ if~$x=y$ or if~$y \in I^\vee(x)$.
According to Theorem~\ref{thmtransitive}, this relation is transitive.
Forming equivalence classes of points~$x,y$ for which~$x \ll y$ and~$y \ll x$,
we obtain the structure of a {\em{partially ordered set}}.
Such a structure was already obtained in the setting of causal fermion systems in~\cite[Sections~5.1 and~5.2]{nrstg}
with a different construction. The method here has the advantage that it is conceptually more convincing and
works in greater generality. We note that if space-time is discrete and the sets~$\{z \in M \:|\: x \ll z \ll y \}$
are finite for all~$x, y \in M$, one recovers the structure of a {\em{causal set}} (see for example~\cite{sorkin}).

We close with two short remarks. We first note that defining the relation~$\ll$ on~$M \times M$
alternatively by the condition~$y \in I^\vee(x)$ also gives rise to transitive causal relations.
But with this alternative definition, the relation~$x \ll x$ need not hold, so that the
above equivalence classes might be empty. We also point out that, in general,
the closed light cones~$J^\vee(x)$ do {\em{not}} seem to give rise to a transitive relation.

\subsection{Definition of Global Retarded Weak Solutions} \label{secdefglobal}
We now introduce the notion of global retarded weak solutions of
the linearized field equations.
A global weak solution~$\v \in L^2_\loc(M, d\rho)$ is defined by the inhomogeneous weak equation
\beq \label{globweak}
\la \Delta \u, \v \ra_{L^2(M, d\rho)} = \la \u, \w \ra_{L^2(M, d\rho)} \qquad
\text{for all~$\u \in \Jvary_0$} \:.
\eeq
 For technical simplicity, for the moment we restrict attention to
inhomogeneities with compact support, i.e.
\beq \label{wcompact}
\w \in L^2_0(M, d\rho) \:,
\eeq
where~$L^2_0(M, d\rho)$ denotes the square integrable jets with essentially compact support
(more general inhomogeneities will be considered in Section~\ref{secpastcompact}).

It remains to make precise what we mean by a {\em{retarded}} solution.
In order to implement the notion that the solution should vanish ``in the distant past''
we again assume that space-time is compactly hyperbolic.
Then we can choose an {\em{exhaustion}} of~$M$ by lens-shaped regions~$(L_n)_{n \in \N}$, i.e.
\beq \label{Lexhaust}
L_1 \subset U_1 \subset L_2 \subset U_2 \subset \cdots \qquad \text{and} \qquad
\bigcup_{n \in \N} U_n = M \:.
\eeq
We take it as a definition that~$\v$ should be a local $L^2$-limit of retarded solutions
in the lens-shaped regions~$L_n$:
\begin{Def} \label{defretarded} Assume that space-time is compactly hyperbolic.
A global weak solution~$\v \in L^2_\loc(M, d\rho)$ of~\eqref{globweak} with compactly supported inhomogeneity~\eqref{wcompact}
is said to be {\bf{retarded}} if there is an exhaustion by lens-shaped regions~$(L_n)_{n \in \N}$~\eqref{Lexhaust}
such that the corresponding weak solutions~$\v_n \in L^2(L_n, d\rho)$ of the
Cauchy problem with zero initial data, i.e.
\beq \label{vnweak}
\la \Delta \u, \v_n \ra_{L^2(L_n, d\rho)} = \la \u, \w \ra_{L^2(L_n, d\rho)} \qquad
\text{for all~$\u \in \overline{\J_{U_n}}^\tmax$}\:,
\eeq
converge in~$L^2_\loc(M, d\rho)$ to~$\v$.
\end{Def} \noindent
We remark for clarity that, since this definition makes a statement on convergence as~$n \rightarrow \infty$,
it suffices to consider the Cauchy problem~\eqref{vnweak} for large~$n$.
In particular, one may restrict attention to the case that the support of~$\w$ is contained in~$L_n$.

\subsection{Constructing Unique Global Weak Retarded Solutions} \label{secglobal}
We now give a procedure for constructing global retarded weak solutions of the linearized field equations
and specify all the necessary assumptions.
In order to keep the setting as simple as possible, we shall make the
following assumption:
\begin{Def} \label{deffiniterange}
The {\bf{Lagrangian}} has {\bf{finite range}} if for every lens-shaped region~$L$
inside~$U$, the set~$U$ can be chosen to be relatively compact.
\end{Def} \noindent
This assumption could be replaced by suitable decay assumptions on the Lagrangian;
for the sake of technical simplicity, we shall not enter such generalizations here.

Let~$\w \in L^2_0(M, d\rho)$ be a compactly supported jet.
Assuming again that~$M$ is compactly hyperbolic
and using that~$M$ is $\sigma$-compact, we can exhaust space-time by
a sequence of lens-shaped regions~$(L_n)_{n \in N}$ inside space-time regions~$(U_n)_{n \in \N}$
with local foliations~$(\eta_{n,t})_{t \in [t_0, \tmax]}$. Moreover, we
choose~$L_1$ such that it contains the support of~$\w$, i.e.
\[ \supp \w \;\subset\; L_1 \subset U_1 \subset L_2 \subset U_2 \subset \cdots \qquad \text{and} \qquad
\bigcup_{n \in \N} U_n = M \:. \]
Using property~(ii) in Definition~\ref{deflocfoliate}, these lens-shaped regions are indeed nested in the
sense of Definition~\ref{defnested}.
Applying the existence result of Theorem~\ref{thmexist} in each lens-shaped region, we
obtain a sequence of solutions with zero initial data.
Working for convenience with the weak equation~\eqref{weak2p}, we obtain
a sequence of weak solutions~$\v_n$ of~\eqref{vnweak}.
Our goal is to show that this sequence of weak solutions converges in a suitable sense
to the desired solution~$\v$. The difficulty is that the solutions~$\v_n$
are not unique (see Proposition~\ref{prpnonunique}), implying that the solutions
do not need to coincide locally. Thus, using again
the notion introduced in Section~\ref{secextend}, we need to control the
shielding.
\begin{Def} \label{defshieldc} Given two lens-shaped regions~$L$ and~$\hat{L}$, we define the
{\bf{shielding constant}} by
\[ \shield(V, L, \hat{L}) = \sup \bigg\{ \frac{\|\u\|_{L^2(V, d\rho)}}{\|\u\|_{L^2(L, d\rho)}} \:\bigg|\:
\u \in  \Big( \chi_L\, \Delta\big( \overline{\J_U}^\tmax \big) \Big)^\perp 
\cap \overline{\text{\rm{span}} \big( \K(\hat{L}), \K(L) \big)} \bigg\} \:. \]
\end{Def} \noindent
If the shielding constant vanishes, we obtain shielding in the sense of Definition~\ref{defshield}.
Therefore, the shielding constant quantifies to which extent shielding is violated.

In the following theorem we control the shielding by a condition which involves both
the shielding constant and the constant~$\Gamma$ in the energy estimate
of Proposition~\ref{prpenes}. In order to get finer control of the dependence 
on the considered space-time region, we introduce the
constant~$\Gamma(L, \hat{L})$ by modifying the inequality~\eqref{hyprev} to
\[ %\beq \label{hypVVh}
\|\u\|_{L^2(L)} \leq \Gamma(L, \hat{L}) \: \|\Delta \u\|_{L^2(L)} \qquad \text{for all~$\u \in \overline{\J_{\hat{U}}}^\tmax$}
\] %\eeq
(thus this is an estimate in~$L$, but the jet space must vanish only
in the future of the bigger lens-shaped region~$\hat{L}$).
Typically, the constant~$\Gamma(L, \hat{L})$ stays finite in the limiting case that~$L$ is fixed
and~$\hat{L}$ exhausts the whole space-time.

\begin{Thm} \label{thmshield}
Assume that the Lagrangian has finite range
and that space-time is compactly hyperbolic.
Moreover, assume that the shielding constant goes to zero so fast that
every~$x \in M$ has an open neighborhood~$V$ such that
\beq \label{Gs}
\sum_{n=1}^\infty \shield(V, L_n, L_{n+1})\: \big( \Gamma(L_n, L_n) + \Gamma(L_n, L_{n+1}) \big) < \infty \:.
\eeq
Then for any compactly supported~$\w \in L^2(M, d\rho)$ there is a global retarded weak solution.
\end{Thm}
\Proof We introduce the subspaces
\[ A := \Delta \big( \overline{\J_{U_n}}^\tmax \big) \:,\;
B:= \text{span} \big( \K(L_n), \K(L_{n+1}) \big)  \;\; \subset \;\; L^2(L_{n+1}, d\rho) \:. \]
The consideration in the proof of Proposition~\ref{prpextend} shows that
\[ \v_{n+1} - \v_n \in A^\perp \cap \overline{B}\:. \]

Given~$x \in M$, we choose an open neighborhood~$V$ such that~\eqref{Gs} holds.
Then, by definition of the shielding constant,
\begin{align}
\| \v_{n+1} - \v_n \|_{L^2(V, d\rho)} &\leq \shield(V, L_n, L_{n+1})\: \| \v_{n+1} - \v_n \|_{L^2(L_n, d\rho)} \notag \\
&\leq \shield(V, L_n, L_{n+1})\: \big( \Gamma(L_n, L_n) + \Gamma(L_n, L_{n+1}) \big)\:
\| \w \|_{L^2(V, d\rho)} \:. \label{shieldes}
\end{align}
The assumption~\eqref{Gs} ensures that the sequence~$\v_n$ converges in~$L^2(V, d\rho)$
to a function~$\v \in L^2(V, d\rho)$.
Since~$V$ can be chosen as a small neighborhood of any point~$x \in M$,
we conclude that~$\v_n$ converges in~$L^2_\loc(M, d\rho)$ to~$\v \in L^2_\loc(M, d\rho)$.

In particular, $\v_n$ converges in~$L^2$ in the lens-shaped region~$L_2$ and therefore in~$U_1$.
Using that the Lagrangian has finite range, we can apply
Lebesgue's dominated convergence theorem to infer that~$\Delta \v_n$ converges in~$L^2$ in
the lens-shaped region~$L_1$. Hence~$\Delta \v_n|_V \rightarrow \Delta \v|_V$ in~$L^2(V, d\rho)$.
Again using that~$V$ can be chosen as a small neighborhood of any point~$x \in M$,
we conclude that~$\Delta \v_n$ converges in~$L^2_\loc(M, d\rho)$ to~$\Delta \v \in L^2_\loc(M, d\rho)$.
To summarize,
\beq \label{vloc}
\v_n \rightarrow \v \text{ in~$L^2_\loc(M, d\rho)$} \qquad \text{and} \qquad
\Delta \v_n \rightarrow \Delta \v \text{ in~$L^2_\loc(M, d\rho)$} \:.
\eeq

Let us verify that~$\v$ satisfies the weak equation~\eqref{globweak}.
Thus let~$\u \in \Jvary_0$. Then there is~$n$ such that~$\supp \u \subset U_n$.
Using again that the Lagrangian has finite range, it follows that~$\u \in \overline{\J_{U_{n+2}}}^\tmax$.
Hence~\eqref{globweak} holds for all~$\v_\ell$ and all sufficiently large~$\ell$.
Using~\eqref{vloc}, we can take the limit~$\ell \rightarrow \infty$ to conclude that~$\v$
satisfies~\eqref{globweak}.
\QED

Let us briefly discuss condition~\eqref{Gs}. As explained after~\eqref{exp}, shielding
takes place on a microscopic length scale~$\delta$. This means that, similar to~\eqref{exp},
the shielding constant~$\shield(V, L_n, L_{n+1})$ should decay exponentially on the scale~$\delta$
if~$n$ is increased. The constant~$\Gamma$ of the energy estimate, however,
increases exponentially on a macroscopic scale. With this in mind, the bound~\eqref{Gs}
seems unproblematic and easy to verify in the applications.

We finally explain in which sense global retarded weak solutions are unique. 
We first recall that the equations~\eqref{vnweak} determine the~$\v_n$ only up to
vectors which are orthogonal to the subspace
\[ \chi_L \,\overline{\J_{U_n}}^\tmax
\;\subset\; L^2(M, d\rho) \:. \]
Similar as explained in Section~\ref{secunique}, this means that
the global retarded weak solutions are unique up to microscopic details which 
we deliberately filtered out by our choice of~$\Jtest$.
Nevertheless, we can hope that, similar as in Theorem~\ref{thmexist}, our
construction gives a distinguished solution, which is determined uniquely by
our construction. This is indeed the case under suitable assumptions, as we now explain.

\begin{Def} \label{defuep}
Space-time has the {\bf{uniform shielding property}}
if every~$x \in M$ has an open neighborhood~$V$ such that for any exhaustion
by lens-shaped regions~$(L_n)_{n \in \N}$ of the form~\eqref{Lexhaust}, the shielding condition~\eqref{Gs} holds.
\end{Def}

\begin{Thm} If space-time has the uniform shielding property, then
the global retarded weak solution of Theorem~\ref{thmshield} does not depend on the choice
of the exhaustion.
\end{Thm}
\Proof We consider two exhaustions by lens-shaped regions~$(L_n)_{n \in \N}$
and~$(\tilde{L}_n)_{n \in \N}$. We iteratively choose subsequences~$(L_{n_k})_{k \in \N}$
and~$(\tilde{L}_{\tilde{n}_k})_{k \in \N}$ such that
\[ L_{n_1} \subset U_{n_1} \subset \tilde{L}_{\tilde{n}_1} \subset \tilde{U}_{\tilde{n}_1} \subset
L_{n_2} \subset \cdots \:. \]
We denote the resulting exhaustion by~$(\hat{L}_n)_{n \in \N}$.
Theorem~\ref{thmshield} gives a corresponding global retarded weak solution~$\hat{\v}$.
This solution coincides with both~$\v$ and~$\hat{\v}$, concluding the proof.
\QED

\subsection{Finite Propagation Speed} \label{secspeed}

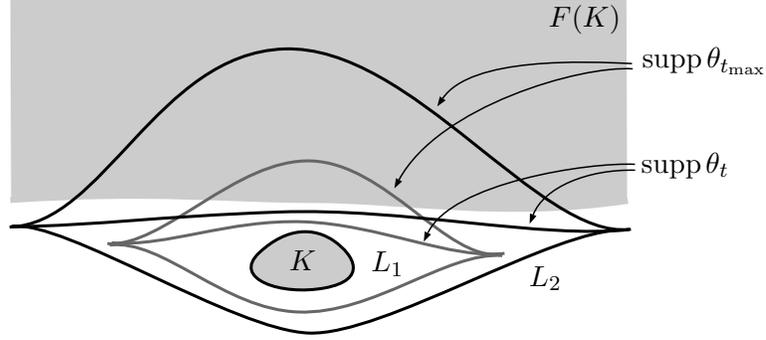
\begin{figure}
% \usepackage[usenames,dvipsnames]{pstricks}
% \usepackage{epsfig}
% \usepackage{pst-grad} % For gradients
% \usepackage{pst-plot} % For axes
% \usepackage[space]{grffile} % For spaces in paths
% \usepackage{etoolbox} % For spaces in paths
% \makeatletter % For spaces in paths
% \patchcmd\Gread@eps{\@inputcheck#1 }{\@inputcheck"#1"\relax}{}{}
% \makeatother
% 
\psscalebox{1.0 1.0} % Change this value to rescale the drawing.
{
\begin{pspicture}(-2.9,-2.2343907)(14.890399,2.2343907)
\definecolor{colour0}{rgb}{0.8,0.8,0.8}
\definecolor{colour1}{rgb}{0.4,0.4,0.4}
\pspolygon[linecolor=colour0, linewidth=0.02, fillstyle=solid,fillcolor=colour0](0.030398808,-0.40562478)(0.6903988,-0.37562478)(1.5903988,-0.3656248)(2.200399,-0.35062477)(2.8053987,-0.40062478)(3.365399,-0.41562477)(4.0403986,-0.42062476)(5.095399,-0.46562478)(5.640399,-0.4906248)(6.535399,-0.53062475)(6.950399,-0.53062475)(7.350399,-0.52562475)(8.200398,-0.44562477)(8.190399,2.2243752)(0.020398807,2.2243752)
\psbezier[linecolor=colour0, linewidth=0.08](0.010398808,-0.43562478)(1.9096414,-0.3081823)(2.8472376,-0.43416065)(4.049286,-0.43562477111816406)(5.2513347,-0.4370889)(6.5495214,-0.69247395)(8.220399,-0.44562477)
\rput[bl](7.1703987,1.7743752){\normalsize{$F(K)$}}
\rput[bl](4.8103986,-1.4456248){\normalsize{$L_1$}}
\psbezier[linecolor=colour1, linewidth=0.04](1.3003988,-1.0156248)(2.0415082,-1.0706204)(3.0110033,0.11913682)(4.010399,0.08437522888183593)(5.009794,0.049613643)(5.8104186,-1.2410601)(6.5203986,-1.1556247)
\psbezier[linecolor=black, linewidth=0.04, fillstyle=solid,fillcolor=colour0](3.3203988,-1.1356248)(2.945184,-1.4843417)(3.6114624,-1.7139652)(4.1903987,-1.6056247711181642)(4.7693353,-1.4972843)(4.5802197,-1.1709303)(4.320399,-0.9856248)(4.060578,-0.80031925)(3.6956136,-0.7869078)(3.3203988,-1.1356248)
\rput[bl](3.720399,-1.3756248){\normalsize{$K$}}
\psbezier[linecolor=colour1, linewidth=0.04](6.550399,-1.1756247)(5.8092895,-1.1206291)(4.8397946,-1.9603864)(3.8403988,-1.925624771118164)(2.8410032,-1.8908632)(2.0403788,-0.9501894)(1.3303988,-1.0356247)
\psbezier[linecolor=black, linewidth=0.04](0.0,-0.7932579)(1.2515082,-0.76875496)(1.9610032,1.5120953)(3.6503987,1.574375228881836)(5.3397946,1.6366552)(7.050419,-0.96290886)(8.240398,-0.82562476)
\psbezier[linecolor=black, linewidth=0.04](8.170399,-0.83562475)(7.2392893,-0.81062907)(4.9797945,-2.2403862)(3.980399,-2.2056247711181642)(2.9810033,-2.1708632)(0.7403789,-0.7001894)(0.030398808,-0.78562474)
\psbezier[linecolor=colour1, linewidth=0.04](1.3603988,-1.0056248)(2.1015084,-1.0606204)(2.911003,-0.6908632)(3.9103987,-0.725624771118164)(4.9097943,-0.76038635)(5.8704185,-1.2310601)(6.580399,-1.1456248)
\psbezier[linecolor=black, linewidth=0.04](0.020398807,-0.7932579)(1.2715083,-0.76875496)(1.8860031,-0.65790474)(3.575399,-0.595624771118164)(5.2647943,-0.5333448)(7.070419,-0.96290886)(8.260399,-0.82562476)
\rput[bl](6.910399,-1.6356248){\normalsize{$L_2$}}
\rput[bl](8.400399,-0.14562477){\normalsize{$\supp \theta_t$}}
\psbezier[linecolor=black, linewidth=0.02, arrowsize=0.05291667cm 2.0,arrowlength=1.4,arrowinset=0.0]{->}(8.314827,0.03933119)(7.6901855,0.069731966)(5.875465,-0.20882921)(5.5059705,-0.9905807312906336)
\psbezier[linecolor=black, linewidth=0.02, arrowsize=0.05291667cm 2.0,arrowlength=1.4,arrowinset=0.0]{->}(8.316354,-0.038951535)(7.6830916,-0.024244277)(7.1029334,-0.2672473)(6.9144435,-0.7522980084544878)
\rput[bl](8.4103985,1.2343752){\normalsize{$\supp \theta_{\tmax}$}}
\psbezier[linecolor=black, linewidth=0.02, arrowsize=0.05291667cm 2.0,arrowlength=1.4,arrowinset=0.0]{->}(8.284827,1.3093312)(7.589023,1.3568523)(5.7075596,0.81141806)(5.1159706,-0.2805807312906336)
\psbezier[linecolor=black, linewidth=0.02, arrowsize=0.05291667cm 2.0,arrowlength=1.4,arrowinset=0.0]{->}(8.274827,1.3793312)(7.6501856,1.409732)(6.1354647,1.4711708)(5.6859703,0.8094192687093664)
\end{pspicture}
}
\caption{The Cauchy separated future.}
\label{figseparate}
\end{figure}

\begin{Def} \label{defsepfut} Let~$K \subset M$ be compact.
We let~$F(K)$ be the set of space-time points~$x$ with the property that
there is an open neighborhood~$U \subset M$ of~$x$ and an exhaustion of~$M$
by lens-shaped regions~$(L_n)_{n \in N}$ such that for every~$n \in \N$ there is~$t \in [t_0, \tmax]$ with
\beq \label{sepfut}
\eta_t|_K \equiv 1 \qquad \text{and} \qquad \eta_t|_U \equiv 0 \:.
\eeq
We refer to~$F(K)$ as the {\bf{Cauchy separated future}} of~$K$.
\end{Def} \noindent
Intuitively speaking, the Cauchy separated future of~$K$ consists of all points which can be separated
from~$K$ by surface layers in exhaustions by lens-shaped regions; see Figure~\ref{figseparate}.
\begin{Thm} \label{thmsepfut}
Let~$\w \in L^2_0(M, d\rho)$ with~$\supp \w \subset F(K)$. Then there is
a global retarded weak solution~$\v$ with~$\v|_K \equiv 0$.
\end{Thm}
\Proof Let~$x \in F(K)$. We choose an open neighborhood~$U$ and an exhaustion~$(L_n)_{n \in \N}$
according to Definition~\ref{defsepfut}. For any~$n$, we choose~$t$ such that~\eqref{sepfut} holds.
Then the subregion~$\tilde{L} := \cup_{\tilde{t} \in [t, \tmax]} \supp \theta_{\tilde{t}}$ is again a lens-shaped region
inside~$U_n$ with local foliation~$(\eta_{\tilde{t}})_{\tilde{t} \in [t,\tmax]}$.
We let~$\v^+$ be the corresponding solution of Theorem~\ref{thmexistpm} with zero initial data at time~$t$
and with inhomogeneity~$\chi_U \w$. Extending this solution by zero to the past gives a solution in~$L_n$ which vanishes identically on~$K$.

We now consider the resulting sequence~$(\v_n)_{n \in \N}$ of solutions.
Using the uniform shielding property (see Definition~\ref{defuep}), we 
conclude that this sequence converges in~$L^2_\loc$. We thus obtain a global retarded
weak solution~$\v$ with inhomogeneity~$\chi_U \w$ which vanishes identically on~$K$.

In order to obtain a corresponding solution with inhomogeneity~$\w$, we use linearity
and a covering argument: We cover~$\supp \w$ by a finite number of open sets~$U_1, \ldots, U_L$
as above and construct corresponding global retarded weak solutions, choosing the inhomogeneity
in the $\ell^\text{th}$ step as
\[ \w_\ell := \chi_{U_\ell \setminus (U_1 \cup \cdots \cup U_{\ell-1})}\, \w \:. \]
Adding these solutions gives the desired global retarded weak solution with inhomogeneity~$\w$
which vanishes identically on~$K$.
\QED

\begin{Def} \label{deffp}
Let~$L$ be a lens-shaped region inside~$U$ with local foliation~$(\eta_t)_{t \in [t_0, \tmax]}$.
The jet space~$\J_U$ is {\bf{future-partitioned}} by the linear operator~$\check{\pi} : \J_U \rightarrow \J_U$
if the following conditions hold:
\begin{gather}
\eta_{t_0}\:(1-\check{\pi}) \equiv 0 \equiv (1-\eta_{\tmax})\:\check{\pi} \label{fp1} \\
\check{\pi}\, \u \in \overline{\J_U}^{t_{\max}} \qquad \text{for all~$\u \in \J_U$} \:. \label{fp2}
\end{gather}
\end{Def} \noindent
These conditions mean in words that the zero boundary conditions at time~$t_{\max}$
can be realized by acting on jets~$\u \in \J_U$ by the operator~$\check{\pi}$.

The {\em{support}} of~$\check{\pi}$ (and similarly~$(\1-\check{\pi})$) is defined by
\[ \supp \check{\pi} := \overline{ \big\{ x \in M \:\big|\: \exists \:\u \in \J_U \text{ with } (\check{\pi} \u)(x) \neq 0 \big\} } \:. \]

\begin{figure}
% \usepackage[usenames,dvipsnames]{pstricks}
% \usepackage{epsfig}
% \usepackage{pst-grad} % For gradients
% \usepackage{pst-plot} % For axes
% \usepackage[space]{grffile} % For spaces in paths
% \usepackage{etoolbox} % For spaces in paths
% \makeatletter % For spaces in paths
% \patchcmd\Gread@eps{\@inputcheck#1 }{\@inputcheck"#1"\relax}{}{}
% \makeatother
% 
\psscalebox{1.0 1.0} % Change this value to rescale the drawing.
{
\begin{pspicture}(-1.5,-2.2241468)(11.187065,2.2241468)
\definecolor{colour0}{rgb}{0.8,0.8,0.8}
\definecolor{colour1}{rgb}{0.6,0.6,0.6}
\definecolor{colour2}{rgb}{0.4,0.4,0.4}
\pspolygon[linecolor=colour0, linewidth=0.02, fillstyle=solid,fillcolor=colour0](0.09039881,-0.41586873)(0.7503988,-0.38586873)(1.6503989,-0.37586874)(2.2603989,-0.36086872)(2.865399,-0.41086873)(3.4253988,-0.42586872)(4.100399,-0.43086874)(5.155399,-0.47586873)(5.700399,-0.50086874)(6.595399,-0.54086876)(7.010399,-0.54086876)(7.410399,-0.5358687)(8.280398,-0.45586872)(8.250399,2.2141314)(0.08039881,2.2141314)
\pspolygon[linecolor=colour1, linewidth=0.04, fillstyle=solid,fillcolor=colour1](0.16706547,0.3724646)(0.40706548,0.2974646)(0.57706547,0.1924646)(0.90206546,0.3124646)(1.3320655,0.5224646)(1.7820655,0.8074646)(2.2120655,1.0424646)(2.6820655,1.2724646)(3.3320656,1.4724646)(4.0020657,1.5574646)(4.6920652,1.4974647)(5.2170653,1.3274646)(5.6320653,1.1324646)(5.9820657,0.9224646)(6.3720655,0.6974646)(6.8370657,0.4624646)(7.3270655,0.2724646)(7.7720656,0.1874646)(8.092066,0.3374646)(7.7170653,0.3924646)(7.1970654,0.5624646)(6.7970653,0.7624646)(6.4470654,0.9624646)(6.0820656,1.1724646)(5.5120654,1.4824646)(5.0570655,1.6724646)(4.6170654,1.8124646)(4.1520653,1.8824646)(3.6070654,1.8524646)(3.1270654,1.7574646)(2.6470654,1.5624646)(2.1320655,1.3124646)(1.6820655,1.0424646)(1.2420654,0.7824646)(0.8770655,0.5924646)(0.49706548,0.4374646)
\pspolygon[linecolor=colour1, linewidth=0.04, fillstyle=solid,fillcolor=colour1](1.8120655,0.1024646)(2.1420655,0.1524646)(2.3870654,0.2324646)(2.6720655,0.3424646)(2.9620655,0.4374646)(3.3170655,0.5474646)(3.7120655,0.6024646)(4.2020655,0.5724646)(4.6170654,0.4974646)(5.0120654,0.3524646)(5.3120656,0.2274646)(5.6520653,0.1074646)(6.0170655,-0.0025354004)(6.3870654,-0.0675354)(6.1370654,-0.1825354)(5.8070655,-0.1025354)(5.4670653,0.0324646)(5.0720654,0.1924646)(4.7220654,0.3374646)(4.3570657,0.4174646)(3.9520655,0.4724646)(3.4370654,0.4424646)(2.9670656,0.3224646)(2.6220655,0.1974646)(2.2770655,0.092464596)(1.9370655,0.0274646)
\psbezier[linecolor=colour0, linewidth=0.08](0.06206548,-0.4475354)(1.961308,-0.32009295)(2.8989043,-0.44607127)(4.1009526,-0.447535400390625)(5.3030014,-0.44899952)(6.601188,-0.70438457)(8.272065,-0.4575354)
\rput[bl](7.2370653,1.7474647){\normalsize{$F(K)$}}
\rput[bl](5.325399,-0.4475354){\normalsize{$L_1$}}
\psbezier[linecolor=colour2, linewidth=0.04](1.5770655,0.1174646)(2.3181748,0.062468898)(2.9176698,0.6622262)(3.9170654,0.627464599609375)(4.916461,0.592703)(5.8270855,-0.13797079)(6.5370655,-0.0525354)
\psbezier[linecolor=black, linewidth=0.04, fillstyle=solid,fillcolor=colour0](3.3870654,-1.1625354)(3.0118506,-1.5112524)(3.6781292,-1.7408758)(4.2570653,-1.632535400390625)(4.836002,-1.524195)(4.6468863,-1.1978409)(4.3870654,-1.0125355)(4.1272445,-0.8272299)(3.7622802,-0.81381845)(3.3870654,-1.1625354)
\rput[bl](3.7870655,-1.4025354){\normalsize{$K$}}
\psbezier[linecolor=colour2, linewidth=0.04](6.5120654,-0.0575354)(5.630956,-0.2125397)(5.001461,-1.937297)(3.9970655,-1.912535400390625)(2.9926698,-1.8877739)(2.3170457,0.13289998)(1.5820655,0.1174646)
\psbezier[linecolor=black, linewidth=0.04](0.0,0.37983146)(1.2515082,0.4043344)(2.22767,1.8318514)(3.9170654,1.8941312662760448)(5.606461,1.9564112)(7.050419,0.23518051)(8.240398,0.3724646)
\psbezier[linecolor=black, linewidth=0.04](8.254218,0.37854236)(7.355557,0.1335966)(4.916858,-2.1943922)(3.9433334,-2.203040409477235)(2.9698088,-2.2116888)(0.8666186,0.3376401)(0.09655591,0.37082332)
\psbezier[linecolor=colour2, linewidth=0.04](1.9103988,0.0124646)(2.6471927,0.070802234)(2.9515717,0.47222617)(3.9451478,0.437464599609375)(4.938724,0.40270302)(5.559048,-0.17297079)(6.178732,-0.1875354)
\psbezier[linecolor=black, linewidth=0.04](0.5603988,0.18649814)(1.538175,0.39766774)(2.286003,1.4551847)(3.9753988,1.517464599609375)(5.6647944,1.5797446)(6.1970854,0.37684718)(7.7670655,0.1674646)
\rput[bl](6.898732,-0.094202064){\normalsize{$L_2$}}
\rput[bl](8.477065,1.2074646){$\supp \check{\pi} \cap \supp (\1-\check{\pi})$}
\psbezier[linecolor=black, linewidth=0.02, arrowsize=0.05291667cm 2.0,arrowlength=1.4,arrowinset=0.0]{->}(8.351494,1.2824205)(7.0886354,1.2620136)(5.6487703,0.8573382)(5.1959705,0.37917530610357064)
\psbezier[linecolor=black, linewidth=0.02, arrowsize=0.05291667cm 2.0,arrowlength=1.4,arrowinset=0.0]{->}(8.341494,1.3524206)(7.5101857,1.3561547)(6.5554647,1.6975935)(6.0459704,1.3091753061035707)
\end{pspicture}
}
\caption{A future-localizing exhaustion.}
\label{figlocalize}
\end{figure}
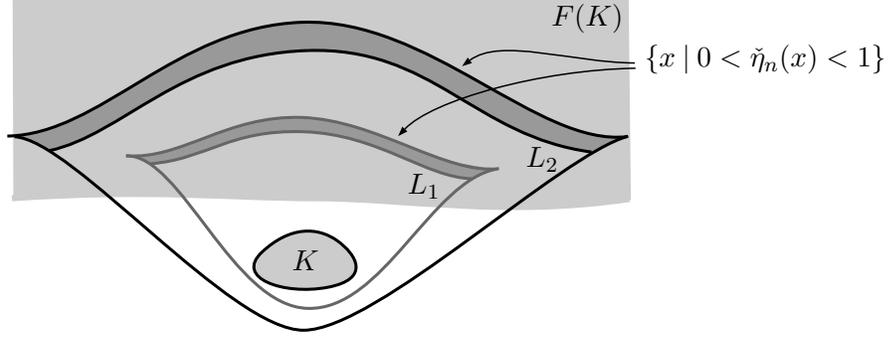

\begin{Def} \label{deffl}
Space-time is {\bf{future localizable}} if for every compact~$K \subset M$ there is
an exhaustion of~$M$ by lens-shaped regions~$(L_n)_{n \in N}$ such that for all~$n \in \N$
the following condition holds: The jet space~$\J_{U_n}$ is future-partitioned by an operator~$\check{\pi}_n$
such that for all~$x \in \supp (\1-\check{\pi}_n)$ and for all~$y \in U_n \setminus F(K)$,
the function~$\L(x,y)$ as well as its first and second derivatives in the direction of~$\Jvary_0$ vanish.
\end{Def} \noindent
Intuitively speaking, space-time is future localizable if there is an exhaustion by lens-shaped regions
such that the future boundaries of the lens-shaped region lie inside and are $\L$-localized in the
separated future of~$K$ (as shown in Figure~\ref{figlocalize};
see also the notion introduced on page~\pageref{cldef}).

\begin{Thm} \label{thmspeed} Assume that space-time is future localizable.
Let~$\w \in L^2_0(M, d\rho)$ be an inhomogeneity with compact support~$K:=\supp \w$.
Moreover, let~$y \not\in J^\vee(K)$. Then there is an open neighborhood~$W$ of~$y$
as well as a global retarded weak solution~$\v \in L^2_\loc(M, d\rho)$ with
\[ \v|_W \equiv 0 \:. \]
\end{Thm}
\Proof Let~$y \not\in J^\vee(K)$ and~$K_y$ a compact set whose interior contains~$y$.
Let~$(L_n)_{n \in \N}$ be an exhaustion of~$M$ by lens-shaped regions chosen
according to Definition~\ref{deffl} for the set~$K_y$.
Choose~$x \in K$. Then from~\eqref{defIJK} we know that~$y \not\in J^\vee(x)$.
Hence, inverting~\eqref{Jquant}, we obtain
\[ \mathop{\text{\Large $\forall$}}_{N \in \N}\:
\mathop{\text{\Large $\exists$}}_{\hat{L} \supset K_N} \:\mathop{\text{\Large $\exists$}}_{L \ll \hat{L}}
\quad \text{with} \quad x \in \overset{\circ}{L} \quad \text{but} \quad y \not\in L \:. \]
Since~$L$ is closed, we can choose open neighborhoods~$U_x$ of~$x$ and~$V_y \subset K_y$ of~$y$ with
\[ \overline{U_x} \subset \overset{\circ}{L} \qquad \text{and} \qquad \overline{V_y} \subset M \setminus L \:. \]
Then the function
\[ \hat{\w} := \frac{1}{\eta_{[t_0, \tmax]}}\: \chi_{U_x}\: \w \]
is bounded (because~$\chi_{U_x}$ vanishes near the boundary of~$L$ where
the function~$\eta_{[t_0, \tmax]}$ is zero). We let~$\hat{\v}$ be the solution of Theorem~\ref{thmexist}
corresponding to the inhomogeneity~$\hat{\w}$. Then the function
\[ \v := \eta_{[t_0, \tmax]}\: \hat{\v} \]
is a weak solution in~$L^2(L, d\rho)$ with zero initial data, i.e.
\[ \la \Delta \u, \v \ra_{L^2(L, d\rho)} = \la \u, \chi_{U_x} \w \ra_{L^2(L, d\rho)} \qquad
\text{for all~$\u \in \overline{\J_U}^\tmax$}\:. \]

Now let~$\u \in \Jvary_0$. By definition, its restriction to~$U$ is in~$\J_U$
(but it does not necessarily vanish at time~$\tmax$). Extending
the function~$\v$ by zero to~$U$ and choosing the operator~$\check{\pi}$ as in Definition~\ref{deffp}
(again for the set~$K_y$), the support of~$\1-\check{\pi}$ does not intersect~$U_x$. Hence,
using that~$\v$ is a weak solution, we obtain
\begin{align}
\la \Delta \u, \v \ra_{L^2(U, d\rho)} &=
\la \Delta ( \check{\pi} \,\u), \v \ra_{L^2(U, d\rho)}
+ \la \Delta \big( (\1-\check{\pi})\, \u \big), \v \ra_{L^2(U, d\rho)} \notag \\
&= \la \u, \chi_{U_x} \w \ra_{L^2(U, d\rho)}
+ \la \Delta \big( (\1-\check{\pi})\, \u \big), \v \ra_{L^2(U, d\rho)} \:, \label{eqres}
\end{align}
where in the last step we used~\eqref{fp1} and~\eqref{fp2}. Since both~$\v$ and~$\chi_{U_x} \w$
vanish outside~$L$, in the last scalar product we may just as well integrate over~$M$,
if all functions are extended by zero from~$U$ to~$M$. Moreover, the last summand in~\eqref{eqres}
is linear in the jet~$\u$ and involves this jet only on the support of~$(\1-\check{\pi})$.
Hence, using that space-time is future localizable (see Definition~\ref{deffl}),
this term can be written as
%since~$\big( (\1-\check{\pi})\, \u$ involves
%\begin{align*}
%\la \Delta \big( (\1-\check{\pi})\, \u \big), \v \ra_{L^2(M, d\rho)}
%&= \int_U \big(\1-\check{\pi}(x) \big) \:\nabla_\u \nabla_\v \ell(x) \: d\rho(x) \\
%&\qquad + \int_U d\rho(x) \int_U d\rho(y)\: \big(\1-\check{\pi}(x) \big)\:
%\nabla_{1,\u} \nabla_{2,\v} \L(x,y) \:.
%\end{align*}
%Again using that space-time is future localizable (see Definition~\ref{deffl}), we can write this equation as
\[ \la \Delta \big( (\1-\check{\pi})\, \u \big), \v \ra_{L^2(M, d\rho)} = - \la \u, \w^\text{err} \ra_{L^2(M, d\rho)} \]
with an ``error jet''~$\w^\text{err}$ supported inside~$F(K_y) \cap U$. 
Therefore, we can apply Theorem~\ref{thmsepfut} to obtain a
global retarded weak solution with inhomogeneity~$\w^\text{err}$ which vanishes on~$K_y$.
Adding this solution to the jet~$\v$, we obtain a global retarded weak solution
which vanishes in~$\overline{V}_y$.

To summarize the result so far, we have shown that for every~$y \not\in J^\vee(K)$
and for every~$x \in K$, there are open neighborhoods~$U_x$ and~$V_y$ as well as a
global weak retarded solution~$\v_x$ of the equation
\[ \la \Delta \u, \v_x \ra_{L^2(M, d\rho)} = \la \u, \chi_{U_x} \w \ra_{L^2(M, d\rho)}
\qquad \text{for all~$\u \in \Jvary_0$} \]
which vanishes identically in~${\overline{V}_y}$,
\[ \v_x|_{\overline{V}_y} \equiv 0 \:. \]
Similar as in the proof of Theorem~\ref{thmsepfut}, we cover~$K$ by a finite number of such
neighborhoods~$U_{x_1}, \ldots, U_{x_L}$.
Summing these solutions and setting~$W = V_{x_1} \cap \cdots \cap V_{x_L}$ 
gives the result.
\QED

Having specified in which sense~$J^\vee(K)$ determines the propagation speed,
we can also generalize other notions familiar from hyperbolic PDEs
(see for example~\cite[Section~8.3]{rendall}).
\begin{Def} The {\bf{domain of influence}} ${\mathscr{I}}(K)$ of a compact subset~$K \subset M$
is defined by
\[ {\mathscr{I}}(K) = J^\vee(K) \cup J^\wedge(K) \:. \]
The {\bf{domain of determination}}~$\D(A)$ of a subset~$A \subset M$ is defined by
\[ \D(A) = M \setminus \bigcup
\big\{ {\mathscr{I}}(K) \,\big|\, K \subset M \text{ compact and } {\mathscr{I}}(K) \cap A = \varnothing \big\} \:. \]
A set~$A \subset M$ is a {\bf{domain of dependence}} for the point~$x \in M$
if~$x \in \D(A)$.
\end{Def}

\subsection{Globally Hyperbolic Space-Times} \label{secglobhyp}
The causal relations introduced in Section~\ref{seccone}
seem somewhat artificial because the condition~$y \in J^\vee(x)$ may depend on the structure of the lens-shaped regions in an arbitrarily distant space-time region. While it is sensible that the condition~$y \in J^\vee(x)$
involves the lens-shaped regions in a sufficiently large region containing~$x$ and~$y$,
this region should nevertheless be compact.
Moreover, in~\eqref{Iqbad} there is the technical complication that even for points~$\tilde{y}$
in a small neighborhood of~$y$, it might be necessary to choose the parameter~$N$
arbitrarily large. Finally, our definitions do not imply that the future cone
is a closed subset of~$M$. We now introduce a setting which avoids all these subtleties.
This definition will not be used in the remainder of this paper, but it might be
a suitable starting point for the future.
\begin{Def} \label{defcs}
Space-time is {\bf{causally simple}} if the following conditions hold:
\bitem
\item[\rm{(i)}] For all~$x \in M$ and every compact set~$K \subset M$
there are compact sets~$K_1$ and~$K_2$ with~$K_1 \subset K_2 \subset M$ such that
\[ J^\vee(x) \cap K = \Big\{ y \in K \:\Big|\: 
\mathop{\text{\Large $\forall$}}_{K_1 \subset \hat{L} \subset K_2} \:\mathop{\text{\Large $\forall$}}_{L \ll \hat{L}} \::\:
x \in \overset{\circ}{L}  \Longrightarrow y \in L \Big\} \:. \]
\item[\rm{(ii)}] For any compact subset~$K \subset M$, the set~$J^\vee(K)$ is closed in~$M$.
\eitem
\end{Def} \noindent
The condition~(i) implies that the parameter~$N$ in~\eqref{Jquant} can be chosen locally
uniformly in~$y$. As a consequence, in~\eqref{Iqbad} the quantifiers may be interchanged to obtain
\[ I^\vee(x) = \overbrace{J^\vee(x)}^{\circ} \:. \]
We thus recover the familiar setting where the open future light cone is the interior of~$J^\vee$,
which in view of~(ii) we can refer to as the {\bf{closed future light cone}}.
Clearly, the name ``causally simple'' is inspired by the related notion
in Lorentzian geometry (see for example~\cite[Section~6.3]{hawking+ellis}
or~\cite[Section~3.10]{minguzzi+sanchez}),
but we point out that the connection between these notions is not more than
a superficial analogy.

Here is another property which does not seem to be satisfied for general minimizers
of causal variational principles, but which seems reasonable to impose
because it holds in Lorentzian space-times (see for example~\cite[Proposition~3.38]{minguzzi+sanchez}):

\begin{Def} \label{defic} The open future cones~$I^\vee(x)$ are~{\bf{inner continuous}}
if for every compact~$K \subset I^\vee(x)$ there is an open neighborhood~$U_x$ of~$x$
such that~$K \subset I^\vee(\tilde{x})$ for all~$\tilde{x} \in U_x$.
\end{Def}

We finally combine previous notions and assumptions to a proposal of what could be a
sensible generalization of the class of globally hyperbolic Lorentzian manifolds
to the setting of causal variational principles:
\begin{Def} \label{defcd} 
\label{defglobhyp}
Space-time is {\bf{globally hyperbolic}} if it has the following properties:
\begin{itemize}[leftmargin=2.5em]
\item[\rm{(i)}] Space-time is compactly hyperbolic (see Definition~\ref{defcompacthyp})
and has the uniform shielding property (see Definition~\ref{defuep} and likewise
for advanced solutions).
\item[\rm{(ii)}] Space-time is causally simple (see Definition~\ref{defcs}).
\item[\rm{(iii)}] The open cones are inner continuous (see Definition~\ref{defic}
and similarly for past cones).
\item[\rm{(iv)}] Space-time is future localizable (see Definition~\ref{deffl})
and similarly past localizable.
\item[\rm{(v)}] Space-time has {\bf{compact diamonds}}, meaning that for all compact~$K, K' \subset M$,
the set~$J^\vee(K) \cap J^\wedge(K')$ is compact.
\eitem
\end{Def} \noindent
We note that here we do not enter the question whether the above assumptions are
all independent. We also remark that condition~(v) is a stronger condition
than imposing that the sets~$J^\vee(x) \cap J^\wedge(x')$ are compact for all~$x,x' \in M$,
similar as explained for smooth manifolds in~\cite[Section~1.1]{bernard+suhr}.
We finally point out that the above assumptions do not rule out the possibility
that the space-time might be non-chronological in the sense that~$x \in I^\vee(x)$
for some~$x \in M$.

\subsection{Global Foliations by Cauchy Surface Layers} \label{secglobfol}
The previous constructions were based on energy estimates
in compact subregions of space-time (more precisely, lens-shaped regions
admitting a local foliation satisfying suitable hyperbolicity conditions).
By extending local solutions we succeeded in constructing global solutions.
But so far we have avoided working with global foliations covering all of space-time.
In a globally hyperbolic Lorentzian space-time, global foliations are known to exist
(see~\cite{bernal+sanchez}). Therefore, it seems an interesting question whether
a globally hyperbolic space-time (see Definition~\ref{defglobhyp}) admits global
foliations by surface layers.
This question is an open problem which goes beyond the scope of the present paper.
But we now give a possible definition of a global foliation and indicate
how a global foliation could be used for constructing global solutions of the Cauchy problem.
We also mention the points which, from our point of view, would be the main difficulties in
carrying out this program. Here is a first suggestion for a definition of a global foliation:
\begin{Def} \label{globfoliate} A function~$\eta \in \C^\infty(\R \times M, \R)$ with~$0 \leq \eta \leq 1$
is called a {\bf{global foliation by Cauchy surface layers}} if the following conditions hold:
\bitem
\item[{\rm{(i)}}] The function~$\theta(t,.) := \partial_t \eta(t,.)$ is non-negative.
\item[{\rm{(ii)}}] The surface layers cover all of~$M$ in the sense that
\[ M = \bigcup_{t \in \R} \overset{\circ}{\supp} \,\theta(t,.) \:. \]
\item[{\rm{(iii)}}] The following hyperbolicity conditions hold:
For every~$T>0$ there is a constant~$C>0$ such that for all~$t \in [-T,T]$,
\beq
(\v, \v)^t \geq \frac{1}{C} \int_M \Big( \|\v(x)\|_x^2\: + \big|\Delta_2[\v, \v]\big| \Big) \: d\rho_t(x) 
\qquad \text{for all~$\v \in \Jvary_0$} \:, \label{hypcondM}
\eeq
where we again use the notation~\eqref{rhot} and~\eqref{jipdef},
writing~$\eta(t,x)$ as~$\eta_t(x)$ and similarly~$\theta(t,x)$ as~$\theta_t(x)$.
\eitem
\end{Def} \noindent
One difficulty is that, since the surface layers are no longer compact,
proving the inequality~\eqref{hypcondM} makes it necessary to control the behavior
of the jets at spatial infinity.
Once this rather subtle issue has been settled, one could follow the strategy in
Section~\ref{sechypsubset} to prove existence and uniqueness, but now globally
in space-time. With this in mind, it would be desirable to work with global foliations.
However, as mentioned above, the existence of global foliations is a challenging
open problem.

\section{Causal Green's Operators and their Properties} \label{secgreen}
Having developed the existence theory for global solutions, we can now
construct advanced and retarded Green's operators and analyze their properties.
In preparation, we extend the existence result for global solutions
of Theorem~\ref{thmshield} to inhomogeneities whose support is not necessarily
compact (Section~\ref{secpastcompact}).
Then the causal Green's operators can be defined in a straightforward way
(Section~\ref{secgreenretarded}).
We finally explain how the difference of the advanced and retarded Green's operator
can be used to describe the homogeneous solution space (Section~\ref{secfundamental}).

\subsection{Past and Spatially Compact Inhomogeneities} \label{secpastcompact}
\begin{Def} \label{defpc} A jet~$\w \in L^2_\loc(M, d\rho)$ is called
{\bf{past and spatially compact}}
if its support lies in the causal future of a compact set~$K$, i.e\
\[ \supp \w \subset J^\vee(K)\:. \]
Similarly, a jet is {\bf{future and spatially compact}}) if~$\supp \w \subset J^\wedge(K)$.
\end{Def} \noindent

\begin{Corollary} \label{corshield}
Under the assumptions of Theorem~\ref{thmshield},
let~$\w$ be a past and spatially compact jet with the property that 
every~$x \in M$ has an open neighborhood~$V$ such that for any exhaustion
by lens-shaped regions~$(L_n)_{n \in \N}$ of the form~\eqref{Lexhaust}, the 
following shielding condition holds:
\beq \label{Gsw}
\sum_{n=1}^\infty \shield(V, L_n, L_{n+1})
\: \big( \Gamma(L_n, L_n) + \Gamma(L_n, L_{n+1}) \big)\: \|\w\|_{L^2(L_n, d\rho)} < \infty \:.
\eeq
Then there is a global retarded weak solution with inhomogeneity~$\w$.
\end{Corollary}
\Proof Since~$M$ is $\sigma$-compact (see the last paragraph of Section~\ref{seccvp}
on page~\pageref{sigmacompact}), we can write~$\w$ as
$\w = \sum_{p=1}^\infty \w^{(p)}$ with compactly supported~$\w^{(p)}$.
According to Theorem~\ref{thmshield}, there are corresponding global advanced
weak solutions~$\v^{(p)}$. Our task is to show that the series~$\sum_{p=1}^\infty \v^{(p)}$
converges in~$L^1_\loc(M, d\rho)$.

Choosing the lens-shaped region~$L_1$ such that it contains the compact set~$K$
with $\supp \w \subset J^\vee(K)$,
we can arrange that all the~$\w^{(p)}$ vanish at initial time~$\tmin$ for all
lens-shaped regions~$L_1, L_2, \ldots$.
This makes it possible to construct
all the solutions~$\v^{(p)}$ with the same series of lens-shaped regions,
exactly as in the proof of Theorem~\ref{thmshield}.
Noting that the estimate~\eqref{shieldes} involves the $L^2$-norm of~$\w^{(p)}$,
the inequality~\eqref{Gsw} ensures convergence of the series~$\sum_{p=1}^\infty \v^{(p)}$.
\QED

\begin{Def} 
The space of all past and spatially compact jets (and similarly future and spatially compact sets)
which satisfy the shielding condition~\eqref{Gsw} is denoted by~$L^2_{\loc, \psc}(M, d\rho)$
(and~$L^2_{\loc, \fsc}(M, d\rho)$).
\end{Def}

Obviously, every jet~$\v \in L^2_0(M, d\rho)$ with essentially compact support
is past and spatially compact as well as future and spatially compact.
Clearly, the converse is true if we assume that the diamonds are compact.
If we assume in addition that space-time has the uniform shielding property
(see Definition~\ref{defuep}), then the shielding condition~\eqref{Gsw} is satisfied for all
compactly supported jets. We thus obtain the following result:
\begin{Lemma} \label{lemmacompact} Assume that space-time is globally hyperbolic
(see Definition~\ref{defglobhyp}). Then
\[ L^2_0(M, d\rho) = L^2_{\loc,\psc}(M, d\rho) \cap L^2_{\loc,\fsc}(M, d\rho) \:. \]
\end{Lemma}

\subsection{Causal Green's Operators} \label{secgreenretarded}
We again assume that the Lagrangian has finite range (see Definition~\ref{deffiniterange}).
Then~$\Delta$ maps compactly supported jets to compactly supported jets,
\[ %\beq \label{Delc}
\Delta \::\: \Jvary_0 \rightarrow L^2_0(M, d\rho) \:.
\] %\eeq
Given~$\w \in L^2_{\loc,\psc}(M, d\rho)$, in Corollary~\ref{corshield}
we constructed a corresponding retarded solution~$\v \in L^2_\loc(M, d\rho)$.
We define the {\em{retarded Green's operator}}~$S^\wedge$ by
\beq \label{Swedge}
S^\wedge \::\: L^2_{\loc,\psc}(M, d\rho) \rightarrow L^2_\loc(M, d\rho) \:,\qquad
\w \mapsto -\v \:.
\eeq
The {\em{advanced Green's operator}}~$S^\vee$ is defined similarly,
\[ S^\vee \::\: L^2_{\loc,\fsc}(M, d\rho) \rightarrow L^2_\loc(M, d\rho) \:. \]

We point out that our construction of causal Green's operators does not rely on
any smoothness assumptions of space-time. For a related construction in
Lorentzian space-times of low regularity see~\cite{hoermann}.

We finally analyze the kernel of the Green's operators. Our starting point is the observation
that in the global weak equation~\eqref{globweak},
the inhomogeneity~$\w$ can be changed arbitrarily by wave functions in the orthogonal
complement of~$\Jvary_0$ (with respect to the scalar product~$\la .|.\ra_{L^2(M, d\rho)}$).
Indeed, denoting this orthogonal complement by~$(\Jvary_0)^\perp$,
the right side of~\eqref{globweak} vanishes identically for any~$\w \in (\Jvary_0)^\perp$.
This suggests that also the Green's operators should vanish on such wave functions.
This is indeed the case, as is shown in the next lemma.

\begin{Lemma}\label{lemmaquotient}
The Green's operators vanish on~$( \Jvary_0)^\perp$ in the sense that
\begin{align}
(\Jvary_0)^\perp \cap L^2_{\loc,\psc}(M, d\rho) &\subset \ker S^\wedge \label{perp1} \\
(\Jvary_0)^\perp \cap L^2_{\loc,\fsc}(M, d\rho) &\subset \ker S^\vee \label{perp2} \:.
\end{align}
\end{Lemma}
\begin{proof} We only prove~\eqref{perp1}, because the proof of~\eqref{perp2} is similar.
Thus let~$\w \in (\Jvary_0)^\perp \cap L^2_{\loc,\psc}(M, d\rho)$.
In view of the limiting procedures in Corollary~\ref{corshield} and Theorem~\ref{thmshield},
it suffices to show that the weak solution constructed in Theorem~\ref{thmexist}
in a lens-shaped region~$L$ vanishes for sufficiently large~$L$.
We choose the lens-shaped region~$L$ so large that~$\w$ vanishes in the initial time surface layer as well as in its past. Then for any~$\u \in \overline{\J_U}^\tmax$,
\[ \la \w, \u \ra_{L^2(L)} = \la \w, \u \ra_{L^2(M, d\rho)} = 0 \]
(here we made use of the fact that~$\u$ vanishes in the surface layer at time~$\tmax$; see~\eqref{oJtestdef}).
Therefore, the linear functional~$\la \w, \u \ra_{L^2(L)}$ vanishes on a dense subset
of the Hilbert space~${\mathcal{H}}$, implying that it is represented by the zero vector~$0=V \in {\mathcal{H}}$.
As a consequence, also the weak solution~$\v:=\Delta V$ vanishes.
\QED

\subsection{The Causal Fundamental Solution and its Properties} \label{secfundamental}
The {\em{causal fundamental solution}} is defined by
\beq \label{Kdef}
G = S^\wedge - S^\vee \::\: L^2_0(M, d\rho) \rightarrow L^2_\loc(M, d\rho) \:.
\eeq
It maps to homogeneous weak solutions of the linearized field equations.

In the above definitions of~$S^\wedge$, $S^\vee$ and~$G$,
we chose the domain of definition as large as possible.
For the applications, however, it is convenient to restrict attention to a smaller domain of ``nice'' jets.
To this end, we define
\begin{align*}
\J^{**}_0 &:= \big\{ \u \in \Jvary_0 \:\big|\: S^\vee \Delta \u = S^\wedge \Delta \u = -\u \big\} \\
\J^*_0 &:= \big\{ \u \in L^2_0(M, d\rho) \:\big|\: S^\vee \u, S^\wedge \u \in \Jvary \big\} 
\Big/ (\Jvary_0)^\perp \\
\J_\sc &:= \big\{ S^\wedge \u_1 + S^\vee \u_2 \:\big|\: \u_1 \in L^2_{\loc,\psc}(M, d\rho) \text{ and } S^\wedge \u_1 \in \Jvary, \\
&\qquad\qquad\qquad\:\,\qquad \: \u_2 \in L^2_{\loc,\fsc}(M, d\rho)  \,\text{ and } S^\vee \u_2 \in \Jvary \big\} \\
\J^*_\sc &:= \big\{ \u_1 + \u_2 \:\big|\: \u_1 \in L^2_{\loc,\psc}(M, d\rho) \text{ and } S^\wedge \u_1 \in \Jvary, \\
&\qquad\qquad\qquad\, \u_2 \in L^2_{\loc,\fsc}(M, d\rho)  \,\text{ and } S^\vee \u_2 \in \Jvary \big\} \Big/ \\
&\qquad\quad \big( (\Jvary_0)^\perp \cap L^2_{\loc,\psc}(M, d\rho) \big) +
\big( (\Jvary_0)^\perp \cap L^2_{\loc,\fsc}(M, d\rho) \big) \:,
\end{align*}
where~$(\Jvary_0)^\perp$ is again the orthogonal complement of~$\Jvary_0$ with respect to
the scalar product~$\la .|.\ra_{L^2(M, d\rho)}$. Working modulo such jets reflects the
general structure of the global weak equation~\eqref{globweak},
where the inhomogeneity~$\w$ can be changed arbitrarily by jets in~$(\Jvary_0)^\perp$.
Working modulo such jets, in what follows we do not need to distinguish between weak and strong
solutions of the linearized field equations, making it possible to work with the relations~$\Delta S^\vee=
\Delta S^\wedge=-\1$. Moreover, in view of Lemma~\ref{lemmaquotient} the causal Green's operators,
and consequently also the causal fundamental solution~$G$, are well-defined on the equivalence
classes obtained by modding out~$(\Jvary_0)^\perp$.
In the definition of~$\J^*_\sc$ we again mod out~$(\Jvary_0)^\perp$,
but this time also respecting the decomposition into the sum of a future and a past spatially compact jet.
In order to avoid confusion, we also point out that, identifying the jet space with their duals
using the pointwise scalar product~\eqref{vsprod}, the space~$\J^*_0$
does in general not agree with~$\Jvary_0$, and~$\J^*_\sc$ does not coincide with~$\J_\sc$.
The above definitions identify the correct dual jet spaces, independent of the
arbitrarily chosen scalar product~\eqref{vsprod}.

It follows immediately from the definitions that~$G$ maps~$\J^*_0$ to~$\J_\sc$
and that~$\Delta$ can be extended to a well-defined operator from~$\J_\sc$ to~$\J^*_\sc$.
\begin{Lemma} \label{lemmaDelJJs}
The operator~$\Delta$ maps~$\J^{**}_0$ to~$\J^*_0$.
\end{Lemma}
\Proof Let~$\u \in \J^{**}_0$. Then the jet~$\v := \Delta \u$ has the property
\[ S^\vee \v = S^\vee \Delta \u = -\u \in \Jvary \:, \]
and similarly for~$S^\wedge \v$. It follows by definition of~$\J^*_0$ that~$\v \in \J^*_0$.
\QED

In the next theorem we combine the properties of the causal fundamental solution
in an exact sequence, similar as obtained for linear hyperbolic PDEs in
globally hyperbolic space-times in~\cite[Proposition~8]{ginouxML}
and~\cite[Theorem~4.3]{baergreen}.

\begin{Thm} \label{thmexact}
Assume that space-time is globally hyperbolic
(see Definition~\ref{defglobhyp})
and that the Lagrangian has finite range (see Definition~\ref{deffiniterange}).
Then the following sequence is exact:
\beq \label{exact}
0 \rightarrow \J^{**}_0 \overset{\Delta}{\longrightarrow} \J^*_0
\overset{G}{\longrightarrow} \J_\sc
\overset{\Delta}{\longrightarrow} \J^*_\sc \rightarrow 0 \:.
\eeq
\end{Thm}
\Proof We proceed in several steps:
\begin{itemize}[leftmargin=2.5em]
\item[(i)] $\Delta : \J^{**}_0 \rightarrow \J^*_0$ is injective: 
Let~$\v \in \J^{**}_0$ with~$\Delta \v=0$. 
Multiplying by~$S^\vee$
and using again the definition of~$\J^{**}_0$, we conclude that~$\v = -S^\vee \Delta \v = 0$.
\item[(ii)] The product~$G \circ \Delta : \J^{**}_0 \rightarrow \J_\sc$ vanishes: Let~$\v \in \J^{**}_0$.
Then, by definition of~$G$ and~$\J^{**}_0$,
\[ G \Delta \v = S^\wedge \Delta \v - S^\vee \Delta \v = -\v + \v = 0 \:. \]
\item[(iii)] If~$G \u = 0$ for~$\u \in \J^*_0$, then~$\u$ can be represented
as~$\u =\Delta \v$ with~$\v \in \J^{**}_0$:
By definition of~$G$ and~$\J^*_0$, we know that
\[ \v:= -S^\vee \u = -S^\wedge \u \in \Jvary \:. \]
Lemma~\ref{lemmacompact} yields that~$\v \in \Jvary_0$.
Finally, the equation~$\Delta \v = \u$ follows by definition of the Green's operators.
\item[(iv)] The product~$\Delta \circ G : \J_0^* \rightarrow \J_\sc^*$ vanishes:
This follows immediately from the definition of the Green's operators.
\item[(v)] If~$\Delta \v = 0$ for~$\u \in \J_\sc$, then~$\v$ can be represented
as~$\v =G \u$ with~$\u \in \J^*_0$: Representing~$\v$ as in the definition of~$\J_\sc$, we obtain
by definition of the Green's operators
\[ \Delta \v = \u_1 + \u_2 \qquad \text{with~$\u_1 \in L^2_{\loc,\psc}(M, d\rho)$ and~$\u_2 \in L^2_{\loc,\fsc}(M, d\rho)$}\:. \]
Hence~$\u_1 = -\u_2 =: - 2 \pi i \,\u$ is compactly supported and~$S^\wedge \u, S^\vee \u \in \Jvary$.
In other words, $\u \in \J^*_0$. Moreover, $G \u = \v$ by construction.
\item[(vi)] The operator~$\Delta : \J_\sc \rightarrow \J^*_\sc$ is surjective:
Let~$\u \in \J^*_\sc$. According to the definition of~$\J^*_\sc$, we can represent~$\u$ as
\[ \u = \u_1 + \u_2 \qquad \text{with~$\u_1 \in L^2_{\loc,\psc}(M, d\rho)$ and~$\u_2 \in L^2_{\loc,\fsc}(M, d\rho)$}\:. \]
Then by definition, the jet~$\v := -S^\wedge \u_1 - S^\vee \u_2$ is in~$\J_\sc$.
Moreover, $\Delta \v = \u$ by definition of the Green's operators.
\eitem
This concludes the proof.
\QED

The image of the operator~$G$ in the exact sequence~\eqref{exact} are the linearized weak
solutions of spatially compact support denoted by
\beq \label{Jlinscdef}
\Jlin_\sc := G  \,\J^*_0  \subset \Jtest \:.
\eeq

\begin{Remark} {\em{ \label{remnodecay}
We note for completeness that it seems reasonable to extend the above construction to
jets which do not have spatially compact support, similarly to the procedure followed in
the study of liner hyperbolic PDEs in globally hyperbolic space-times, cf.~\cite{baergreen}.
However, the construction also involves difficulties, as we now outline:
A jet~$\w \in L^2_\loc(M, d\rho)$ is called {\em{past compact}}
if for any~$x \in M$, the intersection
\[ J^\wedge(x) \cap \supp \w \qquad \text{is compact} \:. \]
{\em{Future compact}} jets are define analogously. A jet is called {\em{timelike compact}}
if it is both future and past compact. Then the goal would be to prove in analogy
to~\eqref{exact} the exact sequence
\beq \label{exact2}
0 \rightarrow \J^{**}_\text{tc} \overset{\Delta}{\longrightarrow} \J^*_\text{tc}
\overset{G}{\longrightarrow} \J
\overset{\Delta}{\longrightarrow} \J^* \rightarrow 0 \:,
\eeq
where the index ``tc'' denotes jets~$\w$ which are timelike compact
and have the property that there are global advanced and retarded weak solutions
with inhomogeneity~$\w$.
The main difficulty in establishing~\eqref{exact2} is that, in order to extend
the existence result of Corollary~\ref{corshield} to jets which do not have
spatially compact support, one would have to get uniform control of our estimates
near spatial infinity. More precisely, the shielding condition~\eqref{Gsw} seems problematic
if~$\w$ grows rapidly at spatial infinity, making it necessary to work out detailed
growth conditions at spatial infinity. Moreover, one would have to make sure that
there is an exhaustion by lens-shaped regions with the property that the support of~$\w$
lies in the future of every surface layer at initial time~$t_0$.
Exactly as explained at the end of Section~\ref{secglobfol}
in the context of global foliations, these are subtle issues which we 
leave as open problems for future research. }}
\QEDrem
\end{Remark}

\subsection{Connection to the Symplectic Form} \label{secsymplect}
In this section we derive an identity involving the causal fundamental solution
and the symplectic form (see Proposition~\ref{prpsigma} below). The analogous formula
in classical field theory is commonly used when quantizing the field in the algebraic formulation.
As we shall see, extending this formula to causal variational principles involves a few subtleties.

The {\em{symplectic form}} is an antisymmetric bilinear form on the linearized solutions
(see~\eqref{I2asymm} or the ``softened version'' in~\eqref{sympdef}).
In~\cite{jet} it is shown that if~$\u$ and~$\v$ are linearized solutions and~$\Omega$ is compact,
then~$\sigma^\Omega(\u,\v)$ vanishes. This gives rise to a conservation law if
one considers the limiting case that~$\Omega$ exhausts the region between two
Cauchy surfaces (for a detailed explanation see~\cite[Section~2.3]{noether} and~\cite[Section~1]{jet}).
In the present more general setting we do not want to assume the existence of a Cauchy surface.
Therefore, we proceed instead as follows.
Let~$\u, \v \in \J^*_0$ be two compactly supported jets. By applying the operator~$G$ in~\eqref{Kdef}
we obtain two linearized solutions~$G \u, G \v \in \Jlin_\sc$ (see~\eqref{Jlinscdef}).
Similar to the procedure in algebraic quantum field theory, we restrict attention to
linearized solutions of this form. We again assume that space-time is globally hyperbolic
and that the Lagrangian has finite range (see Definitions~\ref{defglobhyp} and~\ref{deffiniterange}).
Then we can choose a lens-shaped region~$L$
inside a relatively compact open subset~$U \subset M$ together with an
operator~$\check{\pi} : \J_U \rightarrow \overline{\J_U}^{t_{\max}}$ which is identically equal to one 
in the past of a surface layer which lies to the future of the supports of~$\u$ and~$\v$
(see Figure~\ref{figlens}; for technical details see the proof of
Proposition~\ref{prpsigma} below). 
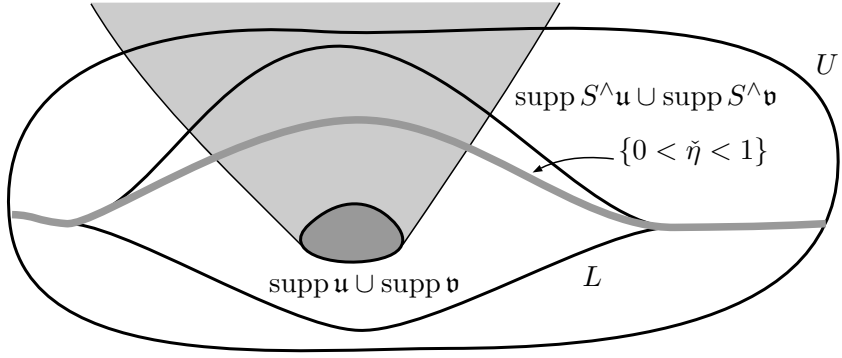
\begin{figure}
% \usepackage[usenames,dvipsnames]{pstricks}
% \usepackage{epsfig}
% \usepackage{pst-grad} % For gradients
% \usepackage{pst-plot} % For axes
% \usepackage[space]{grffile} % For spaces in paths
% \usepackage{etoolbox} % For spaces in paths
% \makeatletter % For spaces in paths
% \patchcmd\Gread@eps{\@inputcheck#1 }{\@inputcheck"#1"\relax}{}{}
% \makeatother
% 
\psscalebox{1.0 1.0} % Change this value to rescale the drawing.
{
\begin{pspicture}(-1,-2.3619804)(17.37521,2.3619804)
\definecolor{colour0}{rgb}{0.8,0.8,0.8}
\definecolor{colour1}{rgb}{0.6,0.6,0.6}
\pspolygon[linecolor=colour0, linewidth=0.02, fillstyle=solid,fillcolor=colour0](1.2252097,2.3469527)(7.4352098,2.3519528)(6.48521,0.8069527)(6.13021,0.28695273)(5.71521,-0.34804726)(5.3652096,-0.85304725)(4.0102096,-0.84304726)(2.7252097,0.41695273)(1.7102097,1.6019528)
\psbezier[linecolor=black, linewidth=0.04, fillstyle=solid,fillcolor=colour1](4.11021,-0.59804726)(3.734995,-0.94676423)(4.4012733,-1.1763877)(4.98021,-1.0680472564697265)(5.559146,-0.9597068)(5.3700304,-0.63335276)(5.11021,-0.44804725)(4.850389,-0.26274174)(4.4854245,-0.24933031)(4.11021,-0.59804726)
\rput[bl](3.5852098,-1.5580473){\normalsize{$\supp \u \cup \supp \v$}}
\psbezier[linecolor=black, linewidth=0.04](0.7952097,-0.5906804)(2.0463192,-0.5661774)(2.755814,1.7146728)(4.4452095,1.7769527435302734)(6.1346054,1.8392327)(7.8452296,-0.76033133)(9.03521,-0.62304723)
\psbezier[linecolor=black, linewidth=0.04](8.96521,-0.6330473)(8.034101,-0.60805154)(5.7746053,-2.037809)(4.77521,-2.0030472564697264)(3.775814,-1.9682857)(1.5351899,-0.49761188)(0.82520974,-0.5830473)
\rput[bl](7.75021,-1.4180473){\normalsize{$L$}}
\psbezier[linecolor=black, linewidth=0.02](4.00521,-0.86304724)(2.9979305,0.14421718)(1.9265878,1.1477886)(1.2152097,2.3419527435302734)
\psbezier[linecolor=black, linewidth=0.02](5.362729,-0.85965747)(6.2100725,0.37796167)(6.8286953,1.3073617)(7.4476905,2.353562964390635)
\rput[bl](6.85521,0.9219527){\normalsize{$\supp S^\wedge \u \cup \supp S^\wedge \v$}}
\psbezier[linecolor=black, linewidth=0.04](0.13520974,-0.5730473)(-0.093619324,1.4867755)(2.0452952,2.123882)(4.21521,1.9769527435302734)(6.385124,1.8300235)(11.046381,2.6967754)(11.145209,0.34695274)(11.244039,-2.00287)(8.005124,-2.2399764)(5.9752097,-2.2430472)(3.945295,-2.246118)(0.3640388,-2.63287)(0.13520974,-0.5730473)
\rput[bl](10.85021,1.4019527){\normalsize{$U$}}
\psbezier[linecolor=colour1, linewidth=0.1](0.16520973,-0.46304727)(0.49520972,-0.47304726)(0.46217275,-0.54590124)(0.88520974,-0.5830472564697265)(1.3082467,-0.6201933)(3.305229,0.8231331)(4.8052096,0.7969527)(6.3051906,0.7707724)(7.673992,-0.64485306)(8.97521,-0.6330473)(10.276427,-0.62124145)(10.325024,-0.6123399)(10.985209,-0.5830473)
\rput[bl](8.32521,0.28195274){\normalsize{$\supp \check{\pi}$}}
\rput[bl](8.52521,-0.2){\normalsize{$\cap \supp (\1-\check{\pi})$}}
\psbezier[linecolor=black, linewidth=0.02, arrowsize=0.05291667cm 2.0,arrowlength=1.4,arrowinset=0.0]{->}(8.16521,0.27695274)(7.9071565,0.30695274)(7.5248556,0.32195273)(7.08521,0.09695274353027344)
\end{pspicture}
}
\caption{Choice of the lens-shaped region~$L$.}
\label{figlens}
\end{figure}%
For the symplectic form we want to take into account the surface layer integral
involving the solutions~$G \u$ and~$G \v$ in the future.
For technical simplicity, it is preferable to work with the ``softened'' surface layer
integral, where for the ``softening'' we work directly with the operator~$\check{\pi}$.
In order to derive the resulting form of the symplectic form, it is helpful to first write the symplectic
form~\eqref{I2asymm} in terms of the linearized field operator~$\Delta$ as
\[ \sigma^\Omega(\u, \v) = \la \chi_\Omega \u, \Delta \big(\chi_{M \setminus \Omega} \v \big) \ra_{L^2(M, d\rho)} 
- \la \chi_{M \setminus \Omega} \u, \Delta \big(\chi_\Omega \v \big) \ra_{L^2(M, d\rho)} \]
(this identity follows from~\eqref{I2asymm} and~\eqref{Lapdef} by a direct computation).
Replacing the characteristic function~$\chi_\Omega$ leads us to the surface layer integral
\begin{align}
\la \check{\pi} &\u, \Delta \big((1-\check{\pi}) \v \big) \ra_{L^2(M, d\rho)} 
- \la (1-\check{\pi}) \u, \Delta \big(\check{\pi} \v \big) \ra_{L^2(M, d\rho)} \label{sy1} \\
&= \int_M d\rho(x) \int_M d\rho(y) \Big( \nabla_{1, \check{\pi} \u} \nabla_{2, (1-\check{\pi}) \v}
- \nabla_{1, (1-\check{\pi}) \u} \nabla_{2, \check{\pi} \v} \Big) \L(x,y) \label{sy2} \\
&\quad\; + \int_M \Big( \nabla_{\check{\pi} \u} \nabla_{(1-\check{\pi}) \v}
- \nabla_{(1-\check{\pi}) \u} \nabla_{\check{\pi} \v} \Big) \ell(x)\: d\rho(x) \:, \label{sy3}
\end{align}
where we used~\eqref{Lapdef} and~\eqref{elldef}.
For clarity, we point out that if~$\check{\pi}$ is replaced by~$\chi_\Omega$, then~\eqref{sy3} vanishes,
whereas~\eqref{sy2} gives back the usual formula for the symplectic form~\eqref{I2asymm}.
We want to evaluate this modified symplectic form for the linearized solutions~$G \u$ and~$G \v$
with~$\u, \v \in \J^*_0$. Using~\eqref{Kdef}, the contributions by the advanced Green's operators
should vanish in the surface layer. In order to avoid the proof of this statement,
we simply {\em{defining}} the symplectic form by taking only the retarded contribution,
\beq \sigma^{\check{\pi}}(G\u, G \v) := 
\la \check{\pi} S^\wedge \u, \Delta \big((1-\check{\pi}) S^\wedge \v \big) \ra_{L^2(M, d\rho)} 
- \la (1-\check{\pi}) S^\wedge \u, \Delta \big(\check{\pi} S^\wedge \v \big) \ra_{L^2(M, d\rho)}\:. \label{sigmadef}
\eeq
The remaining task is to simplify this expression and to show that it is independent of the choices of the
lens-shaped region~$L$ and of the cutoff function~$\check{\pi}$. We begin with a preparatory lemma:

\begin{Lemma} \label{lemmaga} For all~$\u, \v \in \J_0^*$,
\[ \la S^\wedge \u, \v \ra_{L^2(M, d\rho)} = \la \u, S^\vee \v \ra_{L^2(M, d\rho)} \:. \]
\end{Lemma}
\Proof By definition of the Green's operators,
\[ \la S^\wedge \u, \v \ra_{L^2(M, d\rho)} = -\la S^\wedge \u, \Delta S^\vee \v \ra_{L^2(M, d\rho)} \:. \]
Now we can apply the Green's formula (see Lemma~\ref{lemmagreen})
in a sufficiently large lens-shaped region.
Using the support properties of~$S^\vee \u$ and~$S^\wedge \v$, we do not get boundary
terms. Hence
\[ -\la S^\wedge \u, \Delta S^\vee \v \ra_{L^2(M, d\rho)} = -\la \Delta S^\wedge \u, S^\vee \v \ra_{L^2(M, d\rho)}
= \la \u, S^\vee \v \ra_{L^2(M, d\rho)} \:. \]
This concludes the proof.
\QED

\begin{Prp} \label{prpsigma} For all~$\u, \v \in \J_0^*$,
\beq \label{sigmaform}
\sigma^{\check{\pi}}(G\u, G\v) = \la \u, G \,\v \ra_{L^2(M, d\rho)} \:.
\eeq
\end{Prp}
\Proof The first step is to rewrite~\eqref{sigmadef} as a volume integral,
\begin{align*}
\sigma^{\check{\pi}}(G\u, G\v) 
%&= \la \check{\pi} S^\wedge \u, \Delta \big((1-\check{\pi}) S^\wedge \v \big) \ra_{L^2(M, d\rho)} 
%- \la (1-\check{\pi}) S^\wedge \u, \Delta \big(\check{\pi} S^\wedge \v \big) \ra_{L^2(M, d\rho)} \\
&= \la \check{\pi} S^\wedge \u, \Delta S^\wedge \v \ra_{L^2(M, d\rho)} 
- \la  S^\wedge \u, \Delta \big(\check{\pi} S^\wedge \v \big) \ra_{L^2(M, d\rho)} \:.
\end{align*}
Since~$\check{\pi}$ maps to~$\overline{\J_U}^{t_{\max}}$, we can apply the linearized field equations to obtain
\begin{align*}
\sigma^{\check{\pi}}(G\u, G\v) = -\la \check{\pi} S^\wedge \u, \v \ra_{L^2(M, d\rho)} 
+ \la \u, \check{\pi} S^\wedge \v \ra_{L^2(M, d\rho)} \:.
\end{align*}
Using that~$\u$ and~$\v$ vanish on the support of~$\1-\check{\pi}$, it follows that
\[ \sigma^{\check{\pi}}(S^\wedge \u, S^\wedge \v)
= -\la S^\wedge \u, \v \ra_{L^2(M, d\rho)} + \la \u, S^\wedge \v \ra_{L^2(M, d\rho)} \:. \]
Applying Lemma~\ref{lemmaga}, we conclude that
\[ \sigma^{\check{\pi}}(S^\wedge \u, S^\wedge \v)
= - \la \u, S^\vee \v \ra_{L^2(L)} + \la \u, S^\wedge \v \ra_{L^2(L)} 
\overset{\eqref{Kdef}}{=} \la \u, G \,\v \ra_{L^2(L)} \:. \]
Combining this equation with~\eqref{sigmadef} gives the result.
\QED

From~\eqref{sigmaform} one readily sees that the symplectic form does not depend on the
choice of the lens-shaped region~$L$. If one prefers, one can also take~\eqref{sigmaform}
as the {\em{definition}} of the symplectic form.
Obviously, the symplectic form is anti-symmetric in its two arguments,
\[ \sigma^{\check{\pi}}(G\u, G\v) = -\sigma^{\check{\pi}}(G\v, G\u)\:. \]
But we point out that in general it will be degenerate. 
Therefore, in the present context it would be more appropriate to call~$\sigma$
a {\em{presymplectic}} form. It is convenient to also use the standard notation
\[ G(\u, \v) := \la \u, G \,\v \ra_{L^2(M, d\rho)}\:. \]

\section{Discussion and Outlook} \label{secoutlook}
We conclude this paper with a few remarks. The general constructions
of this paper have the purpose of clarifying the underlying analytic and
geometric structures. In order to apply our results in concrete situations, it is a crucial step
to verify the hyperbolicity conditions (see Definitions~\ref{defhypcond} or~\ref{defhypcond2}).
Doing so also involves an appropriate choice of the jet space~$\Jvary$ in~\eqref{Jvarydef}.
Generally speaking, the smaller~$\Jvary$ is chosen, the easier it is to satisfy the hyperbolicity
conditions. The drawback is that the resulting weak solutions are weaker in the sense
that fewer jets are allowed for testing. The correct choice of~$\Jvary$ is not merely a technical
exercise, but it amounts to identifying those degrees of freedom of the system which have
a dynamical behavior in space-time, because only for those degrees of freedom we can hope
to satisfy the hyperbolicity conditions. All the other degrees of freedom must be treated with other,
non-hyperbolic methods. Since these non-hyperbolic methods do not fit to the topic of this
paper, we shall not enter any details but merely illustrate the above considerations by a concrete example.

\begin{Example} {\bf{(treating the scalar component)}} \label{excalzero} {\em{
Suppose we want to apply our methods to electromagnetic fields for Dirac systems
in Minkowski space. In this case, we choose~$\rho$ as the Dirac sea vacuum
regularized on the scale~$\varepsilon$
(for details see~\cite[Section~1.2]{cfs} or~\cite{oppio}).
In~\cite{action} it is shown that the hyperbolicity conditions of Definition~\ref{defhypcond}
are satisfied if we choose~$\Jvary$ for example as the jets~$\J^\text{em}$ generated by smooth electromagnetic
potentials with spatially compact support (for details see~\cite[Section~7]{perturb}).
However, it is has not yet been analyzed whether the scalar jets also satisfy the hyperbolicity
conditions (of either Definition~\ref{defhypcond} or Definition~\ref{defhypcond2}).
Therefore, for the moment the easiest method is to choose~$\Jvary=\J^\text{em}$.
Then our energy methods apply, giving weak solutions~\eqref{weak}.
However, since the test jets are a subset of~$\Jvary$, we are not allowed to test with scalar jets.
In other words, \eqref{weak} does not give us any information on the scalar component of~$\Delta \v$.
This is a major shortcoming, because the scalar component of the linearized field equations is essential
for the conservation laws for surface layer integrals. Therefore, it is important to extend our methods
such as to also satisfy the scalar component of the linearized field equations.

To this end, one can use an {\em{iteration method}}, as we now outline.
The above energy methods gives us a jet~$\v=(0,v)$ with no scalar component.
We now allow for an additional scalar component~$b$ of~$\v$, which we want to choose in such a way that the
scalar component of the linearized field equations holds. Indeed, using the weak EL equations~\eqref{ELtest},
the scalar component of the linearized field equations can be written as
\beq \label{it1}
\int_M \L(x,y)\: b(y) \:d\rho(y) = \int_M D_{2,v} \L(x,y) \:d\rho(y) \:.
\eeq
The integral operator on the left is known to be positive semi-definite (see~\cite[Lem\-ma~3.5]{support}
and~\cite[Remark~4.2]{positive}), and it is strictly positive if restricted to a space
of smooth scalar jets which satisfies~\eqref{Cnontriv}. Then we can invert the integral operator in~\eqref{it1} to determine~$b$.

Clearly, the scalar jet~$b$ also has an effect on the vector component of the
linearized field equations. However, as is worked out in detail in~\cite[Appendix~B.1]{fockbosonic},
both~$b$ and its ``back reaction'' on the vector component of~$\v$ are extremely small because
of scaling factors~$\varepsilon m$ (where~$m$ denotes the rest mass of the Dirac particles).
Therefore, one can apply an iteration method and a fixed-point argument to obtain
the desired weak solution of the linearized field equation~$\v$ for test jets~$\Jtest$
which also include a scalar component and satisfy~\eqref{Cnontriv}.
}} \QEDrem
\end{Example}

We finally discuss the role and significance of {\em{causality}}.
Indeed, in this paper we encountered different notions of causality:
On the level of the causal variational principle, there was a distinction between
timelike and spacelike separation~\eqref{basiccausal}.
When studying the dynamics of linearized waves, on the other hand,
we obtained the structure of past and future cones (see Definition~\ref{defcausal}),
which gave us a transitive relation ``lies in the future of'' (see Theorem~\ref{thmtransitive})
and was compatible with the speed of propagation (see Theorem~\ref{thmspeed}).
This raises the questions: How is this cone structure related to the
causal structure~\eqref{basiccausal}? Are these structures compatible or
are there differences?

The answers to these questions are rather subtle. Before beginning, we point out
that in the so-called continuum limit as worked out in detail in~\cite{cfs},
both the causal structure of~\eqref{basiccausal} as well as the cone structure of
Definition~\ref{defcausal} agree and go over to the causal structure of Minkowski
space. More generally, in~\cite[Section~5]{lqg} it was shown that the
causal structure~\eqref{basiccausal} goes over to the causal structure on a globally
hyperbolic space-time if the ultraviolet regularization is removed by taking the
limit~$\varepsilon \searrow 0$. Therefore, in the limiting case of a classical
space-time in which the linearized field equations go over to linear hyperbolic PDEs,
all the different notions of causality agree.

Clearly, the main interest in the constructions of the present paper lies in the
fact that they also apply to generalized ``quantum space-times'' in which space-time
does {\em{not}} have a manifold structure, and the linearized field equations
can{\em{not}} be expressed in terms of PDEs. In this general setting, the precise connection between
the causal structure in~\eqref{basiccausal} and the cone structures in Definition~\ref{defcausal}
is unclear. We expect that these structures agree ``on the macroscopic scale,'' but at present
there is no mathematically precise formulation of this statement.
In order to explain the connection in some more detail, we note that in the more specific setting of
causal fermion systems, in addition to~\eqref{basiccausal} there is also a
functional~${\mathscr{C}} : M \times M \rightarrow \R$
which distinguishes a time direction (for details see~\cite[\S1.1.2]{cfs})
\[ %\beq \label{tdir}
\left\{ \begin{array}{cl} \text{$y$ lies in the {\em{future}} of~$x$} &\quad \text{if~${\mathscr{C}}(x, y)>0$} \\[0.2em]
\text{$y$ lies in the {\em{past}} of~$x$} &\quad \text{if~${\mathscr{C}}(x, y)<0$}\:. \end{array} \right.
\] %\eeq
Combining this functional with~\eqref{basiccausal}, one could define an alternative cone structure by
\beq \label{coneL}
I^\vee_\L(x) = \big\{ y \in M \:\big|\: \L(x,y)>0 \text{ and } {\mathscr{C}}(x, y)>0 \big\}
\eeq
(where the subscript~$\L$ indicates that this cone structure is induced directly by the Lagrangian).
This definition is easier and more elementary than our previous definition in~\eqref{Iquant}.
However, it is not clear whether it gives rise to transitive causal relations
and whether it is compatible with the propagation speed of linearized solutions.
At present, the only result in this direction are the extensive computations in~\cite{reg}
which indicate that if Dirac sea configurations
in Minkowski space are regularized and the regularization is adjusted such as to satisfy
the EL equations, then the cone structure~\eqref{coneL} does {\em{not}} seem to give rise to
transitive causal relations. But these results seem too special for giving a definitive answer.

The basic difficulty in clarifying the connection between the different cone structures is that
our energy estimates are based on hyperbolicity conditions (see Definitions~\ref{defhypcond}
or~\ref{defhypcond2}) which involve positivity properties of certain surface layer integrals.
These positivity properties should be related to or be a consequence of
the fact that~$\rho$ is a minimizer of the causal variational principle.
But understanding in detail how this connection comes about and how it is related to the
cones in~\eqref{coneL} remains a challenging open problem.

\Thanks{{{\em{Acknowledgments:}} We would like to thank 
Niky Kamran, Igor Khavkine, Johannes Kleiner, Simone Murro, Marco Oppio and Miguel S{\'a}nchez
for helpful discussions. We are grateful to Marco Oppio, Johannes Wurm and the referee for useful comments on the
manuscript.

%\bibliographystyle{amsplain}
%\bibliography{../felix}
\providecommand{\bysame}{\leavevmode\hbox to3em{\hrulefill}\thinspace}
\providecommand{\MR}{\relax\ifhmode\unskip\space\fi MR }
\providecommand{\MRhref}[2]{%
  \href{http://www.ams.org/mathscinet-getitem?mr=#1}{#2}
}
\providecommand{\href}[2]{#2}

\end{document}